\documentclass[ATLAS]{cernphprep}

\usepackage{cernunits,cernchemsym,heppennames2}
\usepackage{cite,fancyvrb}

\usepackage{hyperref}
\graphicspath{{figures/}} 

\usepackage{dcolumn}
\newcolumntype{d}[1]{D{.}{.}{#1}}
\newcolumntype{.}{D{.}{.}{-1}}

\usepackage{stfloats}

\begin{document}
\begin{titlepage}
\PHnumber{2010-024}
\PHdate{29 July 2010}
\DEFCOL{CDS-Library}

\title{Readiness of the ATLAS Tile Calorimeter for LHC collisions}

\Collaboration{The ATLAS Collaboration%
  \thanks{See Appendix~\ref{app_collab} for the list of collaboration members}}
\ShortAuthor{The ATLAS Collaboration}

\begin{abstract}
  The Tile hadronic calorimeter of the ATLAS detector has undergone
  extensive testing in the experimental hall since its installation in
  late 2005.  The readout, control and calibration systems have been
  fully operational since 2007 and the detector has successfully
  collected data from the LHC single beams in 2008 and first
  collisions in 2009.  This paper gives an overview of the Tile
  Calorimeter performance as measured using random triggers,
  calibration data, data from cosmic ray muons and single beam data.
  The detector operation status, noise characteristics and performance
  of the calibration systems are presented, as well as the validation
  of the timing and energy calibration carried out with minimum
  ionising cosmic ray muons data.  The calibration systems' precision
  is well below the design value of 1\,\%. The determination of the
  global energy scale was performed with an uncertainty of 4\,\%.
\end{abstract}
\end{titlepage}

\section{Introduction}
\label{sec:introduction}
The ATLAS Tile Calorimeter (TileCal)~\cite{Ariztizabal1994} is the
barrel hadronic calorimeter of the ATLAS
experiment~\cite{AtlasDetectorPaper} at the CERN Large Hadron
Collider~\cite{LHCpaper}. 
Calorimeters have a primary role in a general-purpose hadron
collider detector. The ATLAS calorimeter system provides
accurate energy and
position measurements of electrons, photons, isolated hadrons, taus and
jets. It also contributes in particle identification and in muon
momentum reconstruction. 
In the barrel part of ATLAS, together with the electromagnetic barrel
calorimeter, TileCal focuses on precise measurements of hadrons, jets,
taus and the missing transverse energy ($E_T^{\mathrm{miss}}$). 
The performance requirements are driven by the ATLAS physics programme: 
\begin{itemize}
\item The energy resolution for jets of $\sigma/E =
  50\,\%/\sqrt{E(\mathrm{GeV})}$ $\oplus 3\,\%$ guarantees good
  sensitivity for measurements of physics processes at the TeV~scale,
  e.g.~quark compositeness and heavy bosons decaying to jets. While
  one cannot separate the individual calorimeter performance issues,
  studies have shown that a random~10\% 
  non-uniformity on the TileCal cells energy response would add
    no more than~1\,\% to the jet energy resolution constant
    term~\cite{TileTDR}.
\item For precision measurements such as the top quark mass, it will be
  desirable to reach a systematic uncertainty on the jet energy scale
  of~1\,\%. Since about a third of the jet transverse energy is deposited in
  TileCal~\cite{tile_dead_cells}, its energy scale uncertainty should
  ultimately be below a 3\% requirement. 
\item The response linearity within 2\,\% up to about 4~TeV is crucial
  for observing new physics phenomena (e.g.~quark compositeness). 
\item A good measurement of 
  $E_T^{\mathrm{miss}}$ is important for many physics signatures, in
  particular for SUSY particle searches and new physics. In addition
  to sufficient total calorimeter thickness and a large coverage in
  pseudorapidity, this very sensitive measurement requires also a
  small fraction of dead detector regions which create fake
  $E_T^{\mathrm{miss}}$. The requirement depends on the signal to
  background ratio of the search.     
\end{itemize}

The Tile Calorimeter has been installed in the experimental hall since
2005 and since then has undergone through several phases of commissioning and
integration in the ATLAS detector system. The main goal of this paper is to
present the outcome of this commissioning phase, at the start of the
LHC collisions 
data-taking. The paper is organised as follows:
Section~\ref{sec:detector} gives a brief description of the Tile
Calorimeter and discusses the overall detector status and the
data-taking conditions after the commissioning was carried
out. Section~\ref{sec:energy_time_reco} presents the method for the
channel signal reconstruction, the overall quality of the detector in
coverage, noise characteristics and conditions stability.
Section~\ref{sec:calibration} shows the details on the three
calibration systems used to set and maintain the cell energy scale and
set the timing offsets, as well as results on the precision and stability of
each system. The related energy scale
uncertainties and the inter-calibration issues are also discussed.  
The last section (Section~\ref{sec:performance}) is devoted to the
validation of the performance using data from cosmic muons produced in
cosmic ray showers in the atmosphere, referred to in short form 
throughout this paper as ``cosmic muons'' or ``cosmic ray muons''.
Results are presented 
on energy and time reconstruction, uniformity across the calorimeter and 
comparison with Monte Carlo simulations. A subsection  
is devoted to the intercalibration of the scintillators that are located in 
the gap between barrel and extended barrels.
%
%

%
%
\section{Detector and data taking setup}\label{sec:detector}

\subsection{Overview of the Tile Calorimeter}\label{sec:overview}

\begin{figure*}[t]
  \begin{center}
    \resizebox{0.75\textwidth}{!}{%
      \includegraphics{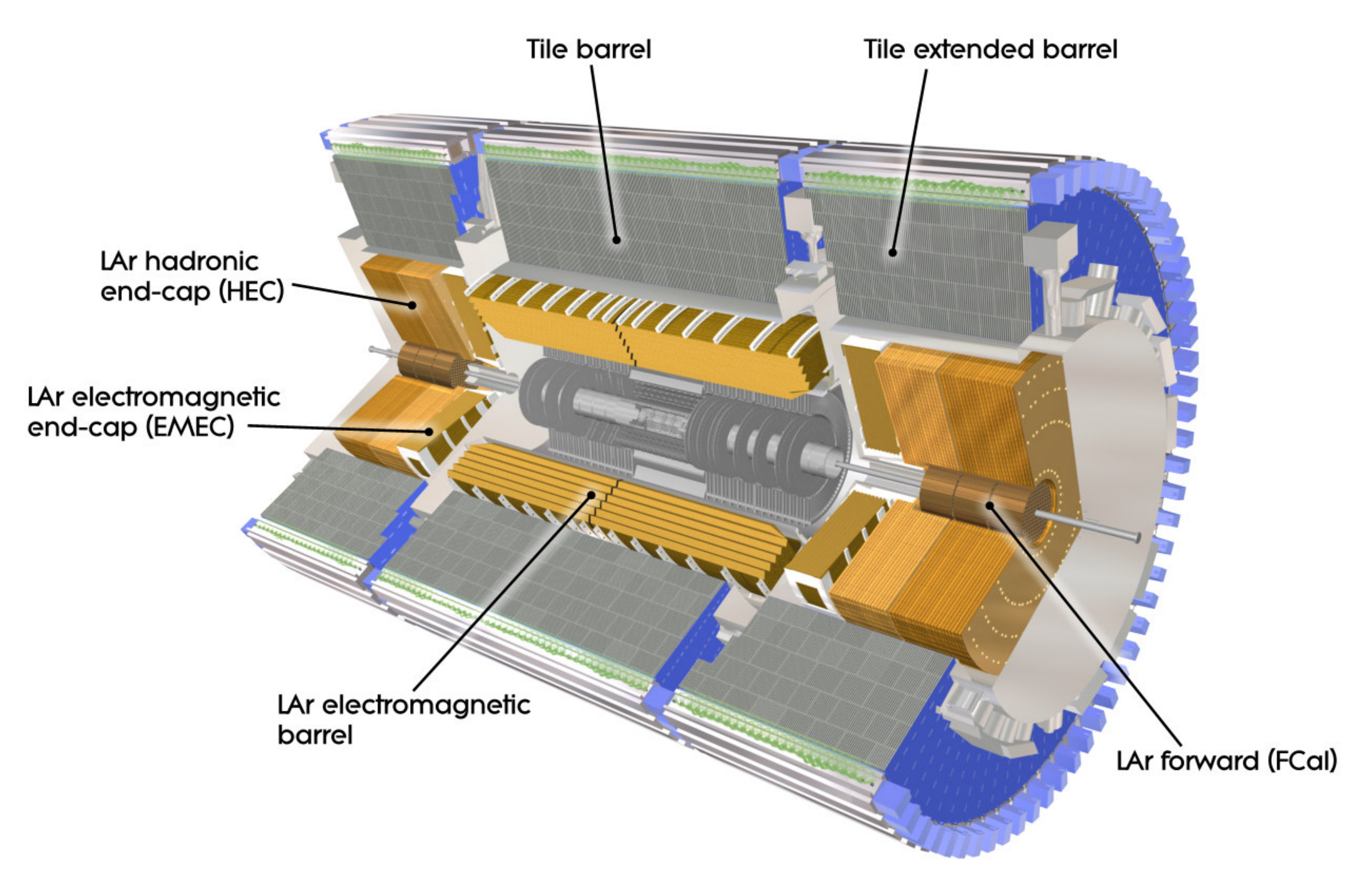}
    }
  \end{center}
  \caption{A cut-away drawing of the ATLAS inner detector and
    calorimeters. The Tile Calorimeter consists of one barrel
    and two extended barrel sections and surrounds the Liquid Argon barrel
    electromagnetic and endcap hadronic calorimeters. In the innermost radii
    of ATLAS, the inner detector (shown in grey) is used for precision
    tracking of charged particles.}
  \label{fig:tile-overview}
\end{figure*}

\begin{figure*}[t]
  \begin{center}
    \resizebox{0.75\textwidth}{!}{%
      \includegraphics{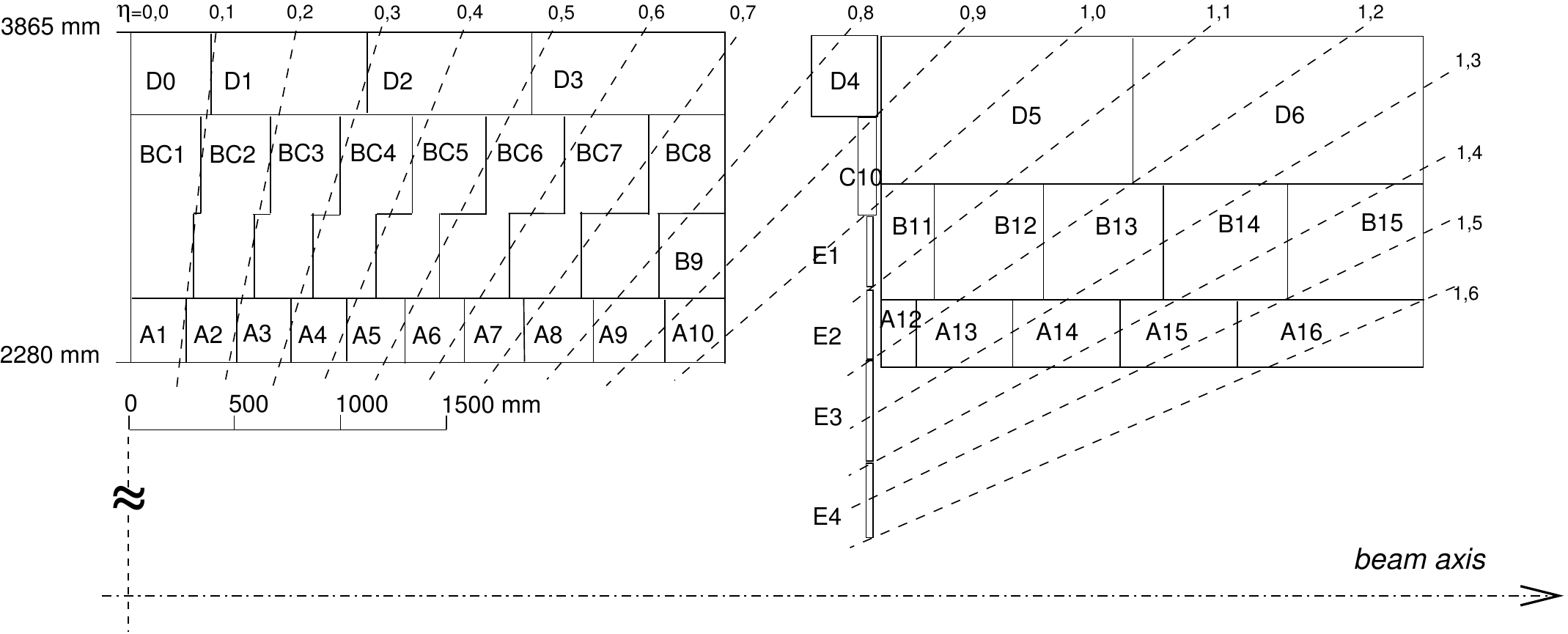}
    }
  \end{center}
  \caption{Segmentation in depth and $\eta$~of the Tile Calorimeter
    modules in the barrel (left) and extended barrel (right). The
    bottom of the picture corresponds to the inner radius of the
    cylinder. The Tile 
    Calorimeter is symmetric with respect to the interaction point. 
    The cells between two consecutive dashed lines form the
    first level trigger calorimeter tower.
  }
  \label{fig:tile-cellmap}
\end{figure*}

\begin{figure}[htb]
  \begin{center}
    \resizebox{0.5\textwidth}{!}{%
      \includegraphics{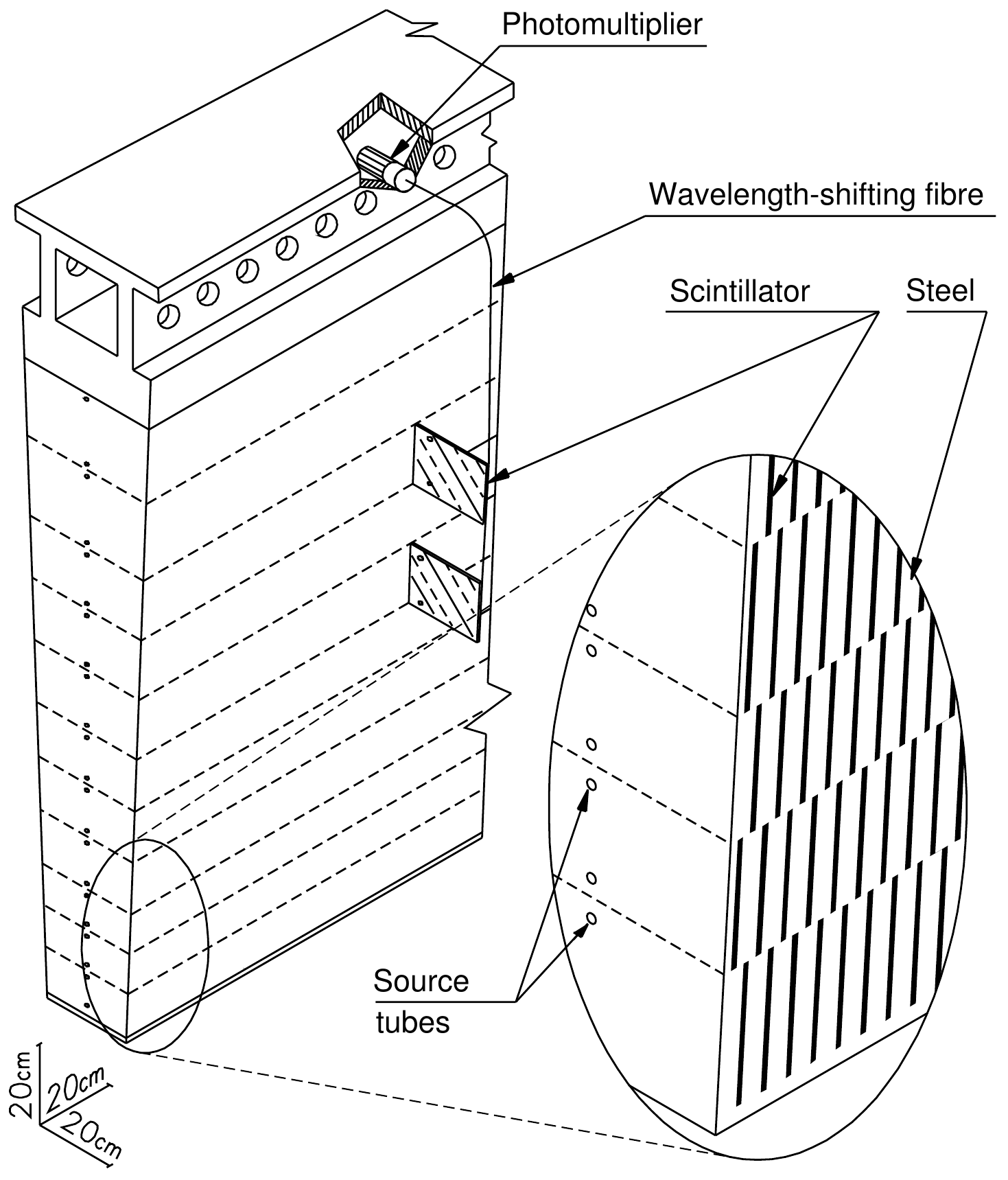}
    }
  \end{center}
  \caption{Schematic showing the mechanical assembly and the optical readout of the Tile Calorimeter, corresponding to a $\phi$ wedge. The various components of the optical readout, namely the tiles, the fibres and the photomultipliers, are shown. The trapezoidal scintillating tiles are oriented radially and normal to the beam line and are read out by fibres coupled to their non-parallel sides.  
  \label{fig:tile-module}}
\end{figure}

TileCal is a large hadronic sampling
calorimeter using plastic scintillator as the active material
and low-carbon steel (iron) as the
absorber. Spanning the pseudorapidity\footnote{The pseudorapidity
  $\eta$ is defined as $\eta = -\ln\left(\tan{\frac{\theta}{2}}\right)$, where
  $\theta$ is the polar angle measured from the beam axis. The
  azimuthal angle $\phi$ is measured around the beam axis, with
  positive (negative) values corresponding to the top (bottom) part
  of the detector.} region
$-1.7<\eta<1.7$, the 
calorimeter is sub-divided into the barrel, also called long barrel
(LB), in the central region ($-1.0<\eta<1.0$) and the two 
extended barrels (EB) that flank it on both sides ($0.8<|\eta|<1.7$),
as shown in Figure~\ref{fig:tile-overview}.  
Both the barrel and extended barrel
cylinders are segmented into 64 wedges (modules) in $\phi$, corresponding 
to a $\Delta \phi$ granularity of $\sim0.1$ radians.
Radially, each module is 
further segmented into three layers which are approximately 
1.5, 4.1 and 1.8 $\lambda$ (nuclear interaction length for protons) thick for the barrel 
and 1.5, 2.6 and 3.3 for the extended barrel. The $\Delta \eta$ segmentation 
for each module is 0.1 in the first two radial layers and 0.2 in the third 
layer (Figure~\ref{fig:tile-cellmap}). 
The $\phi$, $\eta$ and radial segmentation define the three
dimensional TileCal cells. Each cell volume is made of dozens of iron
plates and scintillating tiles. 
Wavelength shifting fibres coupled to the tiles on either $\phi$ edge
of the cells, as shown in figure~\ref{fig:tile-module}, collect the
produced light and are read out via square light guides
by two different photomultiplier
tubes (PMTs), each linked to one readout channel.
Light attenuation in the scintillating tiles themselves would cause a
response non-uniformity of up to 40\,\% in the case of a single
readout, for particles entering at different impact positions across
$\phi$. The double readout improves the response uniformity to within
a few percent, in addition to providing redundancy. 

In addition to the standard cells, the Intermediate Tile Calorimeter (ITC) covers the region $0.8 < \eta < 1.0$ (labelled D4 and C10 in Figure~\ref{fig:tile-cellmap}). To
accommodate services and readout electronics for other ATLAS detector
systems, several of the ITC cells have a special construction: 
per side, three  D4 cells have reduced thickness and eight C10 cells are plain
scintillator plates. Located on the remaining, inner radius surface of
the extended barrel modules, the gap scintillators cover the region of
$1.0 < \eta < 1.2$ (labelled E1 and E2 in the figure), while the
crack scintillators are located on the front of the Liquid Argon
endcap and cover the region $1.2< \eta < 1.6$ (labelled E3 and E4). 
 
In the present (initial) configuration, eight pairs of crack
scintillators have 
been removed to permit routing of signal cables
from the 16 Minimum 
Bias Trigger Scintillators (MBTS), in each side.
Located on the front face of the Liquid Argon end-cap cryostat, the MBTS
span an $\eta$ range of $2.12<|\eta|<3.85$ and are readout by the TileCal
EB electronics. 
They are used mainly for triggering on collisions in the very early
stage of LHC operation and for
rate measurements of halo muons, beam-gas and minimum bias events during
the low-luminosity running.


The Tile Calorimeter readout architecture divides the detector in four
partitions, a definition that is widely used in this paper. 
The barrel is divided in two partitions (LBA and LBC) by the plane
perpendicular to the beam line and crossing the interaction point, and
each of the two extended barrels is a separate partition (EBA and
EBC). 
     
The TileCal readout electronics is contained in ``drawers'' which
slide into the structural girders at the outer radius of the
calorimeter.  Barrel modules are read out by two drawers (one inserted
from each face) and extended barrel modules are read out by one drawer
each. 
Each drawer typically contains 45 (32) readout channels in the barrel
(extended barrel) and a summary of the channels, cells and trigger
outputs in TileCal is shown in
Table~\ref{tbl:number-of-cells}.\footnote{%
  The 16 reduced thickness extended barrel C10 cells are readout by only
  one PMT\@. Two extended barrel D4 cells are merged with the
  corresponding D5 cells and have a common readout.}

\begin{table*}[t]
  \begin{center}
    \begin{tabular}{|c|c|c|c|}
      \hline
      & Channels & Cells & Trigger Outputs\\
      \hline
      Long~barrel     & 5760 & 2880 & 1152  \\
      Extended~barrel & 3564 & 1790 & 768 \\
      Gap~and~crack   & 480 & 480 &  128\\
      MBTS            & 32 & 32 &  32\\
      \hline 
      Total & 9836 &  5182 & 2080 \\
      \hline
    \end{tabular}
  \end{center}
  \caption{Number of channels, cells and trigger outputs of the Tile
    Calorimeter. The gap and crack and MBTS channels are readout
    in the extended barrel drawers.}
  \label{tbl:number-of-cells}
\end{table*}

 
The front-end electronics as well as the drawers' Low Voltage Power
Supplies (LVPS) are located on the calorimeter itself and 
are designed to operate under the conditions of magnetic fields and 
radiation. 
One drawer with its LVPS reads out a region of
$\Delta\eta \times \Delta\phi = 0.8 \times 0.1$ in the barrel and $0.7 \times 0.1$ in the extended barrel.

In the electronics readout, the signals from the PMT are
first shaped using a passive shaping circuit. The shaped pulse is
amplified in separate high (HG) and low (LG) gain branches, with a
nominal gain ratio of 64:1. The shaper, the charge injection
calibration system (CIS), and the gain splitting are all located on a
small printed circuit board known as the 3-in-1 card~\cite{3in1}. The
HG and LG signals are sampled with the LHC bunch-crossing frequency of 40~MHz using a
10-bit ADC in the Tile Data Management Unit (DMU) chip which is
located on the digitiser board~\cite{digitizer}. This chip contains a
pipeline memory that stores the sampled data for up to 6.4~$\mu$s. The
pipeline memory can be adjusted in coarse timing steps of 25~ns. 
The digitisation timing of the ADCs can be adjusted in multiples of
$\sim0.1$~ns so that the central sample is as close to the PMT pulse
peak as possible and to make sure the full extension of the pulse is
sampled. However, this adjustment is possible only for groups of six
channels, so a residual offset remains, that must be dealt with at
the signal reconstruction level (see Section~\ref{sec:ene_time_reco}). 
Due to bandwidth requirements, only seven samples from one gain are
read out from the front-end electronics. A gain switch is used to
determine if the high or low gain is sent. The digitised samples are
sent via optical fibres to the back-end electronics which are located
outside the experimental hall. From the digitised samples, the
back-end electronics determine the time and energy of the channel's
signal as described in Section~\ref{sec:ene_time_reco}.  

In addition to the digital readout of the PMT signal, a millisecond-timescale integrator circuit is also located on the 3-in-1 card. The Tile integrator is designed to measure the PMT current during $^{137}$Cs calibrations (see Section~\ref{sec:calibration}) and also to measure the current from minimum bias proton-proton interactions at the LHC\@. The integration period is approximately 14~ms and a 12-bit ADC is used for the readout. 

Adder boards are distributed along the drawer. Each adder board
receives the analogue signals from up to six 3-in-1 cards
corresponding to cells of the same $\eta$. 
The trigger signal corresponding to a ``tower'' (see
Fig.~\ref{fig:tile-cellmap}) of cells with
$\Delta\eta\times\Delta\phi = 0.1\times 0.1$ is formed by an analogue
sum of the input signals and, together with the signals from the other
calorimeters, are sent via long cables to the Level-1 (L1) calorimeter
trigger system to identify jets, taus, total calorimeter energy and
$E_T^{\mathrm{miss}}$ signatures. 
The signal from all four gap and crack scintillators is also
summed by the adder board and passed to the L1 calorimeter trigger.
A second output of the adder boards (so-called muon output), that
can be used at a later stage to reduce the muon background rates,
contains only the signal from cells of the outermost calorimeter layer.
Presently a fraction of the muon outputs is used
for carrying the MBTS signals to the L1 trigger system.

\subsection{Detector and data taking overview}
\label{sec:datataking}
The detector performance and stability results exposed in this paper are based on calibration systems' data and random triggered events which cover extended periods from mid-2008 up to the end of 2009 excluding the maintenance period between December 2008 and May 2009.  
The results from cosmic muons and single beam are from the autumn 2008
data-taking period, with the exception of the single beam data for
timing studies, for which the winter 2009 and spring 2010 data is also used.  

The Tile Calorimeter at the end of 2008 data-taking period
was fully operational with approximately 1.5\,\% dead cells. The majority of the dead cells were due to three drawers that were non-operational because of 
power supply problems or data corruption, amounting to
60~cells or 1.2\,\%.  The remaining dead cells were randomly distributed
throughout TileCal. During the 2009 data-taking period there were 48 unusable cells, fewer than 1\,\%. 
The number of dead L1 trigger towers is less than 0.5\,\% and they
are uniformly distributed throughout the detector. For details on how
non-operational cells are defined and the breakdown of their problems
for the 2009 data-taking, see
Section~\ref{sec:detector_data_quality_status}.  

The cosmic data used for performance validation was collected mainly between September and
October 2008 using the full ATLAS
detector, including the inner detector and muon systems, with around
one million events used for the present paper. The cosmic
trigger configuration during this run period consisted of L1 triggers
from the muon spectrometer\footnote{See
  Ref.~\cite{AtlasDetectorPaper}, Fig.~1.4, for details on the
  layout.} (both the Resistive Plate Chamber (RPC) and 
the Thin Gap Chambers (TGC)), the L1 calorimeter trigger and the
MBTS\@. For much of the cosmic ray analysis discussed in
Section~\ref{sec:performance}, the data sample was selected by
requiring a L1 trigger and at least one track reconstructed in the
inner detector, from the Pixel, SemiConductor Tracker (SCT) and
Transition Radiation Tracker (TRT).\footnote{See
  Ref.~\cite{AtlasDetectorPaper}, 
  Fig.~1.1, for details on the layout.}
The majority of the events came from the L1 muon spectrometer triggers. 
During this running period, the ATLAS magnets were run in four different configurations; no magnetic field, solenoid magnet on
only, toroid magnet on only and both solenoid and toroid magnets on. The results exposed here were obtained with the full ATLAS fields on. 

From the single beam data used in this paper the "splash" events and "scraping" events are used for time and energy studies. The former term is used for events occurring when the LHC beam hits the closed tertiary collimators positioned 140~m up-stream of the detector and are characterised by millions of high-energy particles arriving simultaneously in the ATLAS detector. The latter occur when the open collimators are scraping the LHC beam, allowing a moderate number of particles to the detector. 

%

%
%

%
%
\section{Detector performance and signal reconstruction}
\label{sec:energy_time_reco}

\subsection{Detector and data quality status overview}
\label{sec:detector_data_quality_status}

The TileCal detector operated at the end of 2009 with 99.1\,\% of cells
functional for the digital readout and 99.7\,\% of trigger towers
functional for the L1. The numbers and fractions of non-operational
cells, channels and trigger towers in the four calorimeter partitions
are shown in Table~\ref{tbl:dead-cells}. 

\begin{table*}
  \begin{center}
    \begin{tabular}{|c|c|c|c|}
      \hline
      Partition & Masked Channels & Masked Cells & Dead Trigger Towers\\
      \hline
      Barrel A-side & 59 (2.05\%) & 23 (1.60\%) & 2 (0.3\%) \\
      Barrel C-side & 58 (2.01\%) & 25 (1.74\%)  & 0 (0.0\%) \\
      Ext.~barrel A-side & 6 (0.29\%) & 0 (0.00\%) & 2 (0.5\%) \\
      Ext.~barrel C-side & 1 (0.05\%) & 0 (0.00\%) & 1 (0.3\%) \\
      \hline 
      Total & 124 (1.26\%) & 48 (0.93\%) & 5 (0.3\%) \\
      \hline
    \end{tabular}
  \end{center}
  \caption{Summary of the number of masked channels and cells in TileCal
    as of November 9th, 2009. The number of dead trigger towers quoted
    is towers that are non-operational due to problems in TileCal's
    front-end electronics, not counting those related to LVPS (18 towers).}
  \label{tbl:dead-cells}
\end{table*}

The problematic channels belong to
two categories: channels with fatal problems and channels with data
quality problems. The so-called fatal problems are channels deemed
unusable and are masked for the offline reconstruction and at the High
Level Trigger (HLT). These channels include: 
\begin{enumerate}
\item 44 cells (88 channels) due to two drawers with non-functional LVPS.
\item 10 channels with no response due to failures of one or more components in the readout chain, such as 3-in-1 cards, PMTs or ADCs.
\item 24 channels with digital data errors (17 channels with a high occurrence rate of corrupted data and 7 with gain switching problems).
\item 2 channels with high noise 
\end{enumerate}
 
\begin{figure}
\centering
\resizebox{0.5\textwidth}{!}{%
  \includegraphics{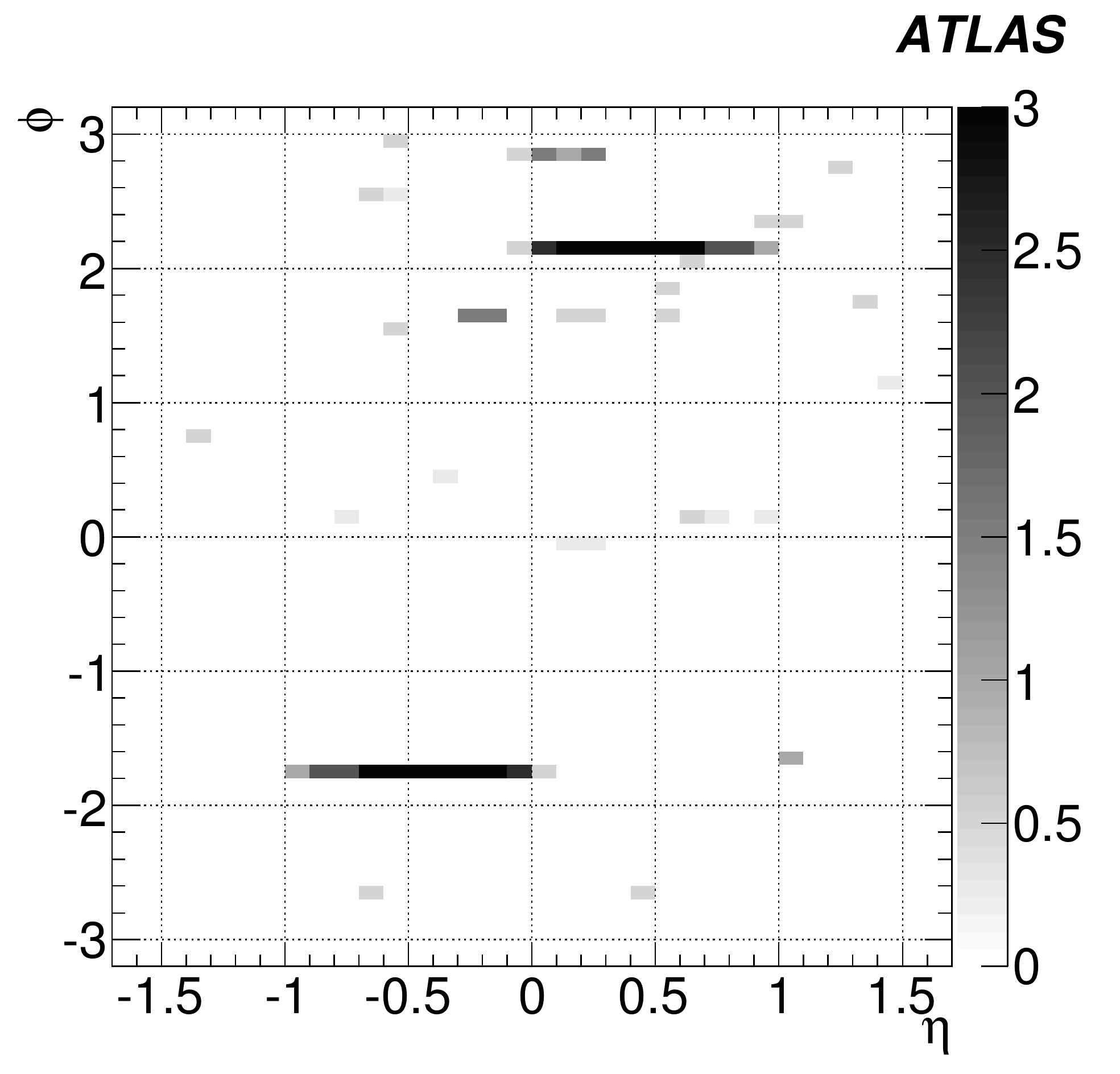}
}
\caption{Position in $\eta$ and $\phi$ of the masked cells representing the status on November 9th, 2009. The colours corresponding to numbers 1,2,3 show the number of layers masked for this ($\eta,\phi$) region. The non-integer numbers indicate that one readout channel of the cell is masked. }
\label{badch_etaphi}
\end{figure}

 
The position in ($\eta,\phi$) as of November 2009 of the unusable
masked cells as described above, are shown in
Figure~\ref{badch_etaphi} and are summarised in
Table~\ref{tbl:dead-cells}. One can notice the majority of the masked
cells concentrated in two non-functional front-end drawers.

Channels with data quality problems are flagged as such for the reconstruction, but they are not masked. These channels include:
\begin{enumerate}
\item Channels with occasional data-corruption problems, mainly due to
  front-end electronics malfunction or bad configuration. These are
  excluded from the reconstruction by checking a quality fragment in
  the data record on an event by event basis. A fraction of the channels
  can be recovered by resetting the front-end between LHC fills. 
\item Channels which cannot be calibrated with one of the calibration
  systems (see Section~\ref{sec:calibration}). These are
  flagged as poorly calibrated channels. 
\item Noisy channels, which are treated by describing appropriately in
  the database their higher-than-average noise level to take into
  account while reconstructing their energy. 
\item Channels where the response varies significantly over
  time. These are also flagged for the offline use as poor quality
  channels but their response can be corrected over time if the source
  of variation is understood. Typical cases include channels with
  varying response due to changes over time of the high voltage
  applied to the photomultipliers. 
\end{enumerate}

\begin{figure*}[t]
\centering
\resizebox{0.49\textwidth}{!}{%
  \includegraphics{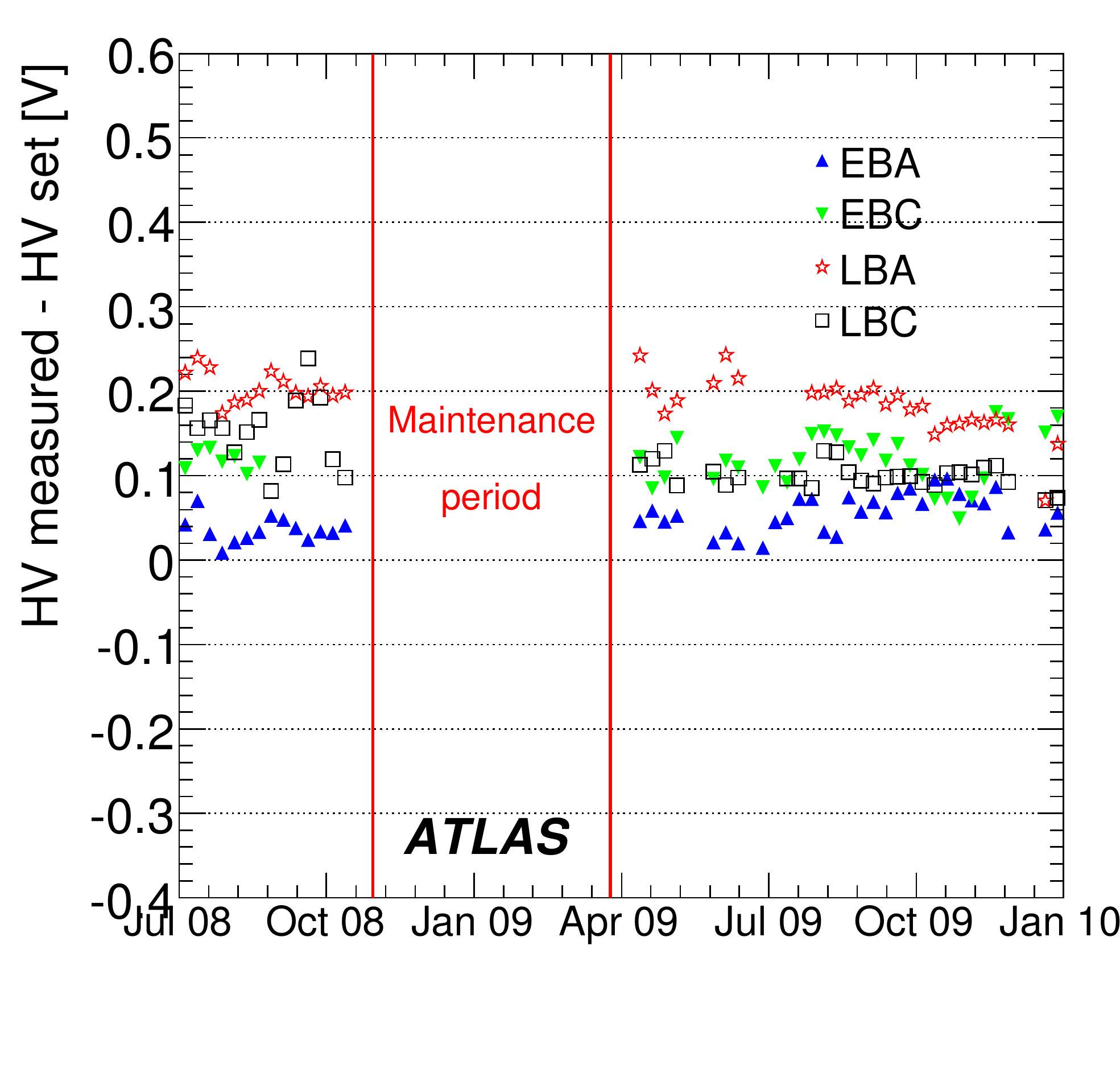}
}%
\resizebox{0.49\textwidth}{!}{%
  \includegraphics{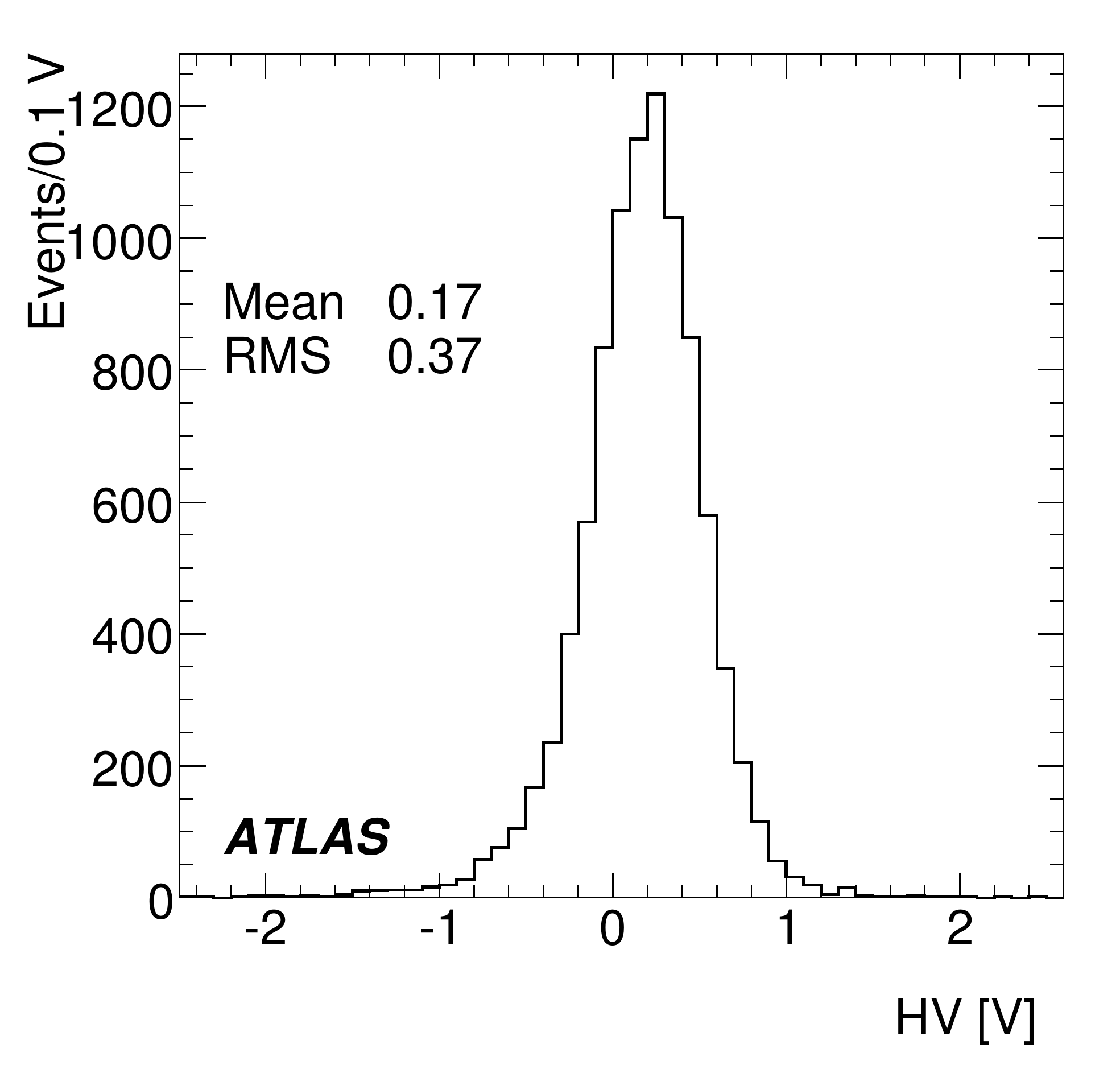}
}
\caption{Stability of the PMT high voltage with respect to its set value,
  averaging over all PMTs for two periods of 3 and 6 months
  (left) separated by the maintenance period. The distribution of the differences of
  the measured and the set HV values for all PMTs over the period considered is also shown (right).} 
\label{HVstab}
\end{figure*}

\begin{figure*}[t]
\centering
\resizebox{0.49\textwidth}{!}{%
  \includegraphics{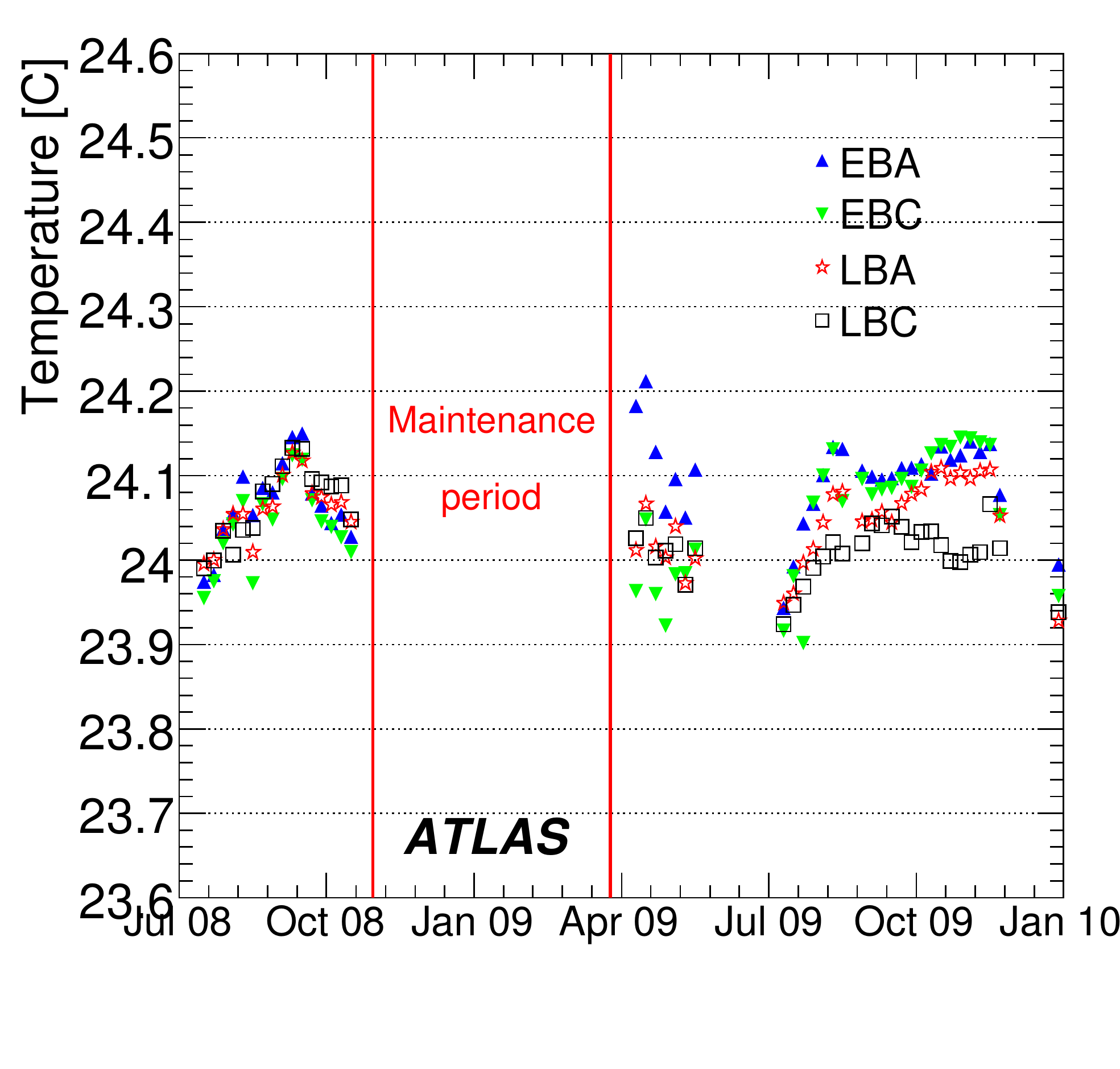}
}%
\resizebox{0.49\textwidth}{!}{%
  \includegraphics{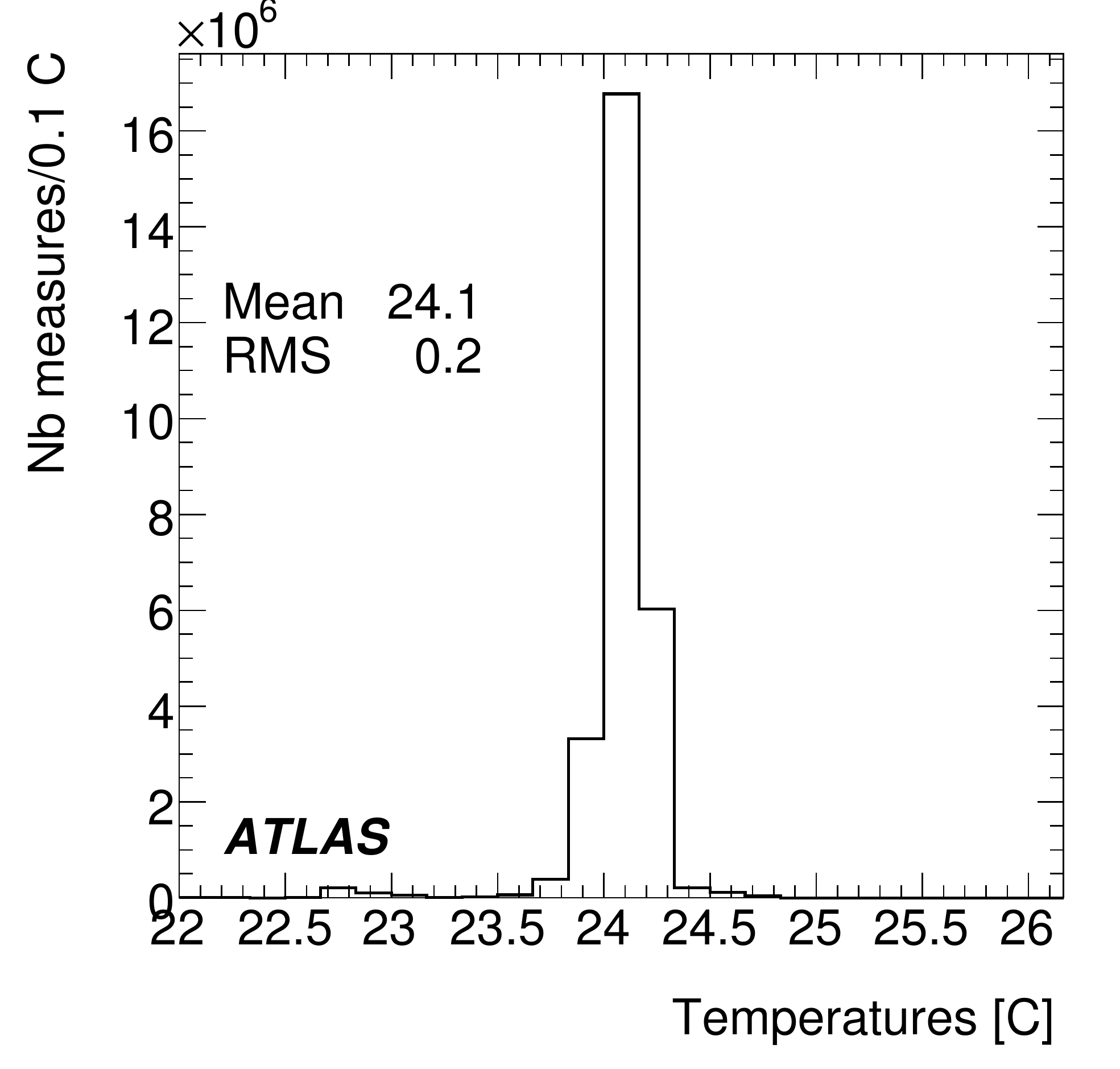}
}
\caption{Stability of the temperature, as measured at one PMT in each drawer, averaging over all drawers and presented for two periods of 3 and 6 months separated by the maintenance period (left). The distribution of the values for individual drawers over the whole period is also shown (right).} 
\label{Tempstab}
\end{figure*}

The parameters that directly affect the measured response of a channel
are the temperature in the drawer and the applied high voltage because the PMT gain depends
on them. The PMT gain $G$ is proportional to $V^7$, where $V$ is the
applied high voltage (HV),
and decreases with temperature by 0.2\,\% per ${}^\circ$C.
The operating conditions of the detector have been constantly monitored online and
recorded by the Detector Control System (DCS). The operating values of voltages,
currents, temperatures at the LVPS and at the front-end have been
very stable. 
Figure~\ref{HVstab} gives a measure of the long term evolution of
the high voltage applied on the PMTs for two periods of 3 and 6 months
separated by the maintenance period. The HV values, which are typically 
close to $\sim670$~V, have shown on
average a difference of 0.17~V with respect to the value set during
intercalibration with an RMS of 0.37~V during the considered
period. This average stability within 0.4~V for the  
whole calorimeter represents a 0.4\,\% reproducibility in the gain of the
PMTs due to this factor alone. Figure~\ref{Tempstab} shows the stability of the temperature
measured by a 
probe installed in one PMT block for the same period as for the HV
measurements. The average over all the calorimeter PMT probes is
$24.1^\circ$C with an RMS of $0.2^\circ$C for a period of 9 months interleaved by the maintenance period.  

\subsection{Energy and time reconstruction} 
\label{sec:ene_time_reco}

\begin{figure}
\centering
\resizebox{0.5\textwidth}{!}{%
  \includegraphics{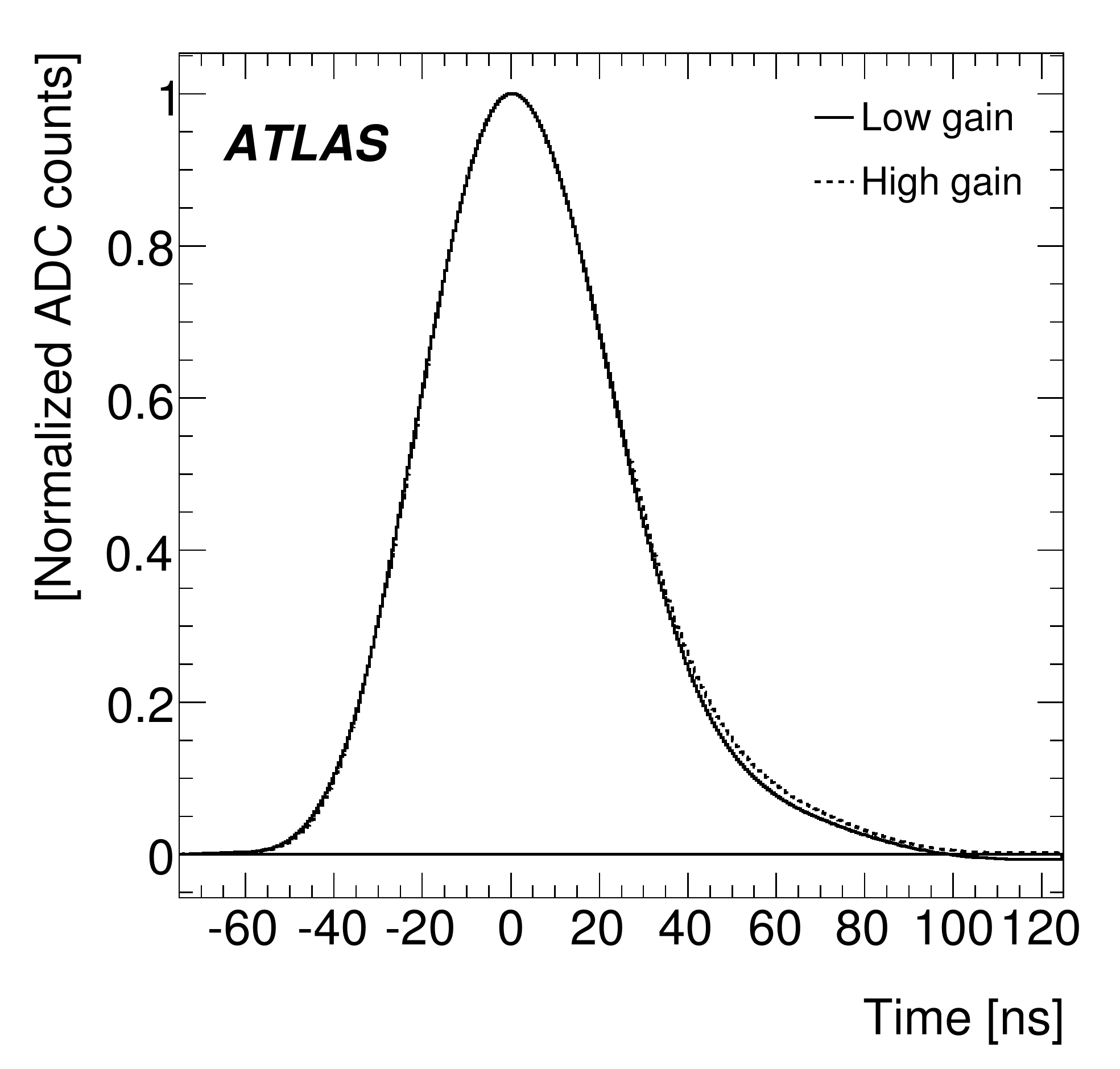}
}
\caption{Pulse shape for high and low gain from testbeam data, used as reference
  for the OF weights calculation.} 
\label{fig:pulse_shape}
\end{figure}

The channel signal properties -- pulse amplitude, time and pedestal --
for all TileCal channels are reconstructed with the Optimal Filtering
(OF) 
method~\cite{OfValidation}, which makes use of
weighted linear combinations of the digitised signal samples (spaced
by 25~ns). Due to the simplicity of its mathematical formulation, OF
is implemented in the Digital Signal Processors (DSPs) of the ReadOut
Driver boards (RODs)~\cite{RodProduction} and therefore provides
energy and time information to the HLT of ATLAS during the online
data-taking. 
At present, since the data-taking rate allows it, the seven digitised
samples are also available offline for all the events together with
the results of the OF reconstruction from the RODs. 
The procedure to compute the energy (given by the amplitude $A$) and time ($\tau$) are given by the equations:
\begin{equation}
\begin{array}{lcr}
\displaystyle A=\sum^{n=7}_{i=1}{a_{i} S_{i}}
&
\,\,\,\,\,\,\displaystyle \tau=\frac{1}{A}\sum^{n=7}_{i=1}{b_{i} S_{i}}
\end{array}
\label{eq:of}
\end{equation}
where $S_{i}$ is the sample taken at time $t_{i}$ ($i=1\ldots n$). The coefficients of
these combinations, $a_{i}$ and $b_{i}$, known as the OF
weights, are obtained from knowledge of the pulse shape 
and noise autocorrelation matrix, and are chosen in such a way
that the impact of the noise to the calorimeter resolution is minimised.
Figure~\ref{fig:pulse_shape} shows the pulse shape extracted from data
taken at the testbeam, selecting a channel with a given value of deposited energy for each gain. This pulse shape is the reference used in the estimation of the OF weights. 
 
The reconstructed channel energy used by the HLT and offline is:
\begin{equation}
E_{channel} = A \cdot C_{ADC \rightarrow pC} \cdot C_{pC \rightarrow GeV} \cdot C_{Cs} \cdot C_{Laser}
\label{eq:echannel}
\end{equation}
The signal amplitude $A$, described in more detail above, represents
the measured energy in ADC counts as in Eq.~(\ref{eq:of}). The factor
$C_{ADC \rightarrow pC}$ is the conversion factor of ADC counts
to charge and it is determined for each channel using a well
defined injected charge 
with the CIS (Charge Injection System) calibration system.  
The factor $C_{pC \rightarrow GeV}$ is the conversion
factor of charge to 
energy in GeV and it has been defined at testbeam for a subset of
modules via the response to electron beams of known momentum in the
first radial layer. This factor is globally applied to all cells after
being adjusted for a dependence on the radial layer
(see Section~\ref{sec:intercalibration}).
%
The factor $C_{Cs}$ corrects for residual non-uniformities after the gain
equalisation of all channels has been performed by the Cs radioactive
source system. 
The factor $C_{Laser}$, not currently implemented, corrects for
non-linearities of the PMT response measured by the Laser calibration
system. The derived time dependence of the last two factors will be
applied to preserve the energy scale of TileCal. The details of the
calibration procedures are discussed in
Section~\ref{sec:calibration}. 

The channel time, $\tau$ in equation~(\ref{eq:of}), is the time difference
between the peak of the reconstructed pulse and the peak of the reference pulse.
The OF weights used in the reconstruction were calculated based on this reference pulse shifted by a time phase that depends on each channel's timing offsets measured
with the calibration systems (and single-beam data), the
time-of-flight from the interaction point to that cell and the
hardware time adjustments mentioned in Section~\ref{sec:overview}. Thus the
reconstructed time $\tau$ should be compatible with zero for energy
depositions coming from the interaction point. 
If the time residual is not well known, for small deviations ($|\tau|
< 15$~ns) the uncertainty of the reconstructed amplitude depends on $\tau$
through a well-defined parabolic function, that can be used for an
energy correction at the level of the HLT or offline reconstruction. 

\begin{figure}
\centering
\resizebox{0.5\textwidth}{!}{%
  \includegraphics{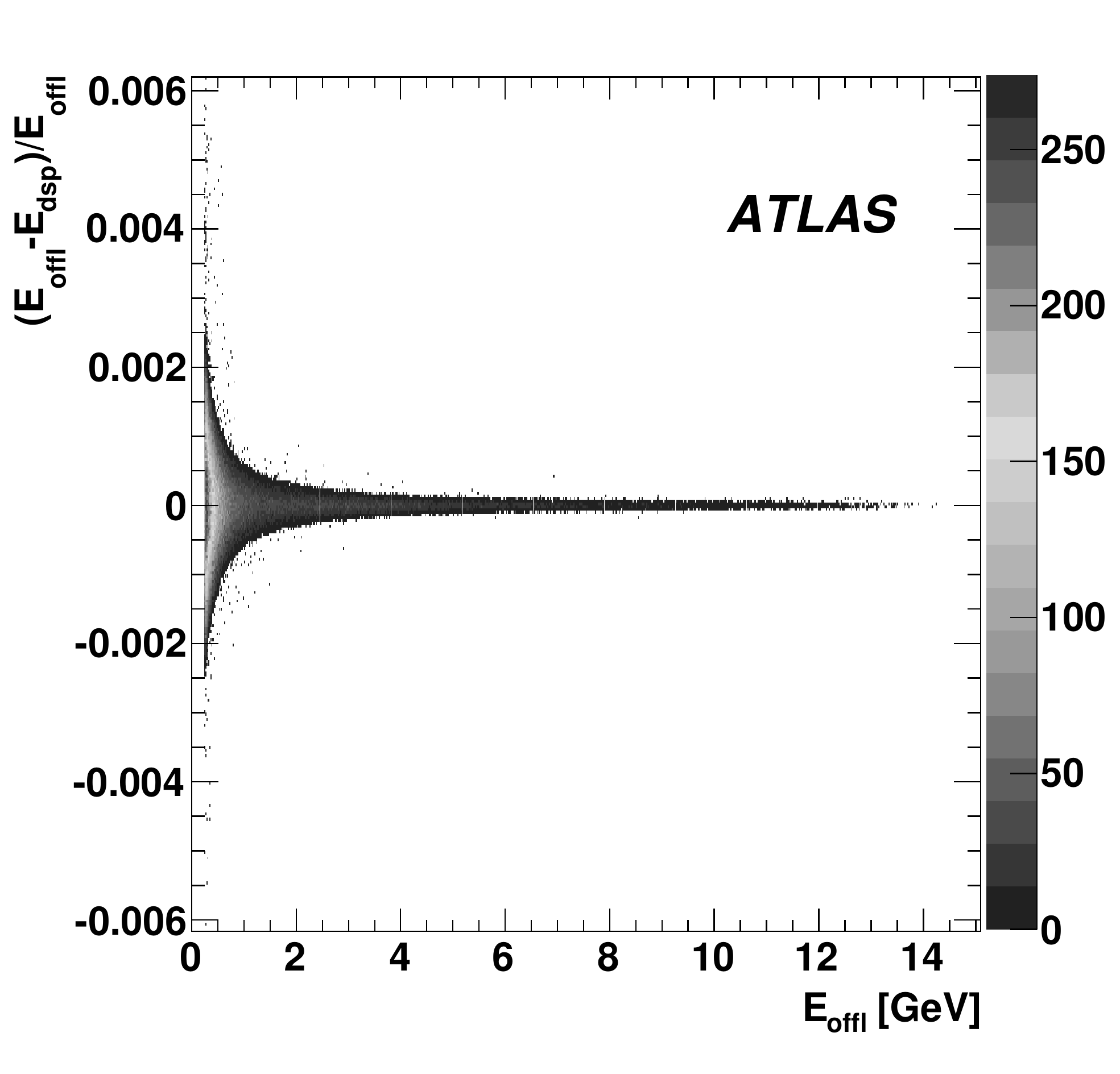}
}
\caption{Difference between the reconstructed offline energy, $E_{offl}$, and the energy given by the DSP $E_{DSP}$ relative to $E_{offl}$ and as a function of $E_{offl}$ (in GeV), extracted from cosmic muon runs.}
\label{fig:plot1}
\end{figure}

The OF results rely on having, for each channel, a fixed and known
time phase between the pulse peak and the 40~MHz LHC clock
signal. This is not the case during the commissioning phase of the
detector, where signals caused by cosmic rays are completely
asynchronous with respect to the LHC clock. 
Nevertheless OF can still be applied in this case and an accurate reconstruction may be 
obtained by applying the proper weights for each event according to the time position of the signal. The estimation of the signal time  
is achieved through an iterative procedure provided by a set of OF weights calculated at different phases
from $-75$~ns to $+75$~ns in steps of 1~ns. 
Figure~\ref{fig:plot1} presents the relative difference between the
reconstructed offline energy and the energy calculated in the DSPs for
cosmic muon data and shows the effect of the limited numerical precision of the DSPs.
The results in the following sections are based on channel
energies reconstructed offline with the iterative procedure to define
the phase.  

The Fit method is another signal reconstruction algorithm. It
is based on a three parameter fit to the known pulse shape function
$g(t)$, as expressed by:
\begin{equation}
S_i = Ag(t_i - \tau) + ped
\label{fit}
\end{equation}
The meaning of the variables $S_i$ and $t_i$ and the parameters $A$ and $\tau$ is the same
as for the OF method, while $ped$ is a free parameter that defines the baseline of the pulse. The Fit method is
mathematically equivalent to OF in the absence of pile-up and noise, but it is not
suitable for fast online signal processing in DSPs. Results
from the Fit and OF methods were compared with testbeam data and were
found to be equivalent~\cite{TileTBpaper}. Since the autumn of 2008
data-taking, the Fit method is used only for 
CIS calibration data, where the pulse is a
superposition of charge-proportional and charge-independent 
components~\cite{TileTBpaper}.

The cell energy is the sum, and the cell time the average, of the
respective measurements by the two corresponding readout channels. In
cases of single readout cells, or if one of the channels is
masked out, the cell energy is twice the energy measured in the single
available channel.  
The measurement of the cell's energy is thus robust to 
failures in a single readout channel.



\subsection{Noise performance}
\begin{figure}
\centering
\resizebox{0.5\textwidth}{!}{%
  \includegraphics{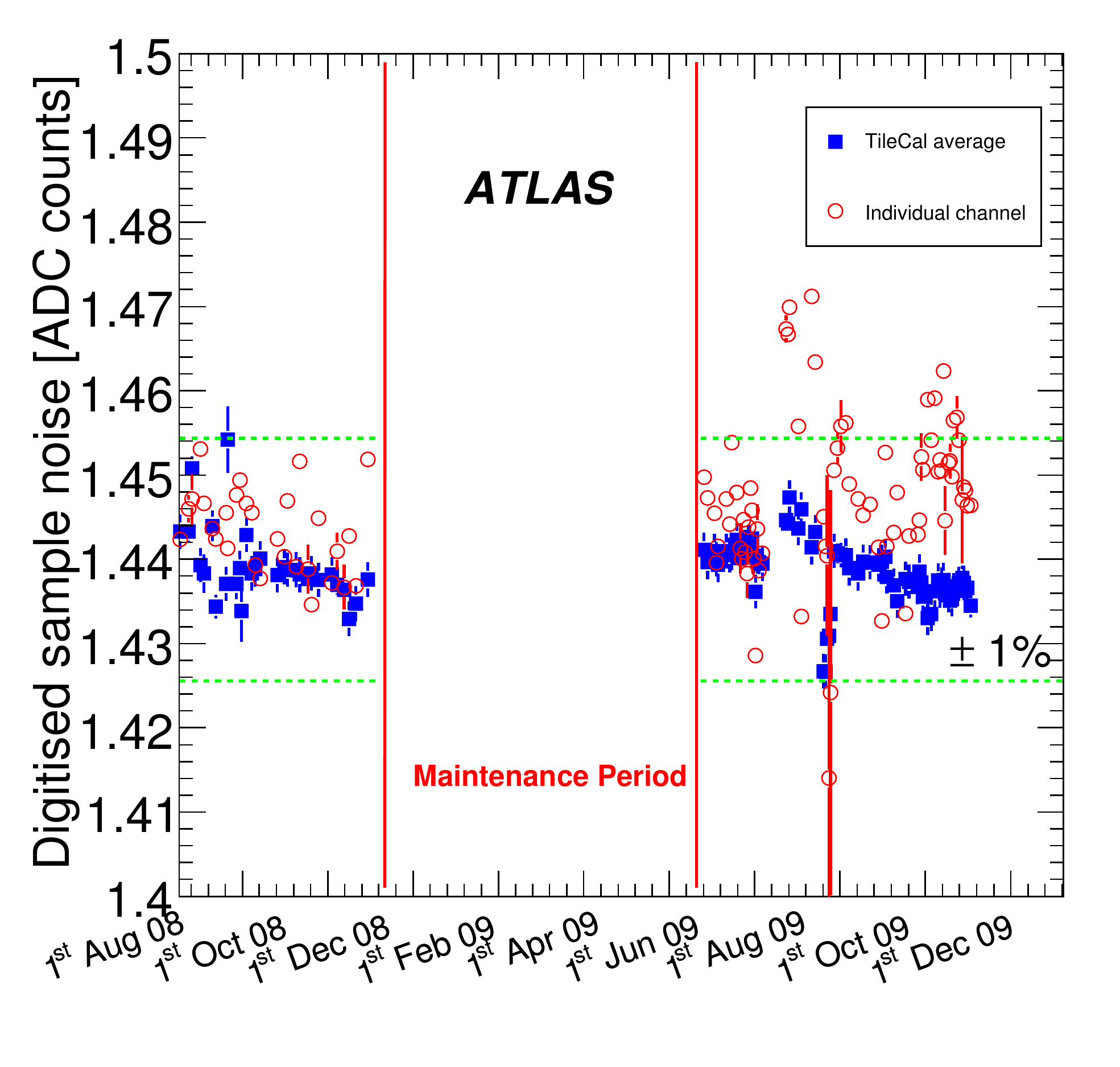}
}
\caption{Stability of average noise (RMS of the single digitised
  samples averaged over events and channels), in ADC counts, for all
  channels and for an individual channel.} 
\label{stabnoise}
\end{figure}

The noise in TileCal was measured in dedicated bi-gain standalone runs
with empty events (often called pedestal runs) and in 
random triggered events within ATLAS physics runs (often called random
triggers). The noise of each channel 
was derived from the seven digitised samples using the same method
that was used for signal reconstruction in cosmic and single beam
events, i.e.~using the OF with 
iterations.\footnote{Note that the level of noise depends on the OF
  method used. The non-iterative OF method results in lower
  noise than the OF with iterations by $\sim 14\,\%$. Note also that
  the non-iterative 
  OF will be applied for the data-taking during the collision phase,
  since the timing will be fixed by the LHC 40~MHz clock frequency.}
In Figure~\ref{stabnoise} the evolution during the
running periods of 2008 and 2009 of the average noise, in ADC counts, is shown
for all channels and for an individual channel. The channel noise is estimated as the RMS of the single digitised samples averaged over the events in dedicated TileCal pedestal runs. The overall stability is
better than 1\,\%.  

\begin{figure}
\centering
\resizebox{0.5\textwidth}{!}{%
  \includegraphics{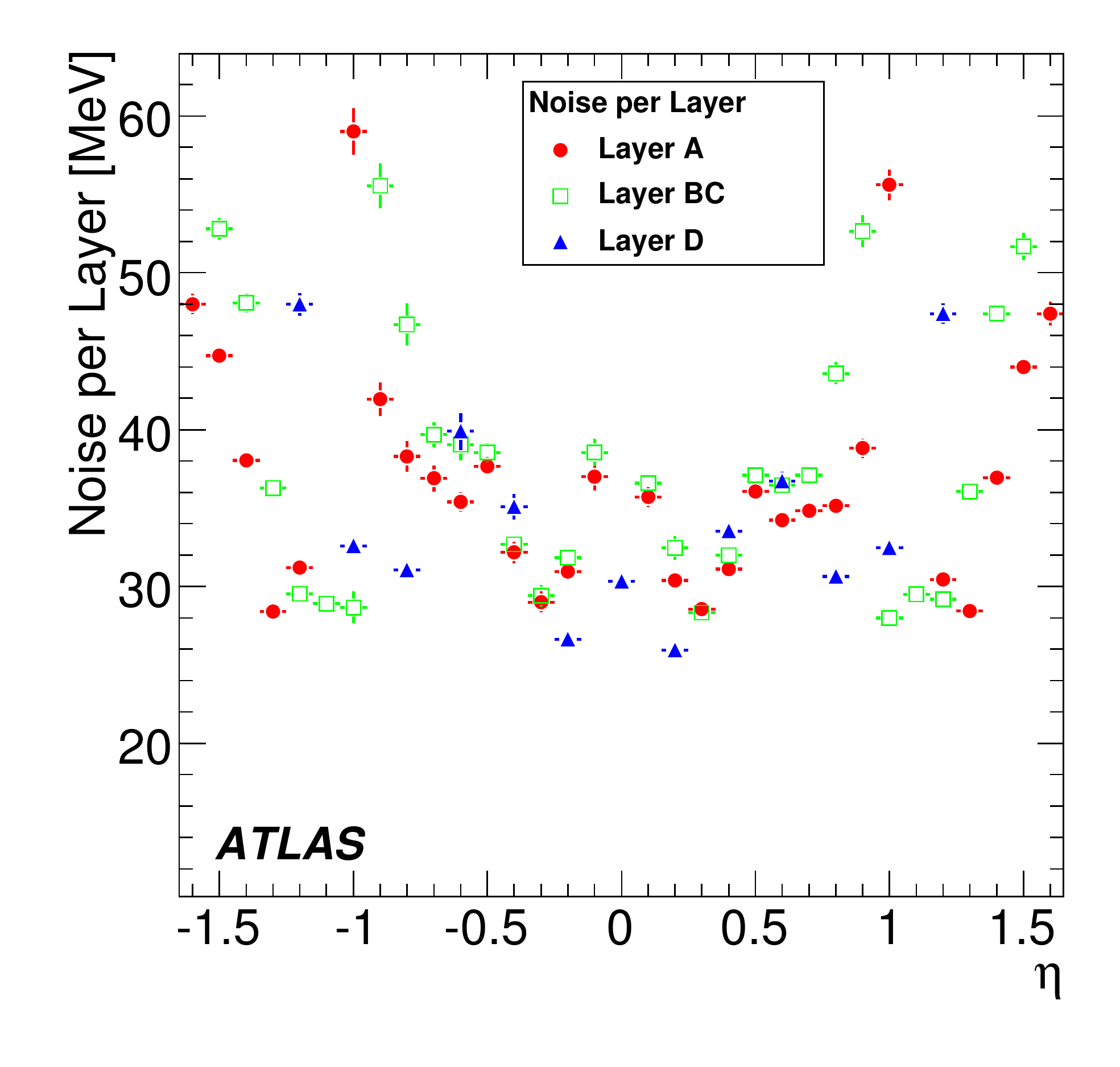}
}
\caption{Average cell noise in random triggered events as a
  function of the cell $\eta$ and radial layer. The noise is
  represented by the RMS of 
  the cell's energy distribution and the error bar shows its spread over
  all cells in the same pseudorapidity bin. 
}
\label{noiseeta}
\end{figure}

The cell noise in MeV as a function of $\eta$ is shown in
Figure~\ref{noiseeta} averaged over all modules in $\phi$ for cells in
a given $\eta$ position. The cell noise is estimated as the RMS of the
cell's energy distribution using the iterative OF signal
reconstruction in random triggered events during a physics run with
LHC single beam in 2008.   
Different
colours are used to indicate cells in different longitudinal
layers. The noise values vary between 30 and 60~MeV.
The channels with higher noise are principally at the proximity of the
LVPS which are located at the outer boundaries of the TileCal barrel
and extended barrel modules.

\begin{figure}
\centering
\resizebox{0.5\textwidth}{!}{%
  \includegraphics{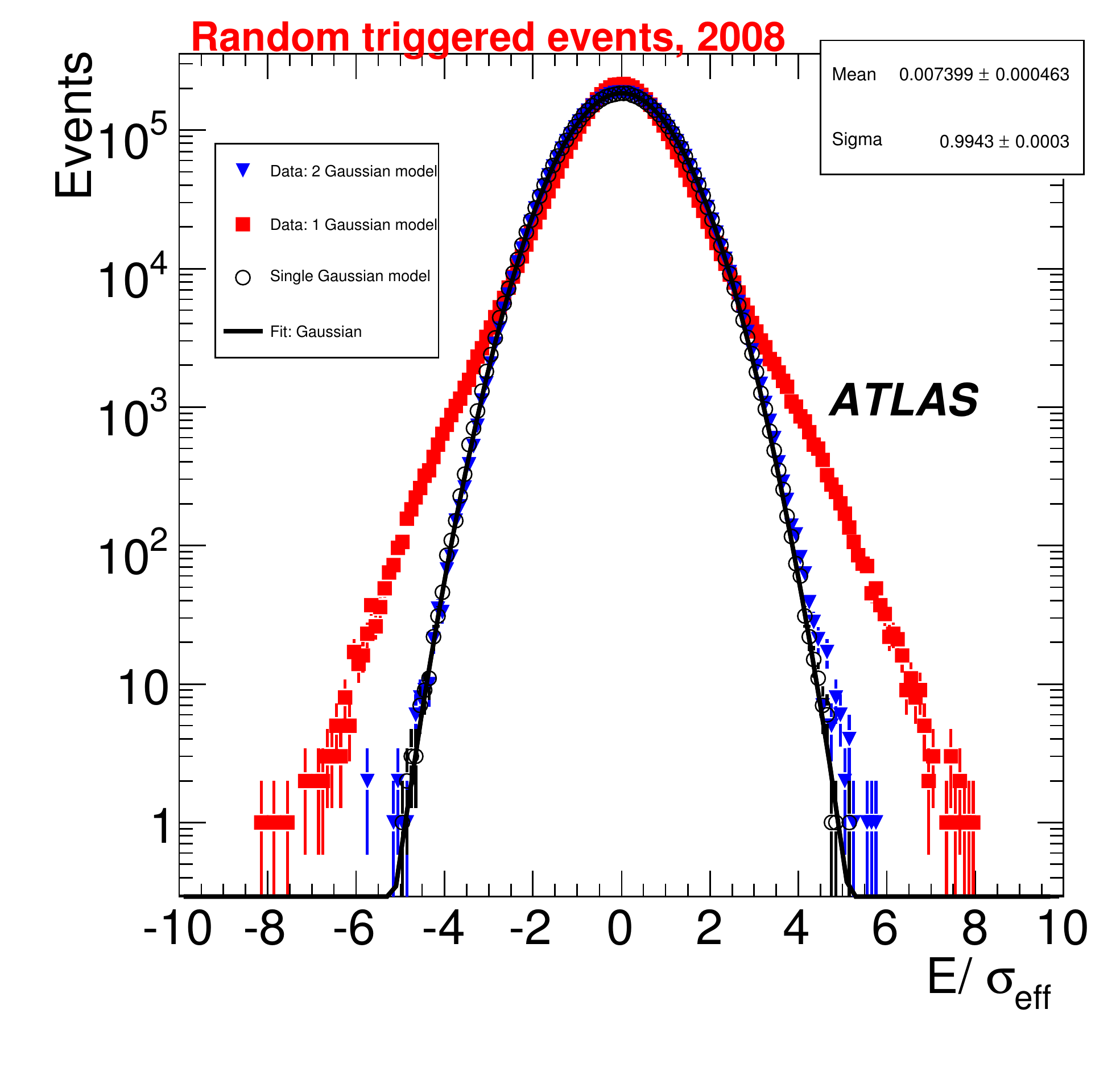}
}
\caption{Significance level of the cell energy as compared to noise
  (Energy/Gaussian $\sigma$) using the single and the double Gaussian
  descriptions of noise in random triggered events.} 
\label{significance}
\end{figure}

The cell noise probability distribution is an important component in the ATLAS calorimeter's energy clustering algorithm.
It is determined from the cell energy in empty events recorded through the standard ATLAS data acquisition chain within physics runs and it is characterised by the $\sigma$ of a fitted single Gaussian to the energy ($E$) distribution. The ratio $E / \sigma$ is used to judge if a cell has a noise-like or a signal-like energy deposition. 
Figure~\ref{significance} shows the ratio $E / \sigma$ for all TileCal cells (squares). One can observe the existence of non-Gaussian tails that could lead to fake signal cells if a criterion of $E / \sigma > 4$ is used. 
However, since a double Gaussian distribution provides a good description of the data, the two Gaussian $\sigma$'s and the relative amplitudes are used to construct a probability 
density function on the basis of which a new ``effective $\sigma$'' ($\sigma_{\mathrm{eff}}$) for every cell is defined at the significance level of 68.3\,\%. The improvement is shown in
Figure~\ref{significance} where the triangles represent the ratio
$E/\sigma_{\mathrm{eff}}$ for all the calorimeter cells. One can
observe that there are no tails when compared to a Gaussian fit (line)
or to a toy Monte Carlo noise 
generator, that randomly attributes to cells energies from a single
Gaussian model (circles). 
Thus the ratio $E/\sigma_{\mathrm{eff}}$ can be safely used to
distinguish signal from noise in a TileCal cell.

%

%
%


%
%
\section{Calibration}
\label{sec:calibration}

\begin{figure*}
 \centering
 \resizebox{0.75\textwidth}{!}{%
     \includegraphics[bb=60 200 655 460]{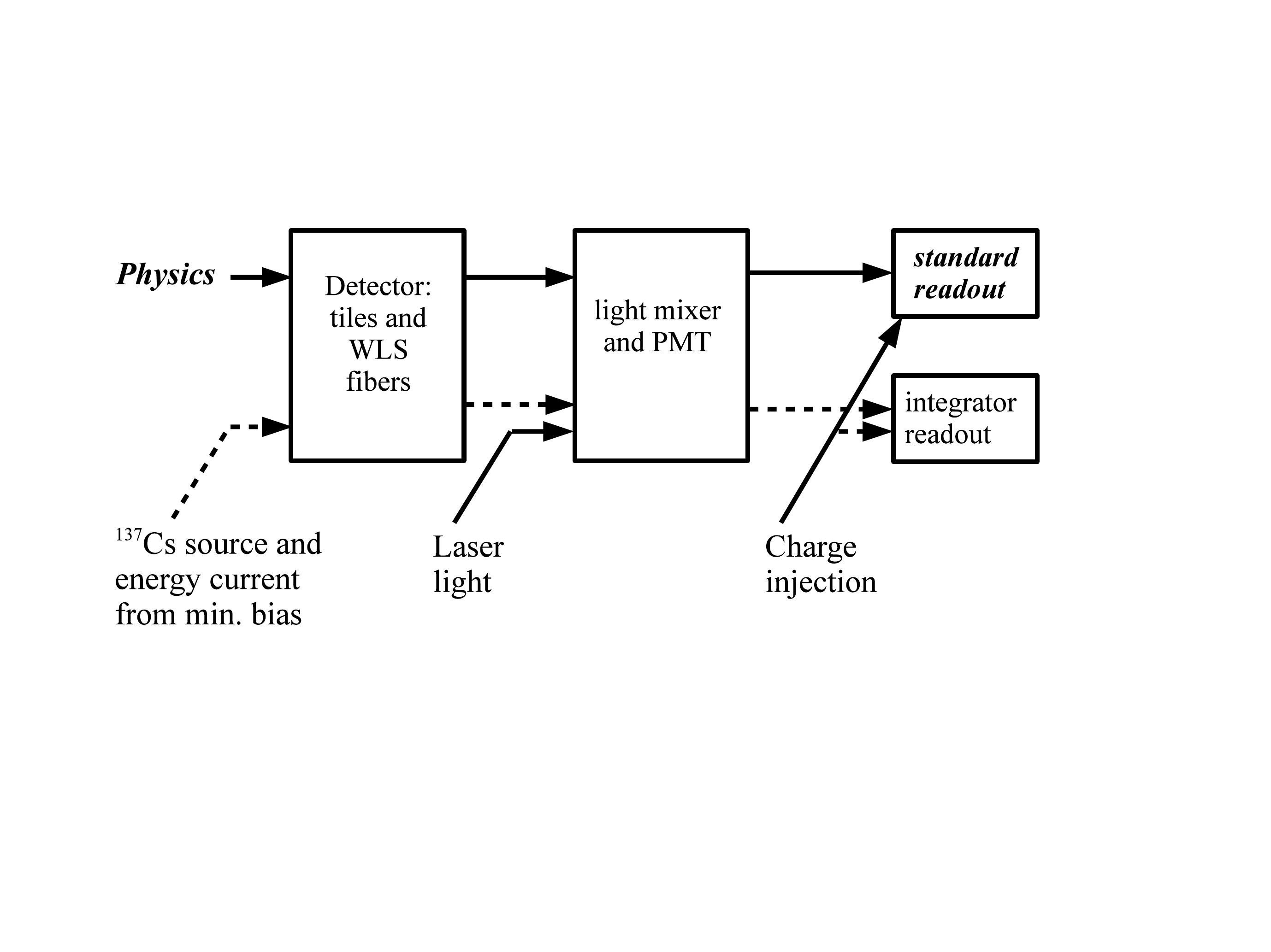}
}
 \caption{Flow diagram of the readout signal paths of the different TileCal calibration tools. The paths are partially overlapping, allowing for cross-checks and an easier identification of component failures.}
 \label{fig:calib-intro-1}
\end{figure*}

This Section describes the calibration procedures and data sets used in TileCal to establish the reference detector response. Furthermore, the calibration results obtained in the years 2008 and 2009, during the commissioning of the Tile Calorimeter in the ATLAS cavern, and the
cross-checks related to the current understanding of its calibration are also discussed. The main objectives of the calibration procedures in TileCal are to:
\begin{itemize}
\item Establish the global electro-magnetic~(EM) scale and the uncertainty associated with it. The EM scale calibration factor converts the calorimeter signals, measured as electric charge in pC, to the energy deposited by electrons, which would produce these signals. 
\item Minimise, measure and correct the cell-to-cell variations at the EM scale.
\item Measure and correct the non-linearity of the calorimeter response.
\item Measure the average time offset between the signal detection and the collision time for every readout channel.
\item Monitor the stability of these quantities in time.
\end{itemize}
The Tile Calorimeter calibrations systems treat different sections of the readout chain as illustrated in Figure~\ref{fig:calib-intro-1}. They provide:
\begin{itemize}
\item Calibration of the initial part of the signal readout path (including the optics elements and the PMTs)
with movable radioactive $^{137}$Cs $\gamma$-sources~\cite{Cs-principles}, hereafter to be called simply Cs.
\item Monitoring of the gains of the photomultipliers by illuminating all of them with a laser system~\cite{TileTDR,laser}.
\item Calibration of the front-end electronic gains with a charge injection system (CIS)~\cite{3in1}.
\end{itemize}

In order to detect non-uniformities or degradation in the detector elements (optical and otherwise), the calibration systems are specified to meet a precision of 1\,\% on the measurement of the response of a cell.

The number of channels that cannot be calibrated by each individual
calibration system is well below 1\,\%. This is additional to the
number of channels that are unusable due to LVPS problems or other
issues not related to the given calibration system. In the following
sections the performance distributions appear sometimes with fewer
channels due to the fact that not all could be available for all the
calibration periods. 

The current calibration protocol includes a number of dedicated calibration 
runs performed with a frequency derived from experience gained during the
detector commissioning. 
The CIS constants are very stable in time and 
are only updated twice per year. For monitoring and identification of bad channels, CIS runs are performed between physics runs twice per week.
For monitoring, laser runs are also performed twice per week.
The resulting laser constants will be used only for monitoring purposes until the 
stability of this calibration system is fully understood.
The Cs scans are performed
outside beam periods, with a periodicity of weeks or months, depending on 
the machine schedule since a full scan takes 6 to 8 hours. Starting from
2010, every Cs run is expected to result in new constants that adjust the global EM energy scale which will be updated accordingly. 
Laser runs accompany Cs runs in order to disentangle between changes related to the optical system and PMTs. Since the laser runs are more frequent than the Cs scans, the former provide information on the PMT gain changes between two Cs scans.

A dedicated 
monitoring system based on slow integrators%
~\cite{3in1} records signals in the Tile
readout channels over thousands of bunch crossings during 
the physics runs and is also a part of the Tile calibration framework. As
this measurement requires experience with collisions it is still being commissioned.

\subsection{Charge injection system and gain calibration in the readout electronics}
\label{sec:calibration:cis}  

The circuitry for the Charge Injection system is a permanent part of each front-end electronics channel~\cite{3in1}
and it is used to measure the pC/ADC conversion factor for the digital readout of the laser calibration and physics data and to determine the conversion factor for the slow integrator readout, measured in 
ohms.



\begin{figure*}
 \centering
 \resizebox{0.49\textwidth}{!}{%
   \includegraphics{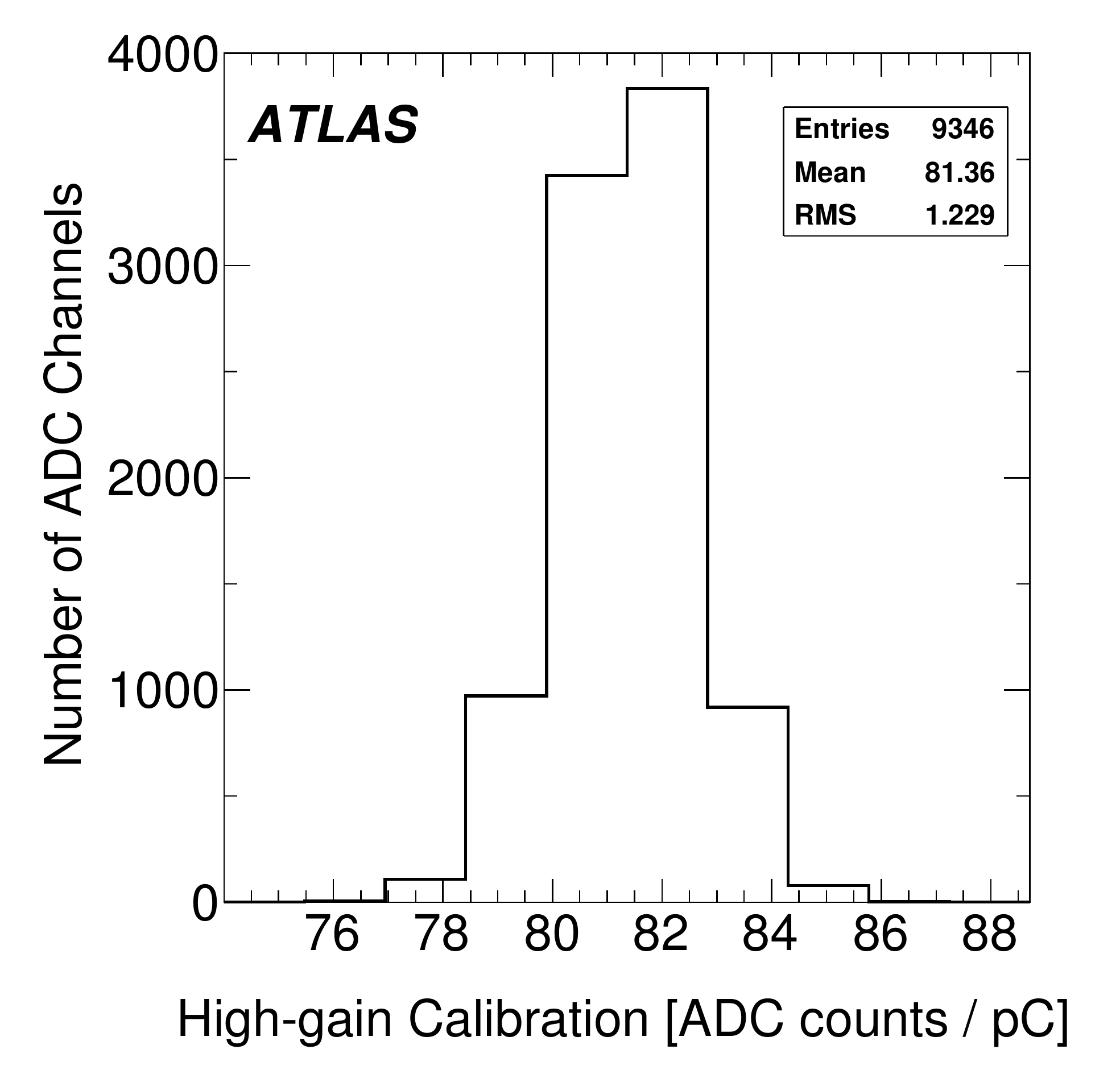}
 }%
 \resizebox{0.49\textwidth}{!}{%
   \includegraphics{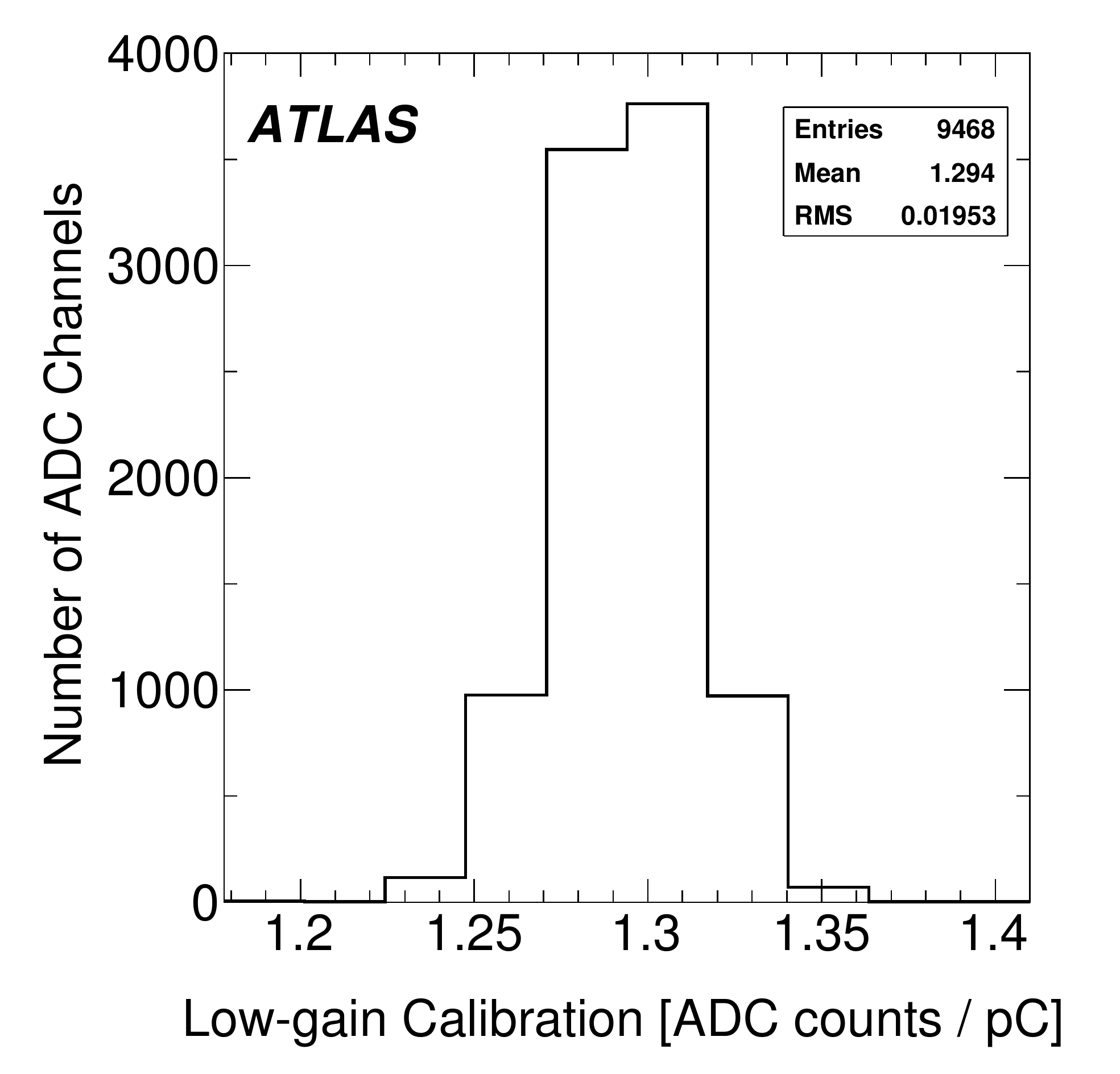}
 }
 \caption{Channel-to-channel variation of the high gain (left) and of
   the low gain (right) readout calibration constants as measured by
   the CIS, prior to any correction. The measured HG/LG gain
     ratio of 62.9 corresponds to the nominal of 64 (see
     Section~\ref{sec:overview}) within tolerances of individual
     electronics components.
   }
 \label{fig:calib-CIS-1}
\end{figure*}

To reconstruct the amplitude for each injected charge, a three-parameter
fit is performed as described  at the end of
Section~\ref{sec:energy_time_reco}, with the amplitude being one of
the parameters of the fit~\cite{TileTBpaper}. To determine the values
of the gains for each channel, dedicated CIS calibration runs are
taken frequently, in which a scan is performed over the full range of
charges for both gains. The  typical channel-to-channel variation of
these constants is measured to be approximately 1.5\,\%, as shown in
Figure~\ref{fig:calib-CIS-1}. This spread indicates the level of
corrections for which the CIS constants are applied. 

\begin{figure*}
 \centering
 \resizebox{0.49\textwidth}{!}{%
   \includegraphics{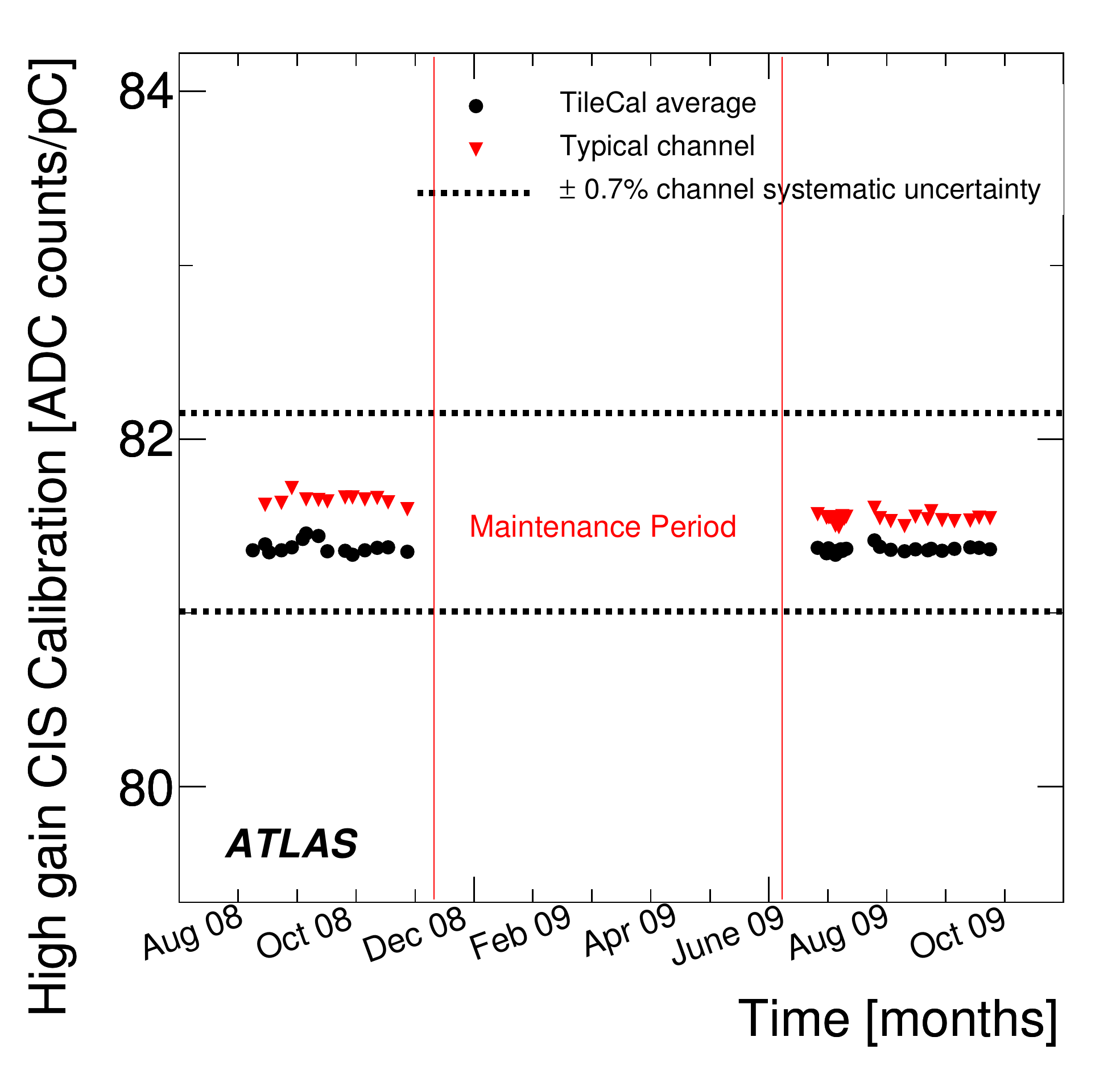}
 }%
 \resizebox{0.49\textwidth}{!}{%
   \includegraphics{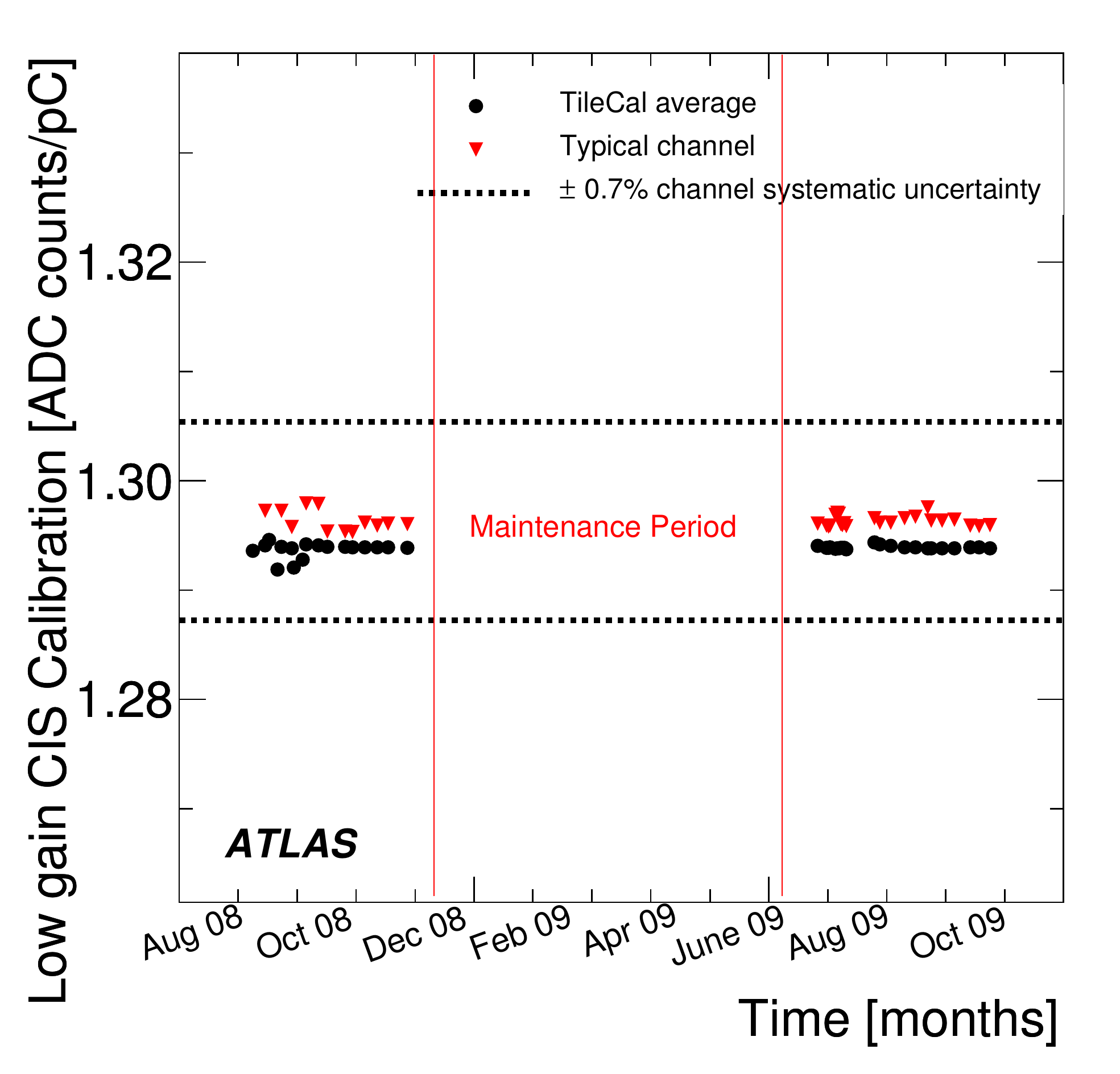}
 }
 \caption{Stability in time of the average high gain (left) and low gain (right) readout calibration constants from August 2008 to October 2009.
   }
 \label{fig:calib-CIS-2}
\end{figure*}

The stability in time of the average high gain and low gain readout calibration constants from August 2008 to October 2009 is shown in Figure~\ref{fig:calib-CIS-2} for
99.4\,\% of the total number of ADCs. 
The time stability of a typical channel is also shown for each gain. Over this period, the RMS variation for the high and low gain detector-wide averages and for the single channels shown, is less than 0.1\,\%. The superimposed bands of $\pm0.7\,\%$ represent the systematic uncertainty for the individual channel calibration constants, mainly due to the uncertainty on the injected charge.

The distributions of high gain and low gain readout calibration constants for individual ADC channels were compared for the sample of channels calibrated during the TileCal standalone testbeam period of 2002 to 2003 and for the full detector in the cavern in 2009. No significant change in the calibration constants was observed, thus limiting the contribution from the CIS calibration to the systematic uncertainty on transferring the EM scale from testbeam to ATLAS to below 0.1\,\%.

\begin{figure*}
 \centering
 \resizebox{0.49\textwidth}{!}{%
   \includegraphics{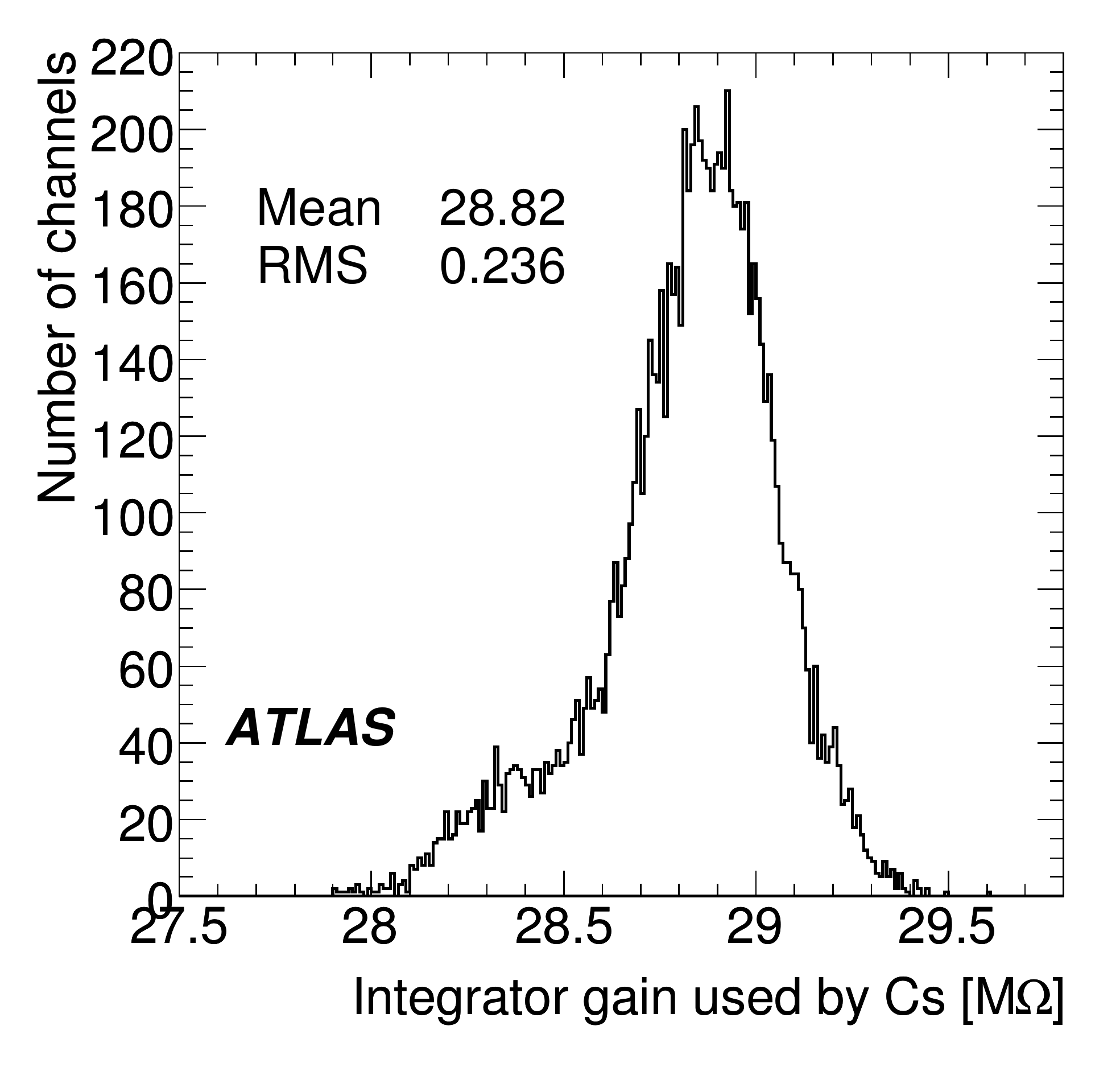}
 }%
 \resizebox{0.49\textwidth}{!}{%
   \includegraphics{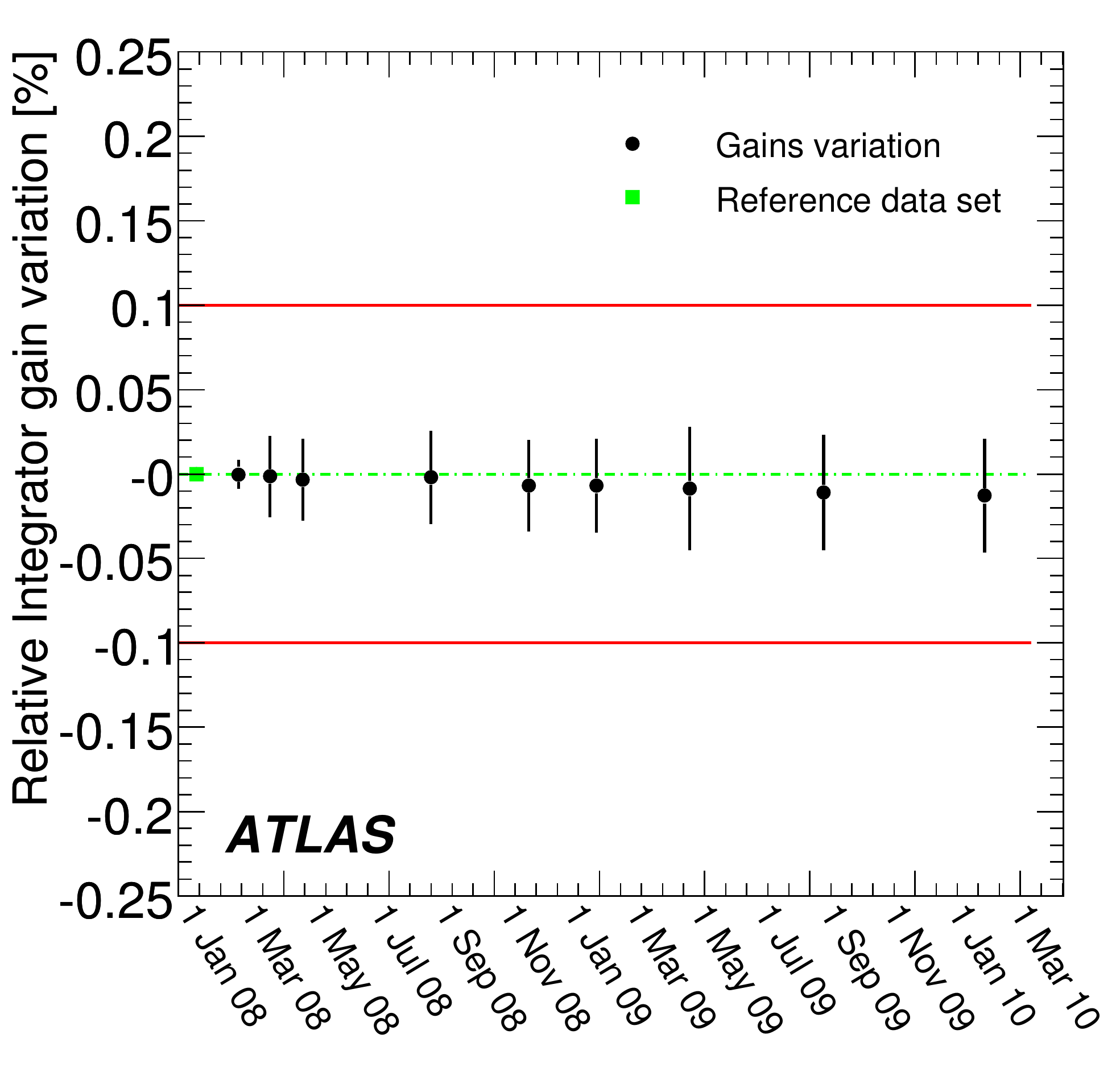}
 }
 \caption{Distribution of the integrator gain used by the Cesium calibration system is shown on the left. Relative stability over twenty-two months of the same integrator gain is shown on the right.
   }
 \label{fig:calib-int-1}
\end{figure*}

To determine the values of the gains for each channel for the current integrator readout, dedicated calibration runs are periodically taken, in which a scan is performed over the full range of currents for all six integrator gains. The channel gain is extracted as a slope from a 2-parameter fit performed on the measured channel response in voltage to each applied current. The typical channel-to-channel variation of these integrator gain constants is measured to be approximately 0.9\,\%, as shown in Figure~\ref{fig:calib-int-1} (left) for the gain used during calorimeter calibration with the 
Cs radioactive source. The 12-bit ADCs used to digitise the PMT currents were produced in two unequal batches with about 2\% difference in amplifier gains, which can be clearly seen in the distribution of the integrator gains in Figure~\ref{fig:calib-int-1} (left).

The relative variation of the integrator gains used by the Cs calibration system is shown in the right part of Figure~\ref{fig:calib-int-1}. The measurements in 
95.9\,\% of the integrators performed at different dates are compared to the reference measurements of January 2008.
The error bars represent the dispersion of the individual channel measurements relative to their reference values in the first run. 
The stability of individual channels is better than 0.05\,\% while the
stability of the average integrator gain is better than 0.01\,\% over
the considered period of time of 26 months.

The variation of the integrator gains for individual channels used in the Cs calibration system readout from 2001 to 2009 was studied on the sample of channels calibrated in both instances. No significant change in the calibration constants was observed over eight years. The contribution from the integrator gain calibration to the systematic uncertainty on setting the EM scale of TileCal in ATLAS as compared to the testbeam was found to be below 0.2\,\%.

\subsection{Calibration with laser system}

The Tile Calorimeter is equipped with a custom-made laser calibration
system~\cite{laser} 
dedicated to the monitoring and calibration of the
Tile photomultiplier properties, including the gain and linearity of each
PMT\@. The frequency doubled infrared laser providing a 532~nm green light
beam is located in the ATLAS USA15 electronics room, 100~m from the
detector. The laser emits short pulses, which reasonably resemble
those from the physics signals, with a nominal energy of a few
mJ. This power is sufficient to simultaneously saturate all Tile
readout channels, and thus to probe their linearity over the full
readout dynamic range. A dedicated set of optical elements insures
proper attenuation, partial de-coherence and propagation of the
original light beam to every photomultiplier used in the Tile Calorimeter
readout. This calibration system was commissioned until
September 2009 and since then it is operating in a stable
configuration. By varying the voltages applied to
the photomultipliers it was shown that the system sensitivity to the
relative gain variations is of 0.3\,\% on data sets recorded over few
hours. The long term stability of the laser calibration system is
under study. 

\begin{figure}
  \centering
  \resizebox{0.5\textwidth}{!}{%
    \includegraphics{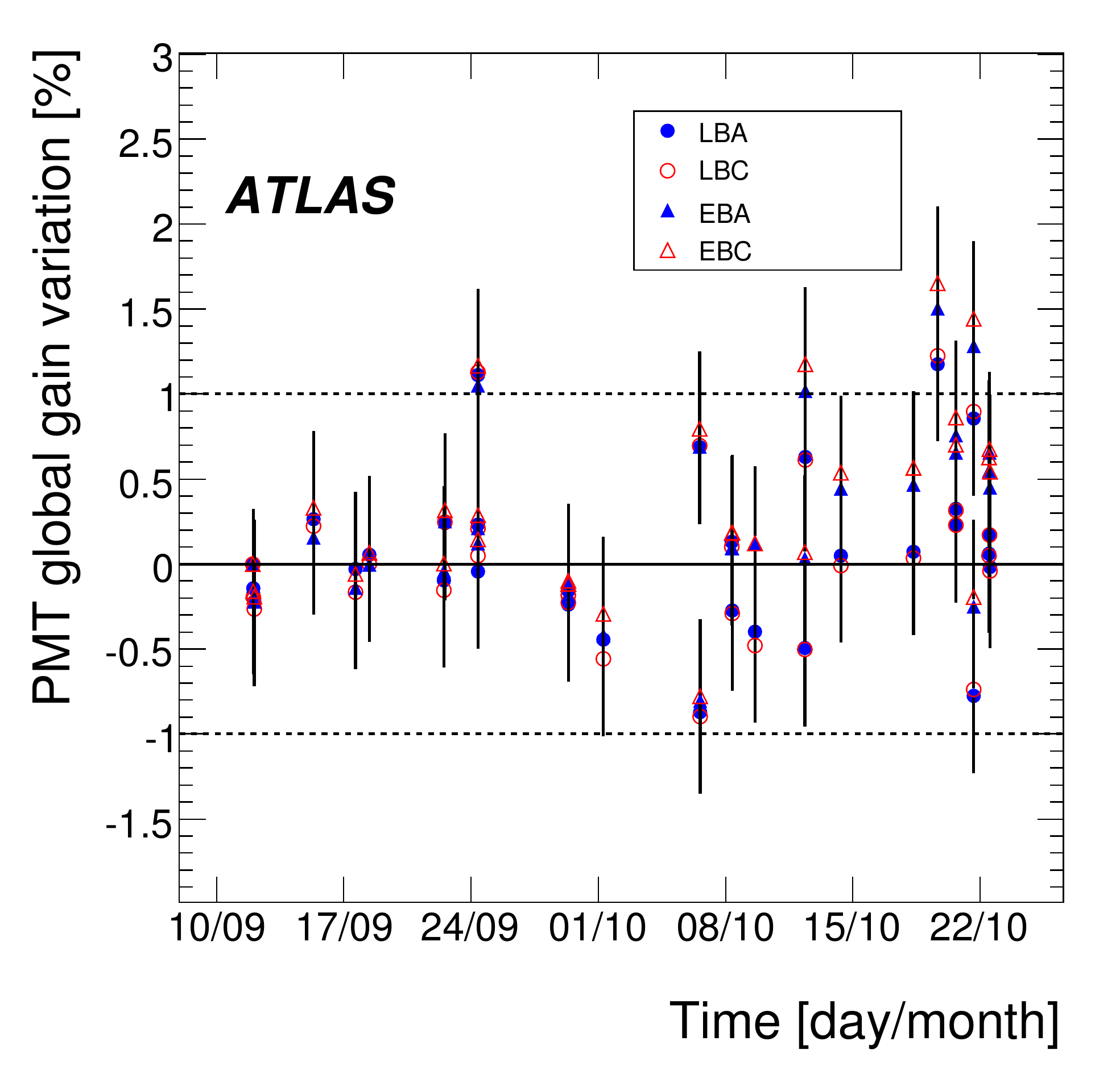}
  }
  \caption{Average PMT gain variation measured by the laser calibration
    system as a function of time over forty days in 2009.
  }
  \label{fig:calib-LAS-1}
\end{figure}

The time stability of the PMT gains was evaluated using dedicated laser runs and averaging over 98.8\,\% of the TileCal channels.
An estimation of the relative gain variation in time was based on the analysis of the shape of the distribution of the PMT responses to the signal induced by the laser system at many instances. The average gain variation as a function of time over 40 days is shown on Figure~\ref{fig:calib-LAS-1}. This variation is found to be within 1.0\,\% over the considered period of time. The displayed error bars of 0.5\,\% account for both the statistical uncertainty and the systematic effects and are entirely dominated by the latter. The systematic uncertainty comes from the limited reproducibility of the light intensity on the photomultipliers downstream of the full optical chain through which the laser beam propagates to the detector. The design goal of the laser system is to monitor the relative gain stability with 0.5\% accuracy for time periods of months to years. The results mentioned above set the precision with which the PMT response stability can be monitored by the laser system between two Cs scans that are typically one month apart and monitor the combined response of the optics elements and PMTs. 

\begin{figure*}
 \centering
 \resizebox{0.49\textwidth}{!}{%
   \includegraphics{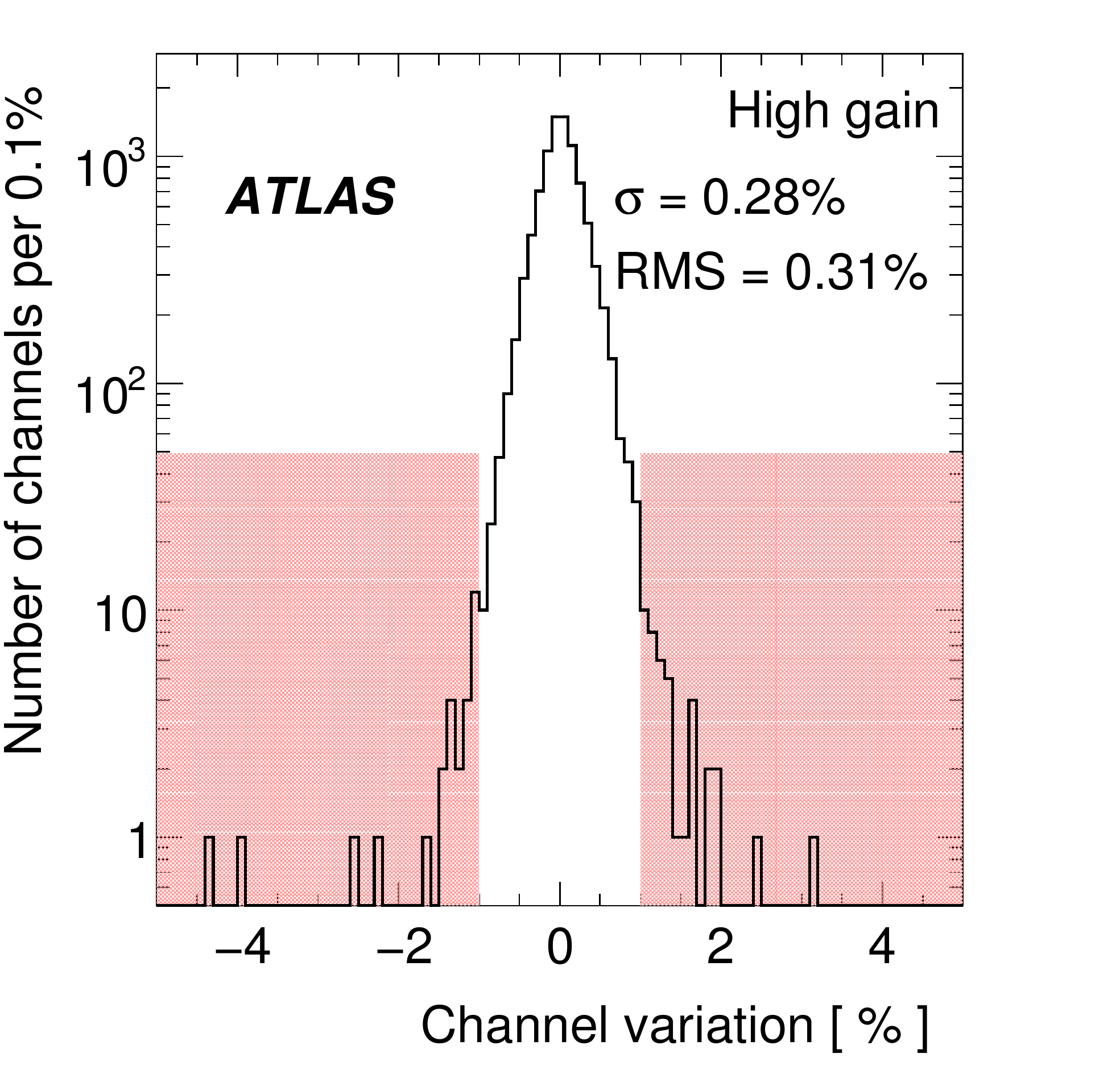}
 }%
 \resizebox{0.49\textwidth}{!}{%
   \includegraphics{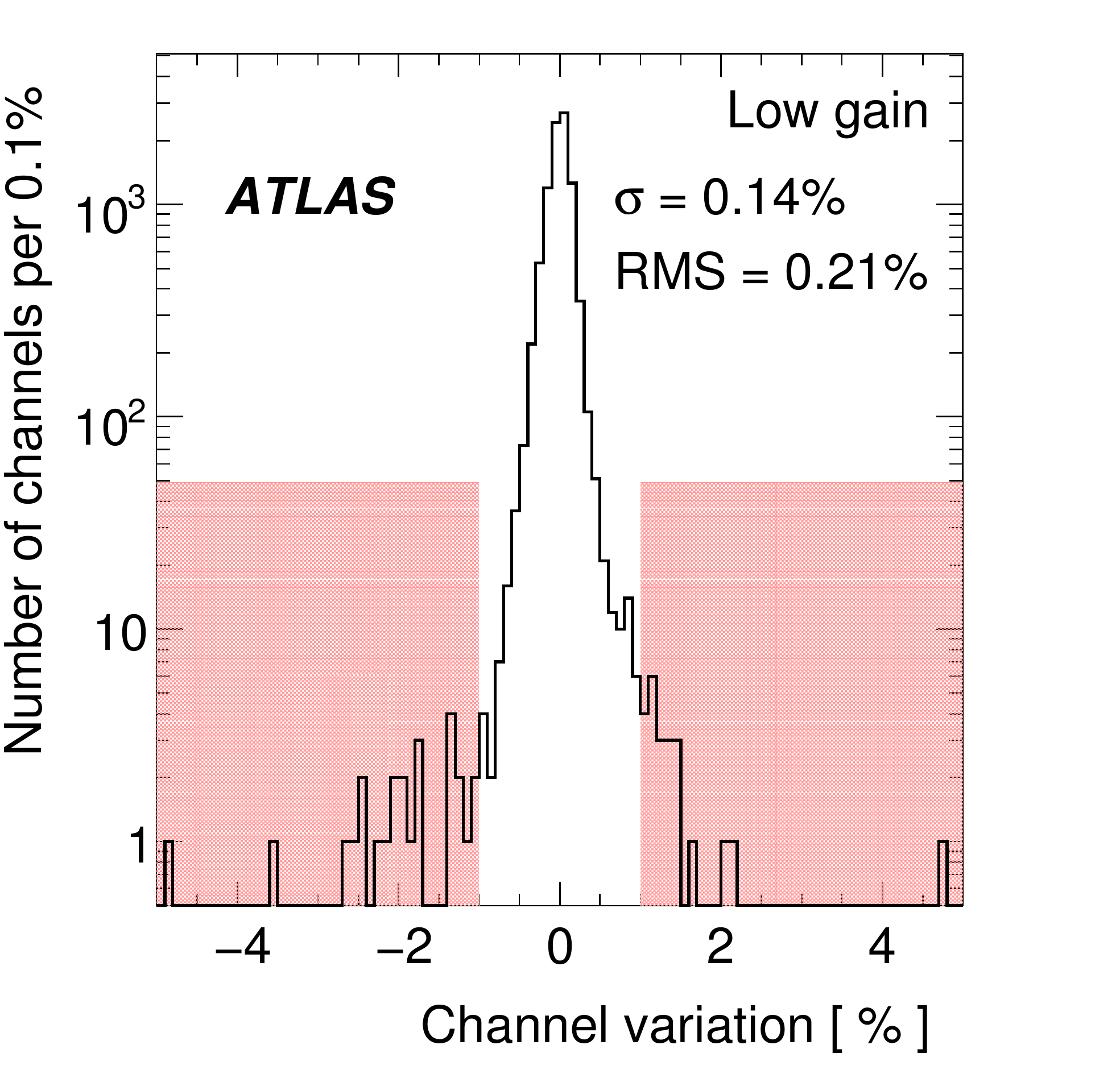}
 }
 \caption{Channel-to-channel variation of the relative gain of the photomultipliers for two Laser calibration runs taken in HG (LG) mode, shown on the left (right).
   }
 \label{fig:calib-LAS-2}
\end{figure*}

Once the global variation of the laser signal is accounted for, the gain stability per individual channel can be studied. A typical channel to channel variation for HG and LG is shown in Figure~\ref{fig:calib-LAS-2}, where the relative gain variations for two laser calibration runs separated by 50 days are presented. The shaded sidebands represent channels with relative variation above 1\,\%. 
The observed RMS of 0.3\,\% (0.2\,\%) in the HG (LG) is a convolution of residual fluctuations of the laser system and variations of the PMT response. Therefore, this RMS can be considered as an upper limit on possible stochastic variations in photomultiplier gains.

Once the intrinsic stability of the laser calibration system is understood, this system will be used to calibrate the gain and linearity\footnote{All the photomultipliers used in TileCal were characterised on their arrival from Hamamatsu at dedicated test benches with LED light sources. No PMT was found with non-linearity worse than 3\,\% up to 800~pC of collected charge.} of each PMT.

\subsection{Calibration based on $^{137}$Cs radioactive $\gamma$-source}
\label{sec:cesium}

The Tile Calorimeter includes the capability of moving through each scintillator tile a 
Cs radioactive $\gamma$-source along the $Z$-direction of the ATLAS detector. Capsules containing 
the Cs sources with activities of about 330~MBq emitting 0.662~MeV
$\gamma$-rays are hydraulically driven 
through a system of 10~km of steel tubes that traverses every
scintillating tile in every module~\cite{Cs-control}. Three
sources of similar intensity are deployed in the three cylinders
of the Tile Calorimeter. When a capsule traverses a given cell, the integrator circuit
located on the 3-in-1 cards (Section~\ref{sec:overview}), reads out the current signal
in the PMTs. 
The total area under the integrator current vs capsule position curve corresponding to the source path length in a cell, is calculated and normalised to the cell size. This estimator of the cell response to Cs is used throughout this Section. 
%
%

Source scans provide the means
to diagnose optical instrumentation defects~\cite{Cs-instrumentation}
and to measure the response of each individual cell.
The precision of the Cs based calibration was
evaluated from the reproducibility of multiple measurements under the same conditions and was found to be about 0.3\,\% for a typical 
cell~\cite{Cs-principles}. The precision is
0.5\,\% for the cells on the edge of the TileCal cylinders and a few
percent for the narrow cells C10 and
D4 in the gap region (see Figure~\ref{fig:tile-module}). As discussed in Section~\ref{sec:itc_calibration}, cosmic ray muons are
used to cross-check the calibration factors for the cells of this type. 

\subsubsection{Intercalibration and EM scale factor via the Cs system}
\label{sec:cs_calib_transfer}

The Cs calibration has proceeded in two distinct phases. 
\begin{itemize}
\item Photomultiplier gain
equalisation to a chosen level of Cs response was performed for every individual channel on the 11\,\% of production TileCal
modules that were tested with particle beams during 
2001-2004~\cite{TileTBpaper}.  
The next step was to measure the numerical value for the fixed EM scale with electron beams.  
With the electrons entering the calorimeter modules at an incidence
angle of $20^\circ$, the average cell response normalised to beam energy was measured to be $(1.050\pm
0.003)$~pC/GeV, defining the TileCal EM scale factor. This factor was determined for the cells of the first layer 
and propagated via the gain equalisation to all the other cells. 
The RMS spread of $(2.4\pm0.1)\,\%$ was found to be due to local variations in individual tile responses and tile-fibre optical couplings. The above two steps effectively
resulted in setting the cell EM scale in the subset of TileCal
modules exposed to electron beams.

\item The second phase in the calibration was to reproduce the above PMT gain equalisation on the full set of the Tile Calorimeter modules in the ATLAS environment and to transfer via the Cs response the EM scale factor as defined in the testbeam. This took place in the second half of 2008. In some cases the PMT gain is intentionally higher by 20\,\% (D0, D1, D2, D3, D4 and C10 cells) in order to improve on signal to noise ratio for the detection of muons (see also Section~\ref{sec:signal_noise}). For the central barrel cells of the third radial layer this improvement will facilitate their possible usage in the L1 muon trigger. The EM scale for these cells is recovered by applying appropriate corrections to the cell energy reconstruction.
\end{itemize}

To set the EM scale as defined at the testbeam, the target
response to Cs for 2008 and 2009 was defined as the response measured at the testbeam scaled by the ratio of
the activities of the testbeam source to the sources used in the cavern. These ratios
were measured by intercalibrating the sources using two TileCal
modules that are kept outside the experimental hall. The source activity decay time between the testbeam and the ATLAS scans was taken into account. By adjusting PMT gains in order to have equal response to Cs between the testbeam and the ATLAS setup, the numerical factor that converts charge to GeV is preserved. It is evident that the comparison of the source activities is of utmost importance in order to preserve the absolute energy scale as set with electrons. 

Five ${}^{137}$Cs
radioactive sources of different ages and activities were used over the last years. Three sources are currently used in the ATLAS cavern and two different sources were used for checks on
instrumentation quality and for the calibration at the testbeam.
In spring of 2009, one long barrel and one extended barrel module
were scanned sixty times under the same conditions with all five
radioactive sources. With the reproducibility of a single measurement
better than 0.1\,\%, a full set of ratios of the source activities was
evaluated with the precision of 0.05\,\%. The results for these ratios
after averaging over all data sets available are shown in the last column of
Table~\ref{tab:csratios}. It should be noted that the third column of the table gives an initial estimation of the activities as measured by the manufacturer with a $\pm\,15\%$ uncertainty. We plan to exchange the sources between the
Tile Calorimeter cylinders in the cavern for future checks on
reproducibility of the responses and also to monitor the ratios of the
source activity in time.

\begin{table*}
  \centering
  \begin{tabular}{|c|c|c|c|}
    \hline
    Source & Location in 2009 & Activity in & Measured activity, \\
      &   & April 2009 ($\pm15\%$) & normalised to RP3713 \\
    \hline
    RP3713  &  Storage & 264 MBq & - \\
    RP4091  &     LB   & 372 MBq & $1.1860\pm 0.0005$ \\
    RP4090  &    EBA   & 363 MBq & $1.1590\pm 0.0005$ \\
    RP4089  &    EBC   & 377 MBq & $1.2180\pm 0.0007$ \\
    RP3712  & Bld. 175 & 319 MBq & $1.2200\pm 0.0005$ \\
    \hline
  \end{tabular}
\caption{Activity of five ${}^{137}$Cs radioactive sources as of April 2009, and ratios with respect to the reference source RP3713 of the measured activities averaged over all data sets collected in the spring of 2009. Source RP3713 was used in calibrations during the test beam period. Source RP3712, kept in Building 175, is used for ageing tests.
}
\label{tab:csratios}
\end{table*}

\subsubsection{Effect of magnetic field}
\label{sec:cs_mfield}

Comparing the EM scale response between the testbeam and full detector, the magnetic field configuration has to be considered.  
During the testbeam no magnetic field was present while during data-taking in ATLAS, TileCal operates in the presence of magnetic
field. The calorimeter iron, mainly the girder volume at the outer
radius, serves as the flux return of the solenoid field. The general
behaviour of iron-scintillator calorimeters in magnetic field is
known from other
experiments~\cite{field-1,field-2,field-3}. A small increase in the
scintillator light yield, which also varies modestly over a broad
range of the applied field is expected. 

\begin{figure*}
 \centering
 \resizebox{0.49\textwidth}{!}{%
   \includegraphics{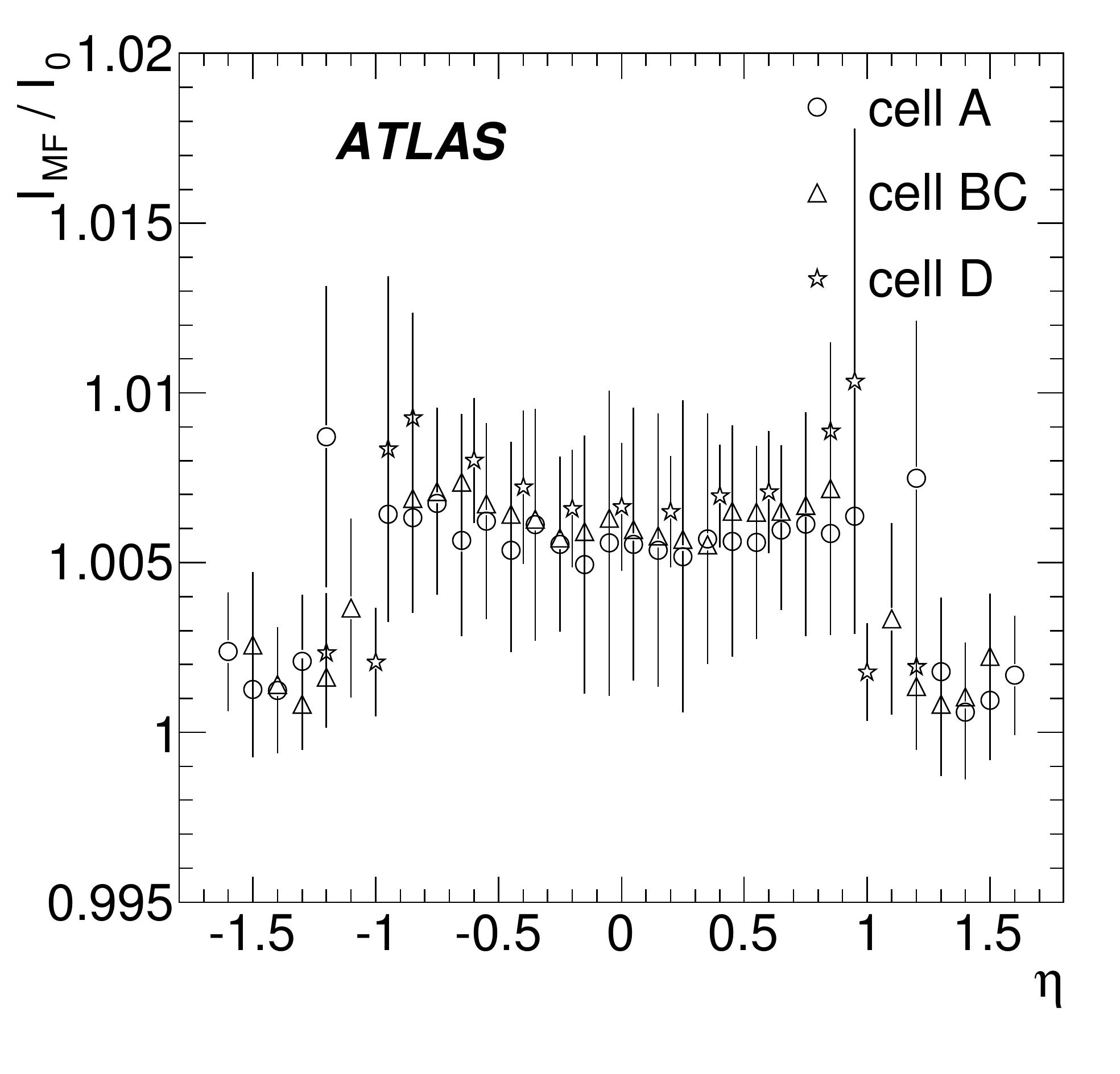}
 }%
 \resizebox{0.49\textwidth}{!}{%
   \includegraphics{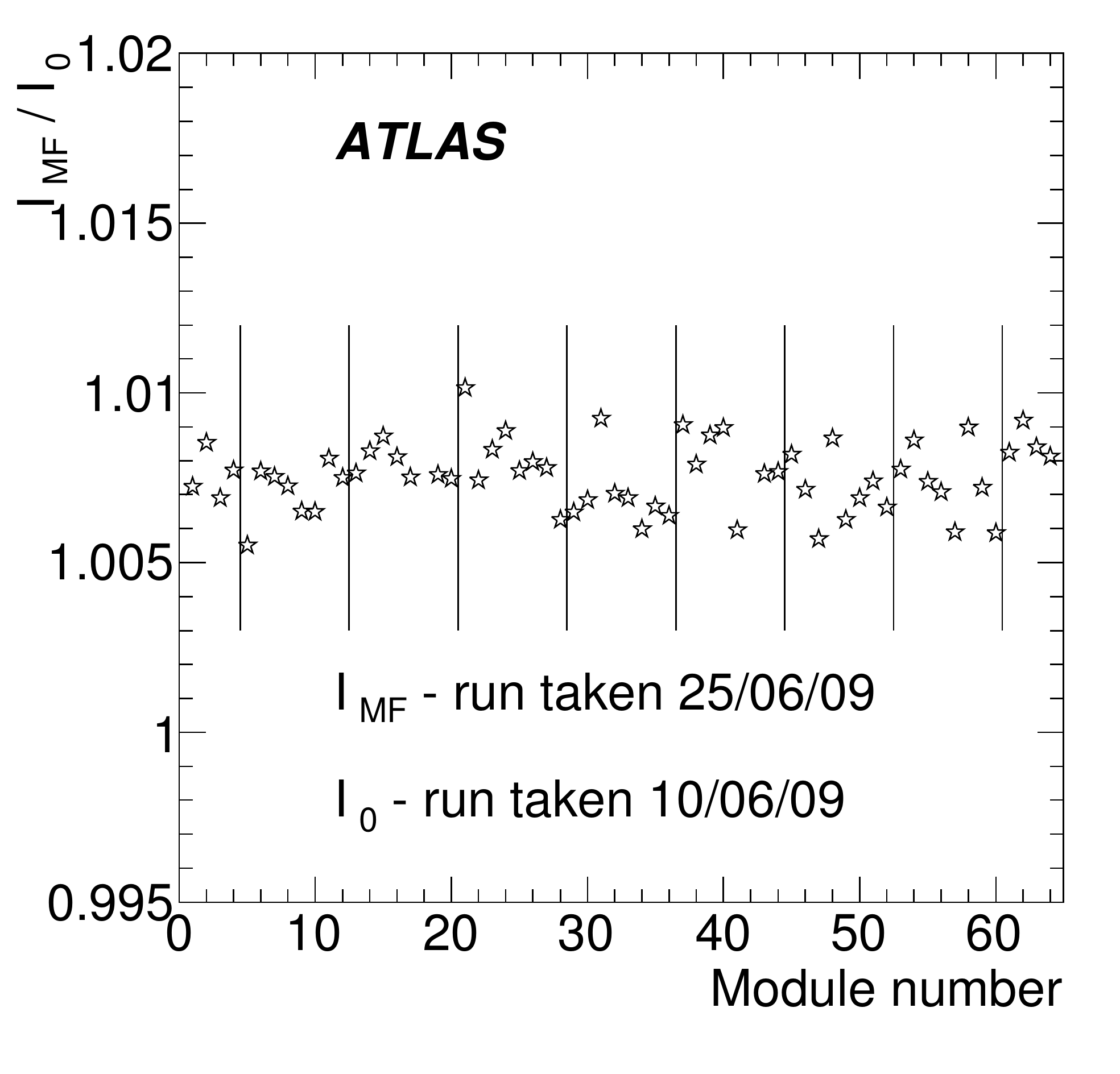}
 }
 \caption{Ratio of the TileCal cell response to the radioactive 
Cs source in full ATLAS magnetic field to the TileCal cell response to the 
Cs source without the field, shown as function of~${\eta}$ (left). Ratio of the TileCal D3 cell response to radioactive 
Cs source in full ATLAS magnetic field over its response to the 
Cs source without the field, shown as function of~${\phi}$
(right). The vertical lines indicate the position of the ATLAS toroid coils.
   }
 \label{fig:calib-CS-6}
\end{figure*}

The impact of the full ATLAS magnetic field on the Tile Calorimeter response was studied using the Cs calibration system. The ratio of the TileCal cell response to a radioactive Cs source in the full ATLAS magnetic field to its response to the Cs source without the field is given in Figure~\ref{fig:calib-CS-6} (left) as a function of~${\eta}$ for two consecutive Cs runs. The cells in individual radial layers are shown with different symbols. The error bars represent the RMS of the above ratio over the sample of the sixty four identical cells in the full~${\phi}$ range. 

As expected, the effect of magnetic field is stronger in the barrel
partitions,
where the flux of the solenoid field return
is the most intense, and where the increase in calorimeter response is
on average $\sim\!0.6\,\%$. A small increase of $\sim\!0.2\,\%$ is observed for the extended barrel.
This increase 
was not fully reproducible in every instance of the magnetic field turn-on in 2008, which contributes~0.5\,\% to the systematic
uncertainty 
of propagating the EM scale from the testbeam to the ATLAS running
configuration. The ratio of the D3 cell\footnote{A cell through which
  the Toroid field return is the strongest.} response to radioactive
Cs source with and without the full ATLAS magnetic field is shown in
the right part of Figure~\ref{fig:calib-CS-6} as function
of~${\phi}$. The vertical lines illustrate the positions
of the Toroid coils. No clear structure in ${\phi}$ is observed,
indicating that in the final ATLAS configuration the full magnetic
field does not significantly affect the Tile Calorimeter response
uniformity in~${\phi}$. Starting from 2010, Cs calibrations will be
exclusively based on the data taken with the full magnetic field. 

\subsubsection{Monitoring with Cs in ATLAS}
\label{sec:cs_inatlas}

Once the EM scale was established and reproduced in ATLAS, periodic scans are performed to monitor the stability of the detector response to the radioactive source in time. This is the final step that insures the monitoring of the known EM scale in time. 

\begin{figure*}
 \centering
 \resizebox{0.75\textwidth}{!}{%
   \includegraphics{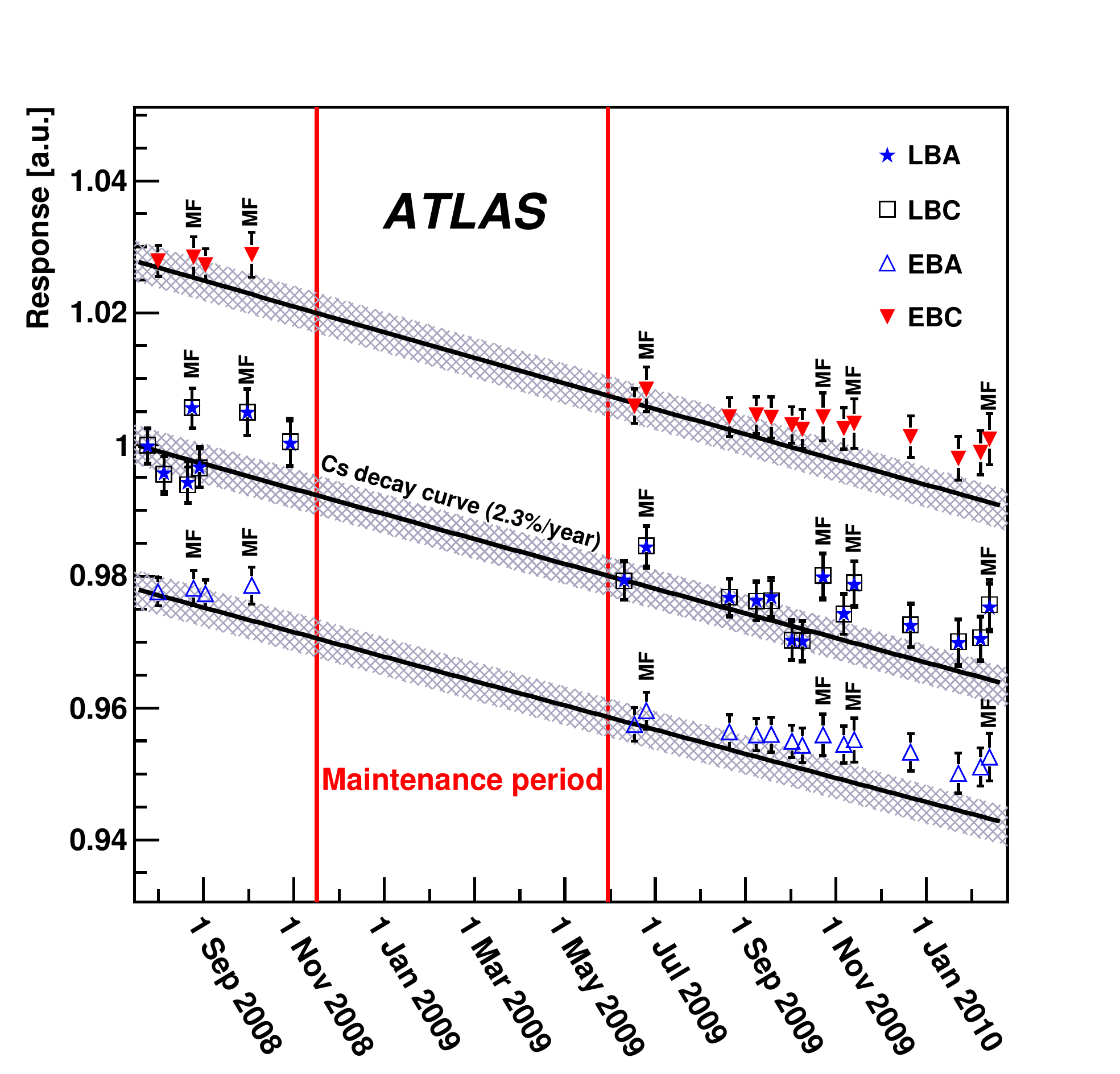}
 }
 \caption{TileCal response to radioactive 
   Cs sources in all four calorimeter partitions
   not corrected for the difference in the source activities as a
   function of time, averaging over all channels in a partition. The error bars represent the RMS spread in the responses
   of the sample of channels used. The ``MF'' symbol stands for
   the Cs calibration data taken with magnetic field on.  
 }
 \label{fig:calib-CS-3}
\end{figure*}

The Tile Calorimeter response to the Cs source as a function of time is shown in Figure~\ref{fig:calib-CS-3}. The first scan was taken approximately two weeks after the original PMT gain equalisation in July~2008. Around 55 calibration runs with the radioactive source are considered for the time period from August~2008 to February~2010.
The maintenance period of six months is indicated by the vertical
lines on Figure~\ref{fig:calib-CS-3} and is excluded from the
studies. The very first points after the maintenance period correspond
to the second gain equalisation to the same target value, corrected for the expected decrease in
the source activity in time, as indicated on the Figure. The average response to the radioactive sources in the four calorimeter partitions is shown by the points of different
colours. Since three sources
with about 3\,\% difference in their activity are used in the barrel
and two extended barrel cylinders, the data points follow three
distinct paths in time. The error bars, which are always below 0.4\,\%,
represent the RMS spread in responses over the full set of channels in
a given partition. The number of cells with unreliable Cs calibration
or with unstable HV level is below 0.2\,\% of the total and they are
excluded from the present study. The shaded bands along the lines
indicate the level of reproducibility of the Cs measurements. The ``MF''
label indicates that the corresponding Cs calibration run was taken with
both the ATLAS toroid and solenoid fields on. The response increase due to  magnetic field is larger in the barrel partitions. Details on the magnetic field effects were already discussed in Section~\ref{sec:cs_mfield}.  

\begin{figure*}
 \centering
 \resizebox{0.49\textwidth}{!}{%
   \includegraphics{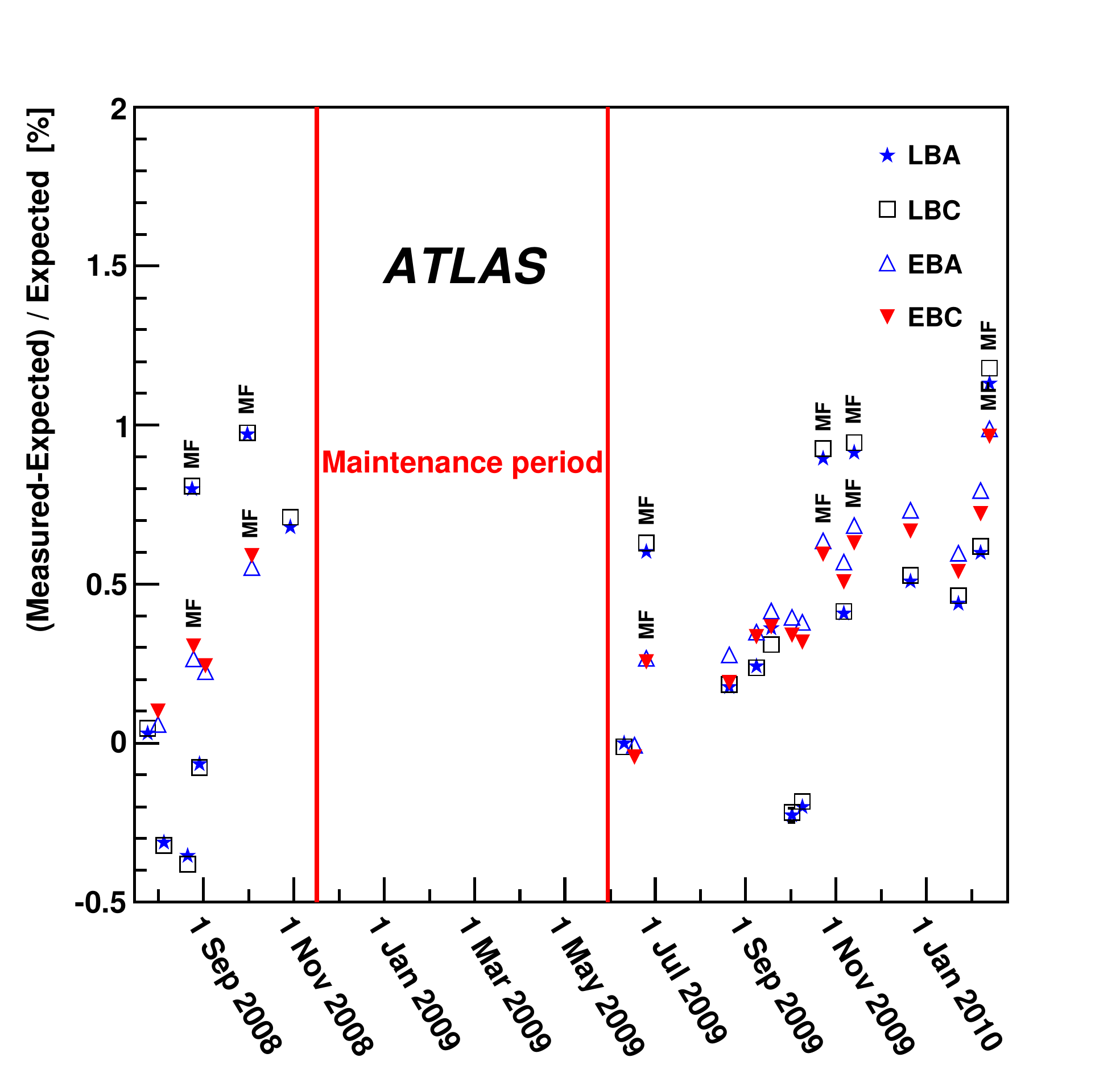}
 }%
 \resizebox{0.49\textwidth}{!}{%
   \includegraphics{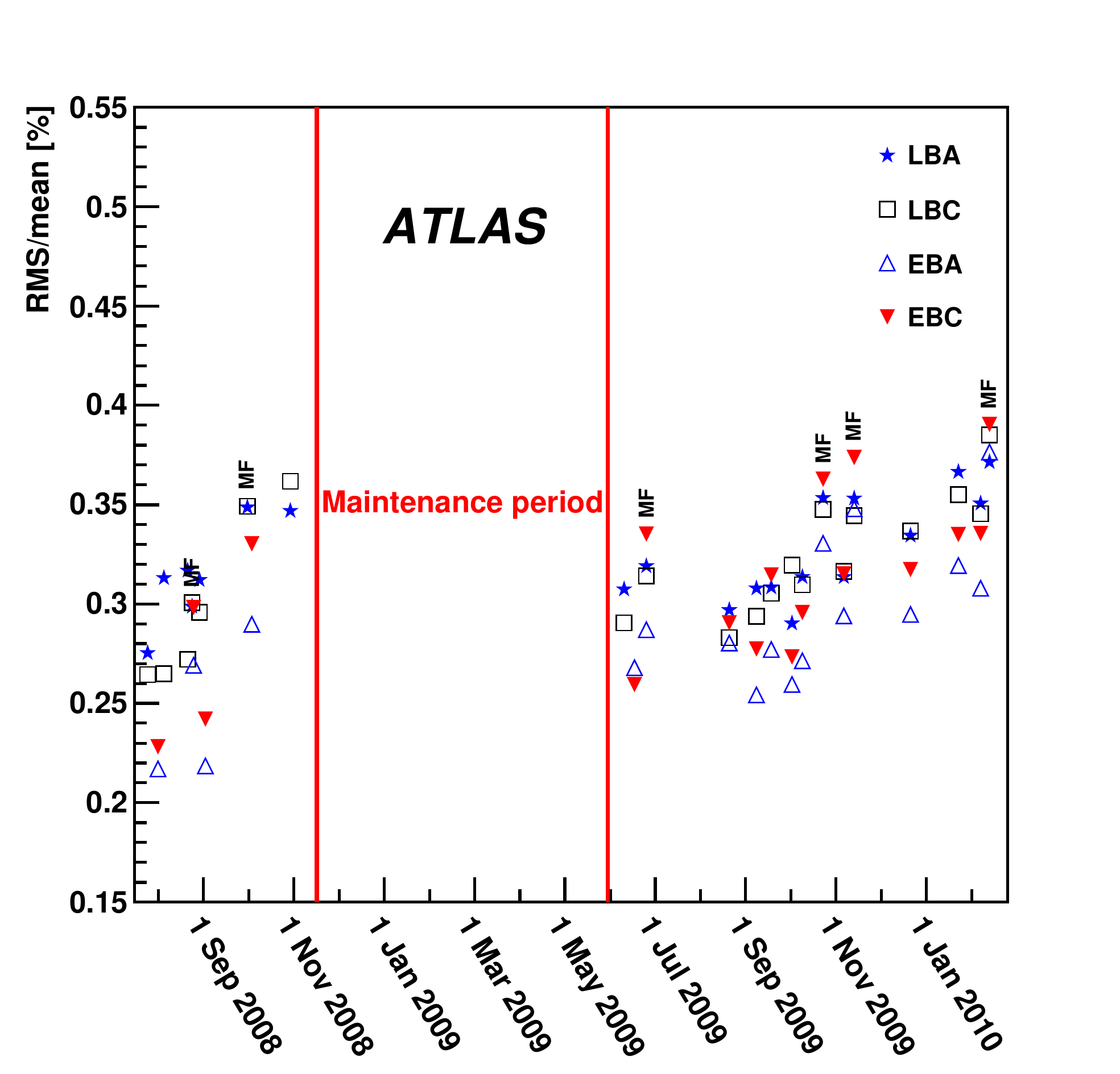}
 }
 \caption{Relative deviations of the TileCal response to 
   Cs sources from the expected value for all four calorimeter
   partitions, shown as a function of time (left). The ratio of
   RMS/mean of the TileCal response to radioactive  
   Cs source in all four calorimeter partitions is shown as function of
   time (right). The ``MF'' symbol stands for the Cs calibration data
   taken in the magnetic field. The response is averaged over the channels of each partition. 
 }
 \label{fig:calib-CS-5}
\end{figure*}

The relative deviation of the measured Cs response from the expected values due to the decrease in the source activity is shown in Figure~\ref{fig:calib-CS-5} (left) for the same set of the calibration runs reported above. Similarly, the maintenance period is excluded and the ``MF'' marks are used when the magnetic field was present during the calibration. The overall TileCal response to the radioactive sources follows the expected Cs decay within 1\,\% when no magnetic field is applied. Within this 1\,\%, there is a visible deviation from the expected decay line with increasing average response over time. 
A study of the Cs calibration procedure has been unable to attribute this increase to any subtle systematic effect, therefore it is attributed to an increase in the detector response and it is under investigation.
A conservative time dependent systematic uncertainty on the calibration of the EM scale of about 0.1\,\% per month is adopted to account for this effect. It is estimated from the Cs data with no magnetic field within two periods of 3 and 7 months in 2008, 2009 and 2010. 
The ratio of RMS/mean of the TileCal response to radioactive 
Cs sources in all four calorimeter partitions is shown as function of time in Figure~\ref{fig:calib-CS-5} (right). The spread in the measured Cs responses stays within 0.4\,\% over seventeen months indicating that the cell-to-cell intercalibration does not significantly change over this period of time. A small effect of the magnetic field on the Cs response spread is also clearly seen. 

\subsection{Calorimeter intercalibration} 
\label{sec:intercalibration}
In this Section the understanding of the cell and layer
intercalibration acquired from the testbeam
and from calibration and single beam data is exposed. The intercalibration as validated by cosmic muons is exposed later in Section~\ref{sec:performance_uniformity}. 

The intercalibration with 
Cs sources in the ATLAS cavern reports channel
response non-uniformities at the level of $\sim 0.4\,\%$ in each
Cs-scan, which is compatible with the precision of this calibration
system. Since the Cs system uses a different readout path than what is used for
the physics signal induced by particles (digital readout), other calibration
uncertainties also have to be considered. 
The charge injection system reports negligible non-uniformity after
the channel-to-channel corrections. The integrators
contribute at the level of $0.05\,\%$, which is negligible.
Altogether, the non-uniformities are given mostly by the Cs system. 
The Cs scans of the whole Tile Calorimeter revealed the same 
level of uniformity among individual optical elements in a cell as was
measured during the optics instrumentation period.

\begin{figure*}
  \centering
  \resizebox{0.49\textwidth}{!}{%
    \includegraphics{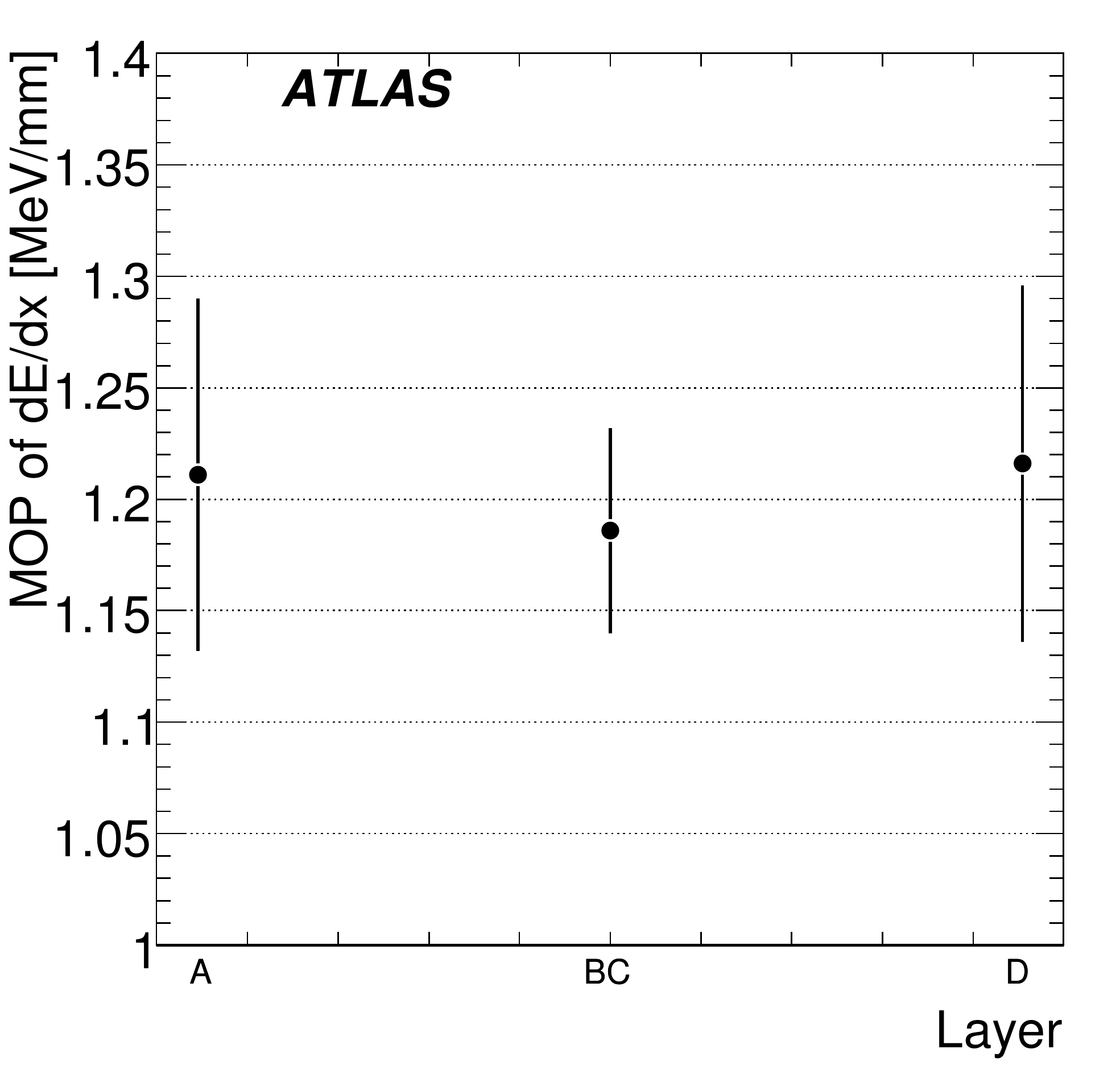}
  }%
  \resizebox{0.49\textwidth}{!}{%
    \includegraphics{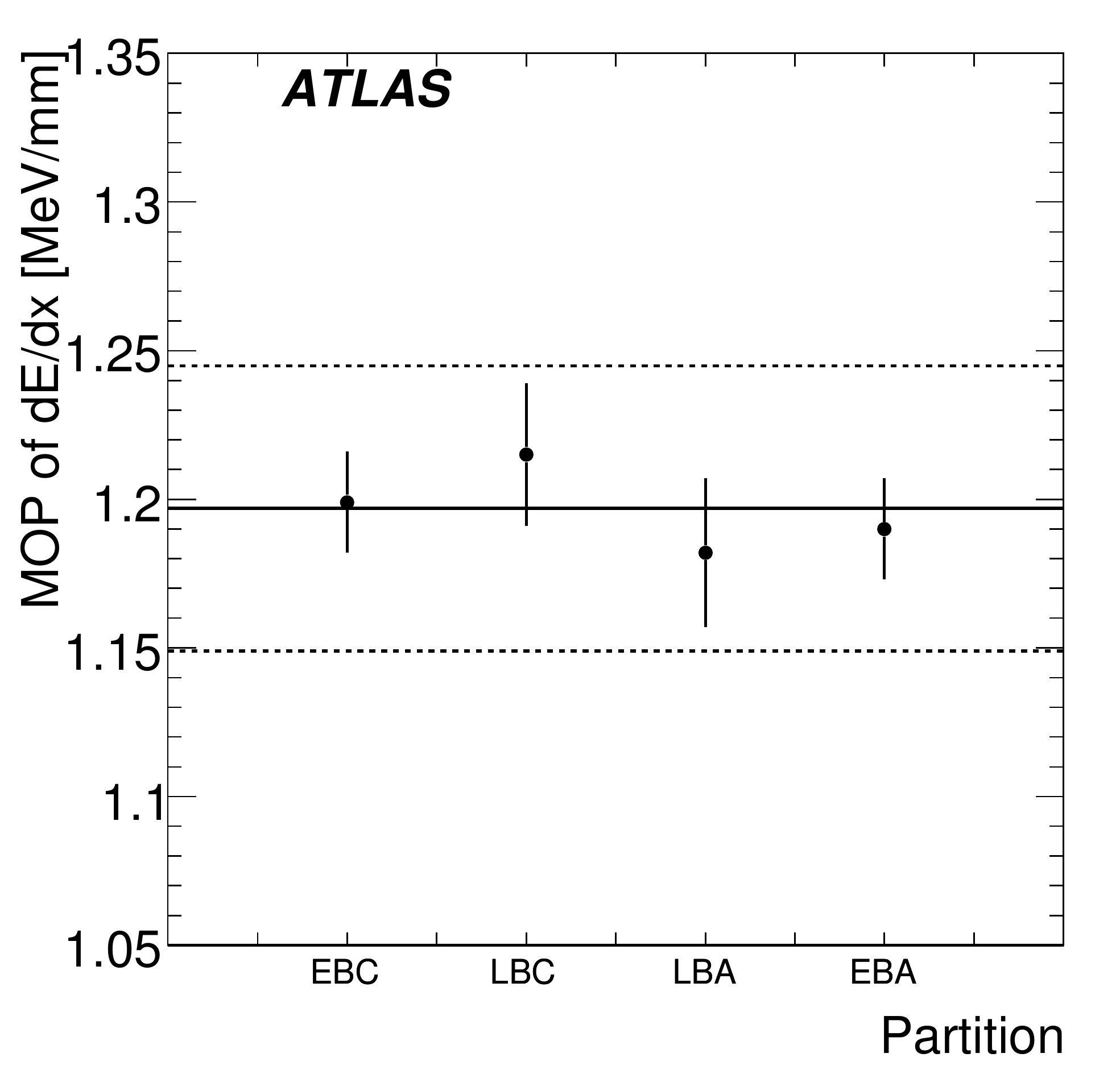}
  }
  \caption{The average energy measured in the single beam
    events recorded in September 2008. Left: average energy measured
    in individual radial layers after the radial
    layer corrections were applied (A is the inner radius layer). Right: the average energy
    measured in individual partitions, demonstrating good
    intercalibration between them.}
  \label{fig:inter_calibration}
\end{figure*}

In the testbeam, a difference in the response to the Cs source and to particles was 
observed, increasing for layers at larger radius~\cite{TileTBpaper}. 
This is due to the increasing size of the scintillator tiles for the external layers, 
and the resulting few percent layer miscalibration is accounted for by applying radial
depth weights in the energy scale calibration. The details of this procedure are
described in Ref.~\cite{note_emscale}. 
%
Figure~\ref{fig:inter_calibration} (left) shows the $\mathrm{d}E/\mathrm{d}x$
for muons crossing the calorimeter parallel to the beam axis along its
whole length from scraping events\footnote{Events produced by the
  proton beam hitting the edge of the collimators located at about 140~m upstream
  ATLAS\@.} in 2008. 
  The $\mathrm{d}E/\mathrm{d}x$ response for the muons from single beam
events was estimated as the peak of the fit to the convolution of a
Landau function with a Gaussian (most probable value, referred 
throughout the paper as MOP). 
  Within a large statistical uncertainty, the response  
vs radial layer is flat. Given the fact that if the radial depth weights had not been applied the ratio of responses between layers A and D would be $1.088$, this observation gives confidence in their use.   
Figure~\ref{fig:inter_calibration} (right) shows the mean response of
the four TileCal partitions to muons. Data are from 2008 single beam
runs. The precision is limited by the systematic uncertainty of $\sim 4\,\%$, while the statistical uncertainty is $\sim 2\,\%$.


\subsection{Uncertainty on the propagation of the EM scale from testbeam} 
\label{sec:emscale}

The EM scale of TileCal in ATLAS is set by adjusting the PMT HV to
reproduce the calorimeter response to the Cs radioactive source to the
level it had during the tests with electron beams, where the EM scale
was determined and measured.\footnote{The modules that were calibrated with
  the beams were carefully chosen to give a representative sample of the full
  TileCal module population.  Thus no significant uncertainty on the EM
  scale is expected to result from data obtained with the electron
  beams.} After correcting for the expected decrease in the Cs source
intensity, the HV levels currently set in ATLAS are expected to
reproduce those used at the testbeam. Any difference in the detector
parameters from that observed at the testbeam, if not fully
understood or disproved and if it affects the EM scale setting, should be
considered as the systematic uncertainty on the EM scale determination. 

The following sources of systematic uncertainties on the EM scale, as discussed in the previous Sections, are only related to the transfer of the EM calibration factor from the testbeam to ATLAS because they originate from differences between the two setups:
\begin{itemize}
\item 0.1\,\% from the calibration of the digital readout (HG, LG) by CIS.
\item 0.2\,\% from the calibration of the Cs readout gains.
\item 0.5\,\% from the non-reproducibility of the calorimeter response after the magnetic field is turned off, as reported by Cs measurements in 2008 (see Section~\ref{sec:cs_mfield}). 
\item 0.3\,\% from the uncertainty to the radial depth weights, briefly
  mentioned in Section~\ref{sec:intercalibration}.
\end{itemize}
The first two uncertainties were evaluated by comparing calibration
results on a fixed sample of channels which were calibrated during the
testbeam and then re-calibrated recently in ATLAS\@. The two latter
are related to observations with limited understanding of the underlying
phenomena. When these uncertainties are combined in quadrature with
the statistical uncertainty on the EM scale derived at the testbeams, the
result is a systematic uncertainty of $\pm0.7\,\%$. 

In addition to the above, there is a systematic uncertainty from the observed increase of the calorimeter response to the Cs source with respect to the expected value by about 0.1\,\% per month as observed during 10 months of frequent monitoring in 2008 and $2009 - 2010$. This is a time dependent uncertainty increasing since the initial EM scale setting in ATLAS in June 2008. 
\begin{itemize}
\item Presently (early 2010) we assign an uncertainty of $-1.5\,\%$ due to
  the increasing response of roughly 0.1\,\%/month as measured by the
  Cs system during 2008 to 2010. 
\item During the data-taking period from which cosmic muon results are
  presented in this paper (September to October 2008), the same
  uncertainty was $-0.8\,\%$. 
\end{itemize}


After setting the EM scale in ATLAS, the high voltage values applied
on the PMTs were compared between the
testbeam periods (2002 to 2004) and June 2008.
While the TileCal response has been calibrated reliably with the Cs
system to match the response measured during beam calibrations and hence to
transfer EM scale to the ATLAS cavern, the PMT high voltages
for the LB partition in June 2008
had to be lowered on average by $(6.5\pm0.2)$~V compared to
those used during testbeam calibrations. This was due to the fact that
the Cs system measured an increased response in June 2008 for the beam
calibrated modules with respect to their response in testbeams.
%
%
If this response increase had not been a detector effect
but an artifact of the Cs calibration system, a corresponding bias of
-5.3\,\% (the true energy being higher than the measured one) would have to
be considered as an uncertainty for the cosmic data taken in autumn 2008.
This would be added to the uncertainty from the observed increase of
roughly 0.1\,\% per month since June 2008, as mentioned above.

The energy response from muons is a handle to assess this uncertainty
or bias. A full description on the energy scale analysis with cosmic and
testbeam is given in 
Section~\ref{sec:energy_performance}. The comparison between the
testbeam and ATLAS EM scale is performed via the double ratio of
$\mathrm{d}E/\mathrm{d}x$ Data/MC ratios of cosmic over testbeam
muons for LB modules. In other words, the agreement of data to the MC
energy scale between testbeam and ATLAS is compared.
Table~\ref{tab:dedx_truncated_mean} presents the values and the uncertainties of the
above mentioned double ratio per layer. 
Among the calibration related uncertainties, the contributions from the non-reproducibility of the response increase due to magnetic field and from the unexplained response increase measured by the Cs during 2008 are comprised. The reported ratios show an agreement of the EM scale
set in 2008 and the expected scale as it was transported from the
testbeam within the uncertainty range. However, the possible calibration bias
mentioned in the previous paragraph, that would be represented by a double ratio of $0.95$, can be excluded only at 
a $\lesssim 2 \sigma$ level.

If the uncertainty coming from the reduced high voltage settings with
respect to the testbeam is not taken into account, the overall
estimate of the EM scale systematic uncertainty from the calibrations
is $(-1.7\,\%,+0.7\,\%)$ in early 2010.\footnote{This uncertainty is
  $(-1.1\,\%,+0.7\,\%)$ for October 2008, the period in which the
  cosmic muon data of this paper were collected.}

\subsection{Timing calibration} 
\label{sec:timing_calibration}

To allow for optimal reconstruction of the energy deposited in the
calorimeter by the OF signal reconstruction method (see
Section~\ref{sec:energy_time_reco}), the time difference between the
digitising sampling clock and the peak of the PMT pulses must be
minimised and measured with a precision of 1~ns. To achieve this, the
clock phases in the DMUs in the front-end hardware (see
Section~\ref{sec:overview}) are adjusted in multiples of
0.1~ns. Ideally all PMT signals would be sampled at the peak but
several factors limit the ability to do this. First, the clock phase
is defined per digitiser board which corresponds to six readout
channels. Second, only one clock phase can be defined for both gains
and there is a 2.3~ns difference between the HG and LG pulse
peaks. Therefore in the front-end hardware, the accuracy of phase
synchronisation for individual channels is limited to be within
3~ns. Any residual time differences between the clock phase and the
pulse peak are measured for each channel and accounted for in the OF
signal reconstruction algorithm.  

The time phase and the residual offsets for all channels can be
measured using the laser calibration system, cosmic-ray events, beam
splash and collision events.  What is exposed in this Section is the
procedure to only pre-set the timing in order to synchronise the
detector with the trigger signals and with the other detectors prior
to the final detailed adjustments, to be carried out with collisions
data. 

Prior to beam, the laser was the primary source used to measure the
channel timing. Since the laser light is asynchronous with respect to
the clock, a single reference channel in each partition was selected
and all other channels' timing was defined with respect to that
reference~\cite{note_timing_laser}. The timing precision for channels
in the same module is 0.6~ns for 99\,\% of the Tile Calorimeter
readout channels. In addition, the mean time difference between the HG
and the LG was measured to be $(2.3\pm0.4)$~ns. One limitation in the
laser system for timing calibration is understanding the propagation
time in the laser fibres from the laser source to the PMTs. For this
reason, the inter-partition timing and global timing with 
respect to the rest of ATLAS were coarsely set using cosmic-ray data
and more accurately using 2008 beam data.  

\begin{figure*}
 \centering
 \resizebox{0.49\textwidth}{!}{%
   \includegraphics{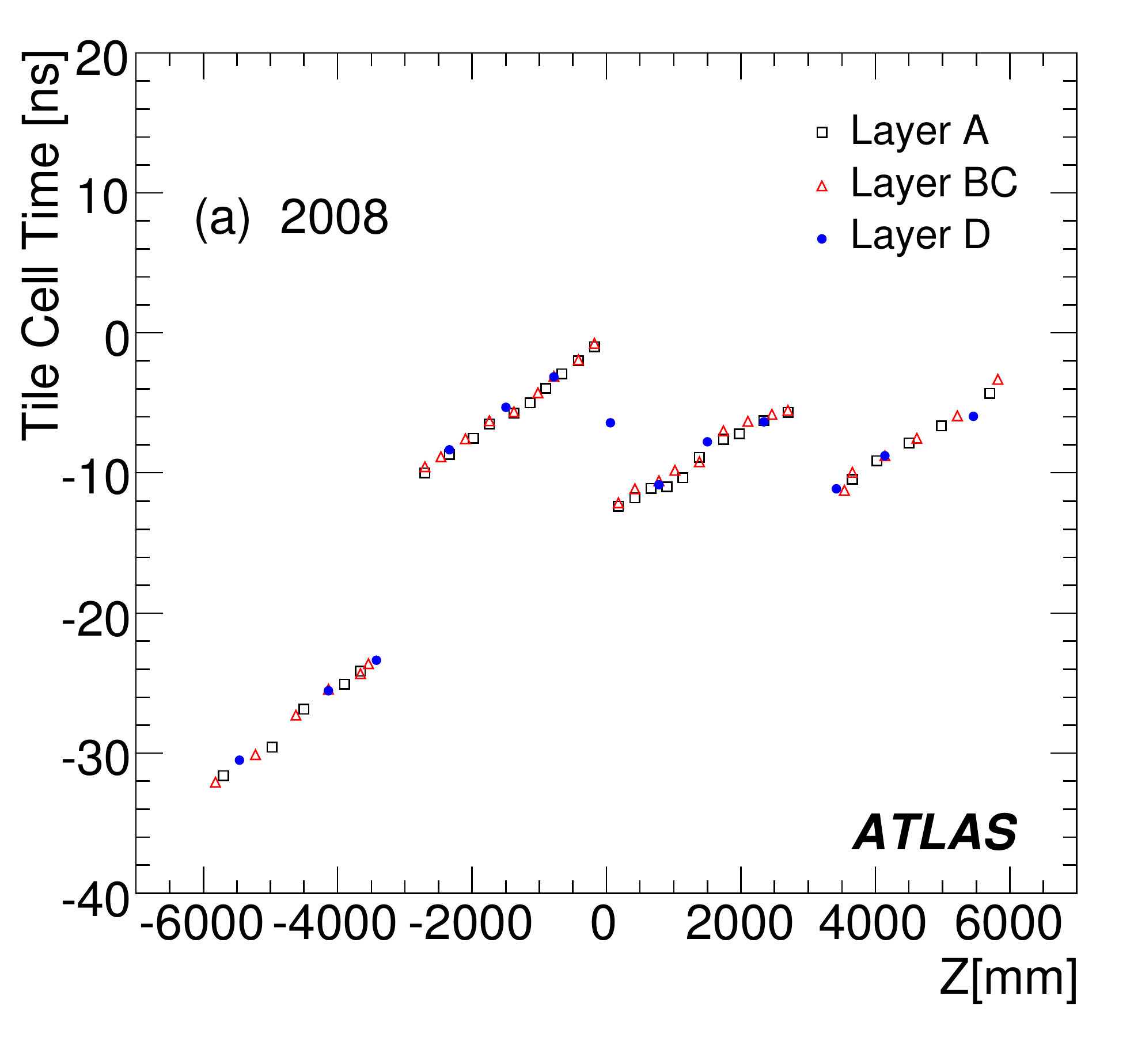}
 }%
 \resizebox{0.49\textwidth}{!}{%
   \includegraphics{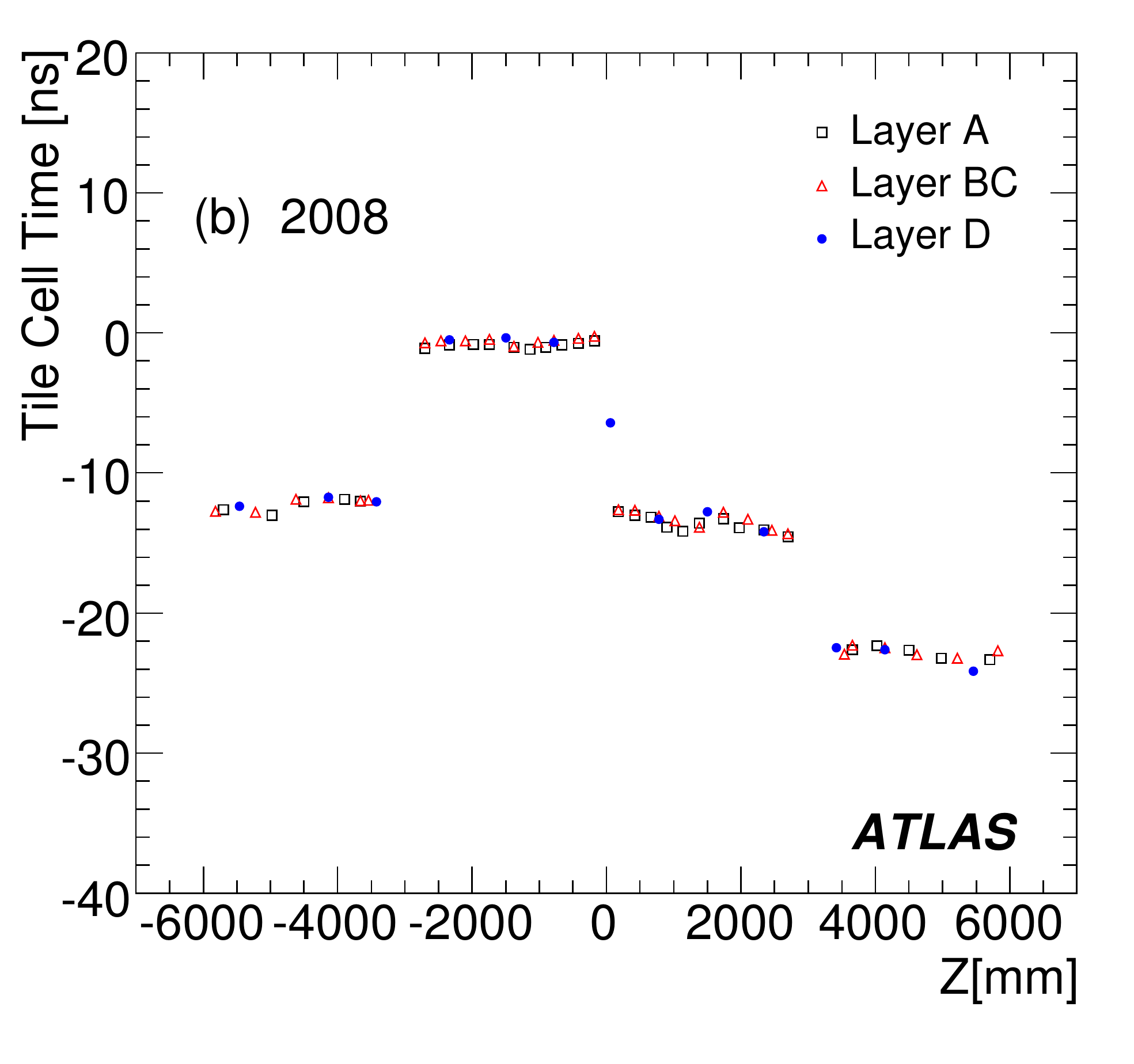}
 }
 \resizebox{0.49\textwidth}{!}{%
   \includegraphics{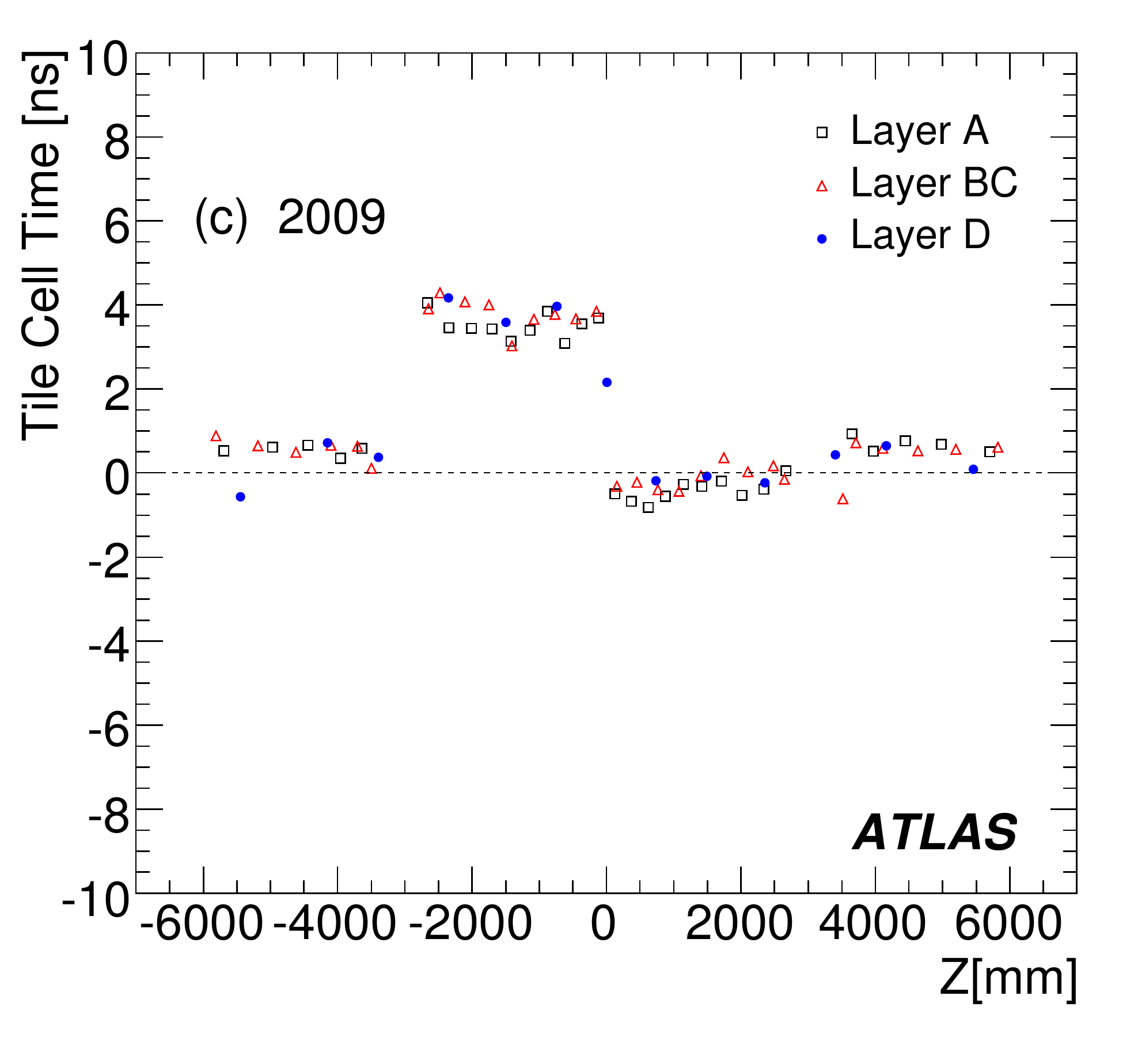}
 }%
 \resizebox{0.49\textwidth}{!}{%
   \includegraphics{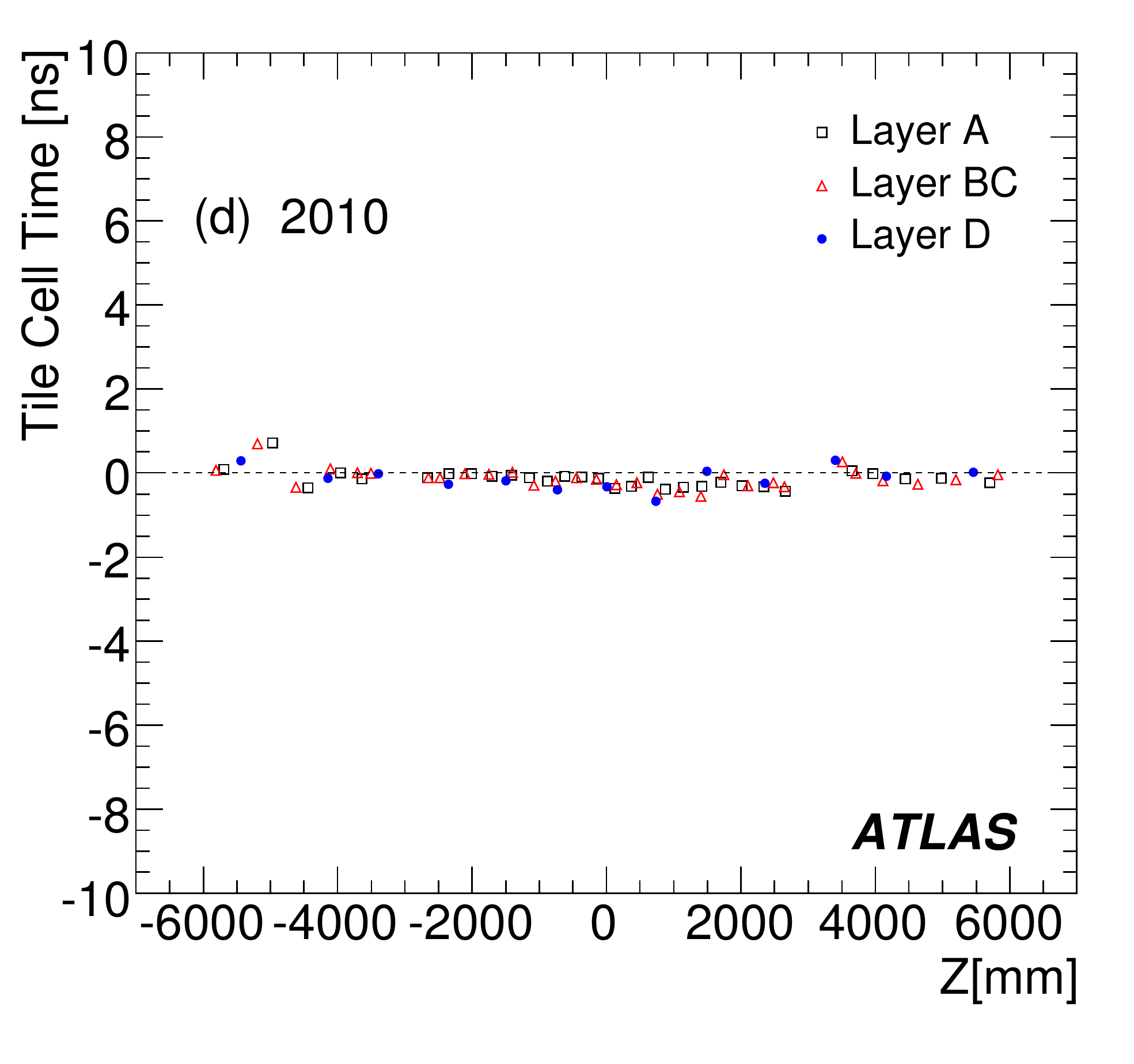}
 }
  \caption{Timing of TileCal signals recorded with single beam data in
    September 2008 (a and b), November 2009 (c) and February 2010 (d). The
    time is averaged over the full range of the azimuthal angle $\phi$
    for all cells with the same $Z$-coordinate
    (along beam axis), shown separately for the three radial layers.
    Corrections for the muon time-of-flight along the z axis are applied in the 
    b), c) and d) figures, but not on the top left (a).
  }
  \label{fig:time-1}
\end{figure*}

The timing calibration based on laser data was validated using beam
splash events.
These events contain millions of high-energy
particles arriving simultaneously in the ATLAS detector. Since the total
deposited energy is large, it is only possible to study the
timing response in the LG\@.  Using these events, the time
intercalibration of individual channels in the same module was
confirmed to be 0.6~ns. 
  
Figure~\ref{fig:time-1} shows the cell time measured in beam splash
events, averaged over the full range of the azimuthal angle $\phi$
for all cells with the same $z$-coordinate of ATLAS (along the beam
axis). The visible discontinuities at $Z = 0$ and  $Z = \pm 3000$~mm
for the 2008 
data are due to the uncorrected time differences between the four
TileCal partitions.  These were calculated using the 2008 data and
adjusted for the 2009 running period. After the muon time-of-flight
corrections (b), the timing shows an almost flat
distribution within 2~ns in each partition, confirming a good
intercalibration between modules with the laser system. The residual
slopes, present in all modules, were
corrected for by comparing the 2008 single beam data to the laser data
and optimising the effective speed of light in the calibration system
optical fibres. Consequently, in 2009, the TOF-corrected timing
distribution (c) is even more uniform. In preparation for the 2010 run,
the 2009 single beam results were used to provide the offsets for all cells and,
as is shown in Figure~\ref{fig:time-1}~d) for the 2010 single beam results, all remaining
disuniformities were corrected for.  The spread of the TileCal cell timing distribution
 at the start of the 7~TeV collisions is of 0.5~ns\footnote{This value takes into account 97\,\%
 of the TileCal channels. The timing for the remaining outliers was adjusted offline.}.

%

%
%

\hyphenation{ca-lo-ri-me-ter}
%
%
\section{Performance with cosmic ray muons}
\label{sec:performance}
The calorimeter response to muons is an important issue since isolated
muons will provide a signature of interesting physics events in the 
LHC collisions phase. For example, semileptonic $t\bar{t}$ decays, the
so-called ``gold-plated'' Higgs decay channel $H\rightarrow
Z^{0}+Z^{0}$ and some SUSY processes involve high-$p_\mathrm{T}$ muons
in their final states, while low-$p_\mathrm{T}$ muons originate from
$B$-meson decays~\cite{ATLAS_CSC}. In addition, since the
interaction of muons with matter is well understood, the prediction of
this response is reliable, and its investigation with data can provide
information on the detector performance and intercalibration.

The TileCal energy response performance was studied using cosmic muon data 
collected in 2008, with the goal of verifying the calibration in terms of EM scale and its uniformity over the whole calorimeter.
After an initial comparison of the muon energy signal and the
corresponding noise in the same set of cells (in
Section~\ref{sec:signal_noise}), the methods and results
of the studies of muon response versus path length are described.
These studies were based on the extrapolation into TileCal of cosmic
muon tracks reconstructed by the Inner Detector, which is described in
Section~\ref{sec:ID_tracks_event_selection}. The performance of the
energy response to 
testbeam muons 
was also checked at low energy, for comparison.

Muon response results and comparison to Monte Carlo simulations are
presented in Section~\ref{sec:energy_performance}. This Section
focuses on several key issues: 
the response uniformity versus radial layer, $\eta$
and $\phi$, the propagation of the EM
scale from testbeam to the full detector configuration in
the ATLAS cavern,  
and a discussion on the systematic uncertainties, 
such as the
ones arising from possible biases of the muon response estimation with
the muon momentum and path length. 
A separate Section~\ref{sec:itc_calibration} is devoted to 
calibration of special TileCal cells (ITC, gap and crack
scintillators). 

The measurement of the time-of-flight of particles in TileCal can
be used either for background removal (cosmic and
non-collision events) or physics
analyses~\cite{note_Rupert}.
A good synchronisation of the TileCal cells is important for that, and
its validation with cosmic ray muons is described in Section~\ref{sec:performance_time_reco}.

\subsection{Muon response compared to noise}
\label{sec:signal_noise}
The TileCal readout system is designed so that even small signals induced
by muons are well separated from the noise. This feature has been
demonstrated with testbeam data~\cite{TileTBpaper}. 
Nevertheless the performance has to be confirmed
with data taken with the full ATLAS detector, since the environment is
more noisy and 
changes to the powering system have been made.

This exercise was performed on a large statistics run, with the data
sample described in Section~\ref{sec:datataking}: events 
from various first level triggers were required to have at least one 
reconstructed Inner Detector track. However, these tracks were not
used in any further event or cell selection, for this study. Instead,
a different method was used, based on track reconstruction using only
TileCal data. This algorithm, named TileMuonFitter, was developed for
the data analysis and monitoring of TileCal in the cosmic muon
commissioning phase~\cite{proc_hough,hough2}. It uses no
external tracking information and uses the set of TileCal cells with
energy above a 250~MeV threshold to fit a straight line from the top
to the bottom cells (it therefore also ignores the track curvature
inside the solenoid magnetic field).  In order to reproduce as closely
as possible the signal from muons originating in physics collisions, a
loose projectivity requirement was imposed. Tracks were selected
according to the coordinates of their intersection with the horizontal
plane (within $\pm 400$~mm) and to their angle with respect to the
vertical, corresponding to a pseudorapidity range of $0.3 < |\eta| <
0.4$.

\begin{figure*}
  \centering
 \resizebox{0.49\textwidth}{!}{%
   \includegraphics{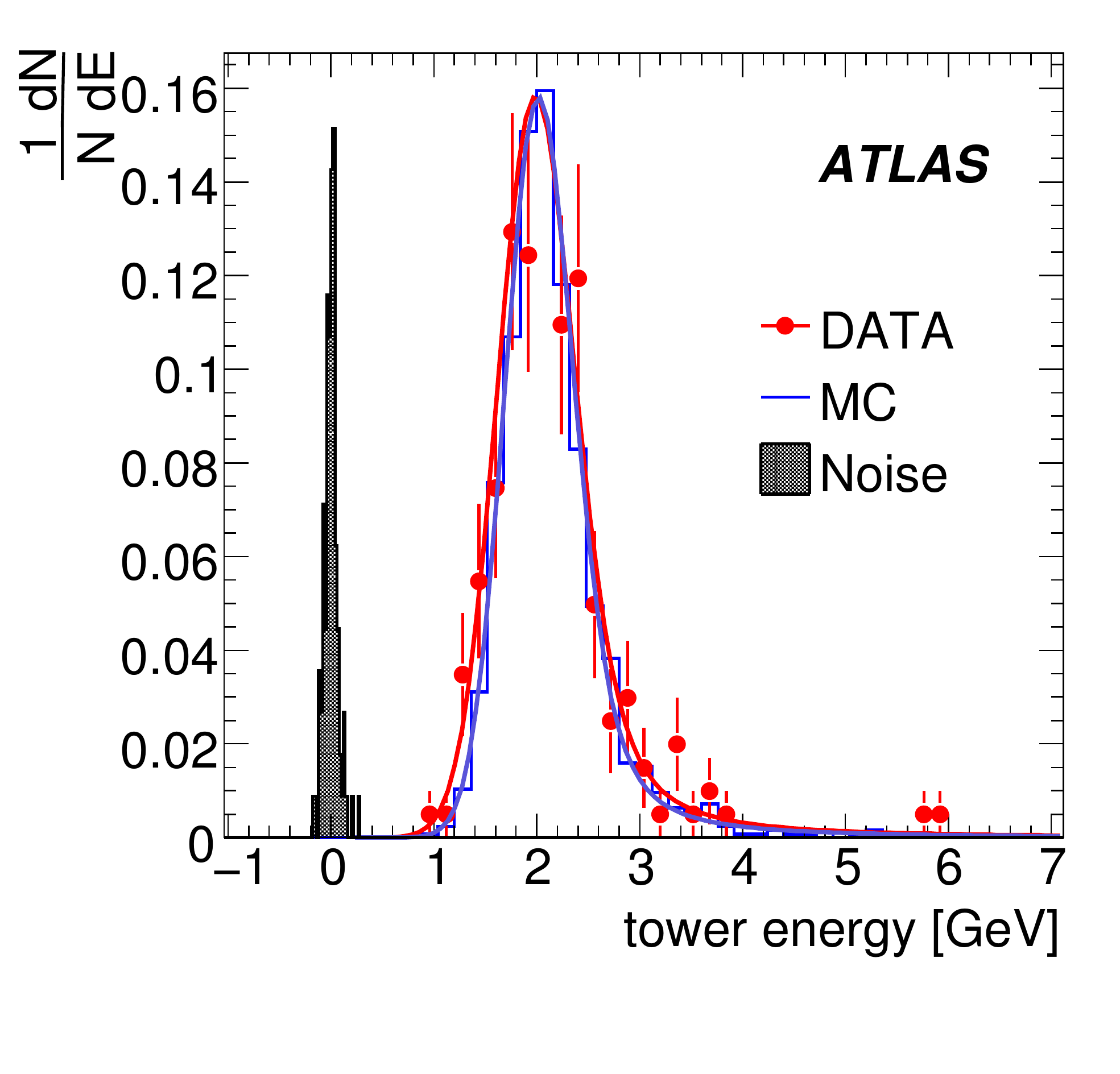}
 }%
 \resizebox{0.49\textwidth}{!}{%
   \includegraphics{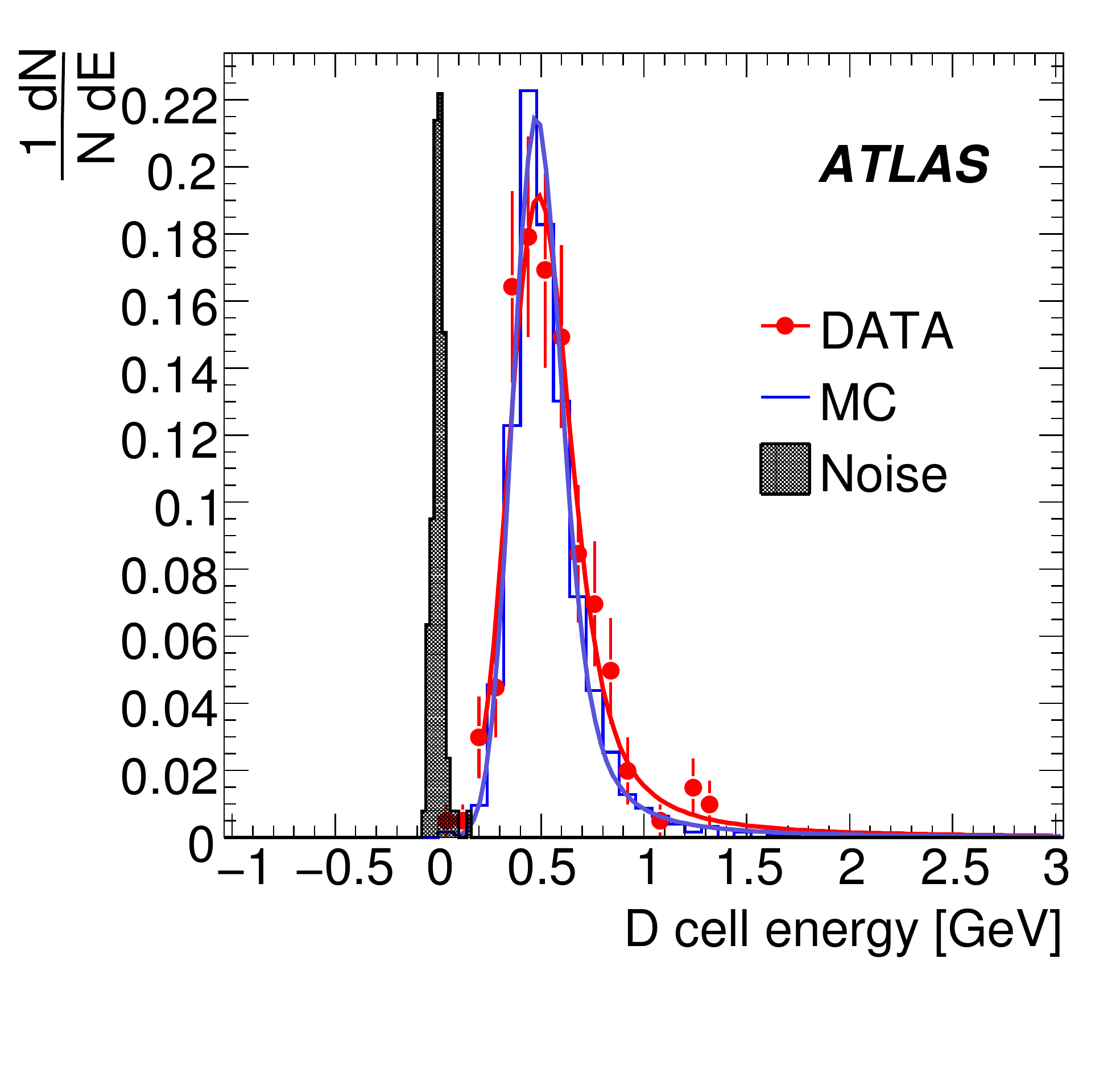}
 }
  \caption{Example of the muon signal and corresponding noise for
    projective cosmic muons entering the barrel modules at
    $0.3 < |\eta| < 0.4$. Top and bottom 
    modules are treated separately and the momentum range of the
    cosmic muons was restricted to be between 10 and 30~GeV/$c$. Left: the
    total energy summed up over selected cells. Right: the similar distribution 
    of last radial compartments that can be eventually used to assist
    in muon identification. The signal (red) comes from the
    cosmic muon data sample (see text), the
    corresponding noise (black) is obtained with the random trigger sample.
    }
  \label{fig:TMF_signal_to_noise_projective}
\end{figure*}

The signal is either the total energy in TileCal summed up over cells 
selected by the TileMuonFitter algorithm, or the
response in the last radial compartment for the D-cells selected by
that algorithm. 
The noise is evaluated from random triggers using the
same cells as for signal.
The results are shown in Fig.~\ref{fig:TMF_signal_to_noise_projective} for tracks entering barrel modules within the pseudorapidity range $0.3 < |\eta| < 0.4$. Top and bottom module responses are considered as two independent entries, so the signal corresponds to that of one module.
 The signal and noise distributions
are well separated for both the total calorimeter response and the last
radial layer signal.

In order to estimate the signal-to-noise ratio, the energy
distribution is fit to the convolution of a Landau function with a
Gaussian. Considering the peak of that convolution fit 
as the signal, and the RMS of the random trigger
distribution as the noise, the signal-to-noise ratio is then $S/N =
29$ for the total response and  $S/N = 16$ for D cells. Since muons
are the smallest
energy signals that TileCal will measure, these values show a good
performance of the calorimeter.
The obtained values are lower than for testbeam\footnote{In
  testbeam~\cite{TileTBpaper}, muon beams at a nominal energy of
  180~GeV were used for this study. Taking into account the 20~GeV to
  180~GeV response ratio, the testbeam $S/N$ ratios at 20~GeV for the
  tower and the D cells should amount to 42 and 17 respectively.},
but the difference is consistent with a higher noise level in the
ATLAS cavern and with a higher number of cells being summed.


\subsection{Methods for muon response studies}
A brief overview of the analysis methods applied to investigated data
samples is provided in this Section.
First, we briefly describe the dedicated testbeam (TB) studies with
low-energy muons (Section~\ref{sec:tbmuons}). The algorithms and
event selection used in the cosmic data analysis are then reported in
Section~\ref{sec:ID_tracks_event_selection}.  

\subsubsection{Analysis of low energy testbeam muons}
\label{sec:tbmuons}
The TB setup, operating conditions and results are summarised in
Ref.~\cite{TileTBpaper}.  
Since most of the previous muon TB results were obtained with 
180~GeV beams and this energy is too high for the comparison with 
cosmic ray data, a dedicated study was performed with
low-energy muons selected from a pion beam at a nominal energy of
20~GeV. These muons originate from pion decay, the distribution of their
momenta is calculated to range from 11.5~GeV/$c$ to 20~GeV/$c$, peaking
at around 17~GeV/$c$.  
Data was collected from ten runs with pion beams impinging on
one barrel module at different projective incidences, from 
$-0.65\leq \eta \leq -0.15$ and $0.15\leq \eta \leq 0.45$. 

Two sets of cuts were applied to select muons from the nominal
pion beam:
\begin{itemize}
\item Single particle events were selected by requiring a MIP-like
  response in the beam scintillators upstream of the calorimeter
  modules. Particles with large angle with respect to the beam axis
  and/or halo particles were removed by applying suitable cuts in the
  upstream beam chambers.
\item Contrary to muons, pions produce hadronic showers that leave
  signal also in towers 
  surrounding the one hit by the beam. This feature is exploited in the
  muon/pion selection -- events with signal above noise
  ($E \gtrsim 3\sigma_\mathrm{noise}$) in
neighbouring towers were considered pions and were removed from
  further analysis.
  Moreover, an upper limit on the response in the 
  impact cell in the first calorimeter radial layer was imposed, in order to
  avoid pion showers with large electromagnetic shower fraction, whose
  typical lateral (in $\eta\times\phi$) size is smaller than that of a
  cell. 
\end{itemize}
As the projective beams hit the centre of the given calorimeter tower,
the muon response was
summed up only from cells in the impact tower. The selection
criteria mentioned above guarantee a muon to impinge on the selected tower,
therefore no further cut to reject noise events was needed. 

The muon track length in the given cell was considered as the 
radial size of that cell divided by the cosine of the beam incident
angle. This approach is fully adequate
for projective muons entering the calorimeter at a cell's centre in
both $\eta$ and $\phi$ direction, see also Fig.~\ref{fig:tile-cellmap}.

The Monte Carlo simulation of the TB setup takes into account the 
detailed detector and beam geometry as well as the momentum
distribution of the incident muons.

\subsubsection{Analysis of the cosmic ray muons with tracks reconstructed by
  the Inner Detector} 
\label{sec:ID_tracks_event_selection}

The performance of the calorimeter was analysed by taking 
advantage of the information provided by the central tracking. 
This is an important handle  for the study of the  calorimeter cell
response which is  sensitive to the muon path length and momentum. 
  
\paragraph{Track extrapolation and event selection\\}
Events  were triggered at the first level trigger by RPC and TGC\@.
The tracking information is obtained from the Inner Detector
reconstruction, without further contribution from the Muon Spectrometer.
Selected events are required to have one reconstructed track in the
SCT volume. 
Events with reconstructed multiple tracks  are  rejected.
Tracks  in the TRT do not have  $\eta$ information and  are not used
in the  study.
The quality of the  tracks is enhanced by requiring at least eight hits 
in the silicon detectors (Pixel and SCT).
The tracking requirements introduce some cut-off in the  distributions 
of transverse and longitudinal  impact parameters. 
These are $|d_0|\leq 380$~mm  and $|z_0| \leq 800$~mm,
respectively.\footnote{The transverse impact parameter is defined
  as the distance to the beam axis of the point of the closest
  approach of the track to the coordinate origin. The longitudinal
  impact parameter is the $z$-coordinate (along the beam axis) of the
  same point.}

The tracks  are extrapolated through the volume of the calorimeters
using the tool described in Ref.~\cite{AtlasExtrapolator}, which uses
propagation of the track parameters and covariances that
take into account material and magnetic field. Extrapolation is
performed in  both directions, along the muon momentum and opposite to
it. This allows to study  the  response of  modules in the  
top and bottom part of the detector. Since the track parameters are
measured in the centre the method could be sensitive to systematic
differences top/bottom.    

\begin{figure}
  \centering
  \resizebox{0.5\textwidth}{!}{%
    \includegraphics{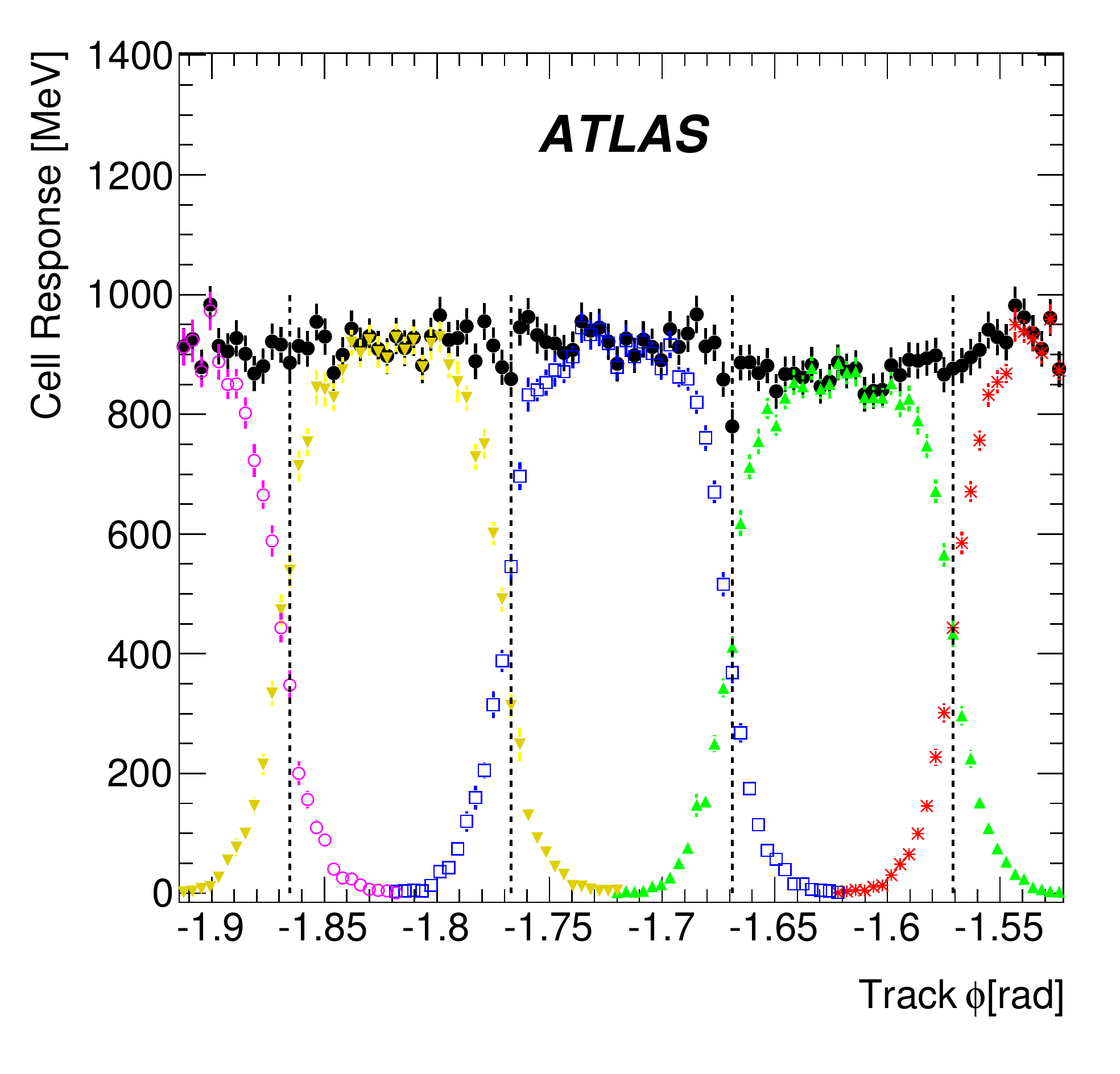}
  }
  \caption{Mean response of cells in the second layer as a function of
    track $\phi$-coordinate for the bottom central region of the
    calorimeter. 
    Tracks with $10 < p < 30$~GeV/$c$ were selected.
    The average response over all central region cells in
    the given module is shown by symbols of different colours/styles,
    whereas the total response summed over all modules is shown with 
    black full circles. Vertical lines denote nominal edges of the modules.
  } 
  \label{fig:module_prof}
\end{figure}

Figure~\ref{fig:module_prof} demonstrates the correct TileCal cell geometry
description. It shows the response of cells in the second  
layer as a function of the $\phi$-coordinate measured at the
inner-radius impact point in the given cell. The cells' response average is 
computed over tracks along the $\eta$ directions in
the central barrel region. The responses corresponding to cells of
individual modules 
(width of $\Delta\phi\approx 0.1$) are shown with symbols of different
colours/styles. The match with the nominal position of the cell edges,
displayed by vertical lines, is evident. The total response summed over
all modules is superimposed as well and it is reasonably uniform
across $\phi$.

The alignment between Tracker and Calorimeter  was investigated using
tracks with a limited transverse impact parameter ($|d_0| <
100$~mm). 
The alignment between tracks and
nominal cell edges in the second layer of TileCal is within the
selected bin size 
($\sim5$~mm). This precision is fully adequate for the correct
identification of the cells under study  and computation of
quantities  relevant to the analysis.  

One of the key parameters of the track is the path length through a
given cell. The track extrapolation provides crossing points of the
muon track in each radial layer. Additional linear
interpolations are performed using the detailed cell geometry to
define the entry and exit points for every cell. The track path
length is then evaluated as the distance between the entry and exit
points for every cell crossed by the muon. 
In the analysis we consider, for each event, only cells with path length $L >
20$~cm. 

 An upper limit of 30~GeV/$c$ on the muon momentum is used in the analysis
in order to restrict the muon radiative energy losses which show considerable 
fluctuations and can have an impact on data/MC comparisons.
In a small fraction of events the cell response is compatible 
with the pedestal level  although the cells should  be hit  by a muon. 
The muon  actually hits a neighbouring   module.  This  is consistent
with the expected deviation from the muon trajectory  due to multiple
scattering. In order to limit this effect we restrict the analysis to
muons with momenta as measured in the Inner Detector larger than
10~GeV/$c$ and apply a fiducial volume cut requiring the  track  to be
well within  the module (that has a half width of $\Delta\phi = 0.049$):  
\begin{equation}
|\phi_{track}-\phi_{cell}| <0.045.
\label{eq:IDTrackCut}
\end{equation}
In order to remove residual noise contribution, a cell energy
cut of 60~MeV is applied.

Muon tracks close to the vertical direction  are badly  measured in
the Tile  Calorimeter due to the  
strong sampling fraction variation caused by the vertical orientation
of the scintillating tiles. 
To ensure more stable results, tracks are required  
to enter in the cells with a minimal angle with respect to $\eta=0$ direction. 
Given the crossing  points at the inner and outer cell radial edges
we require   
\begin{equation}
  \label{eq:dz_inner_outer_edge}
  |z_{inner}-z_{outer}| \ge 6~\mathrm{cm}. 
\end{equation}
This cut has an appreciable effect only on 
very central cells, within the vertical coverage of the ID. 

Approximately 100~k data events satisfied the above mentioned
selection criteria and were further analysed. The corresponding
statistics available in the MC sample was about twice higher.
 
 \paragraph{Performance checks\\}
 \begin{figure}
   \centering
   \resizebox{0.5\textwidth}{!}{%
     \includegraphics{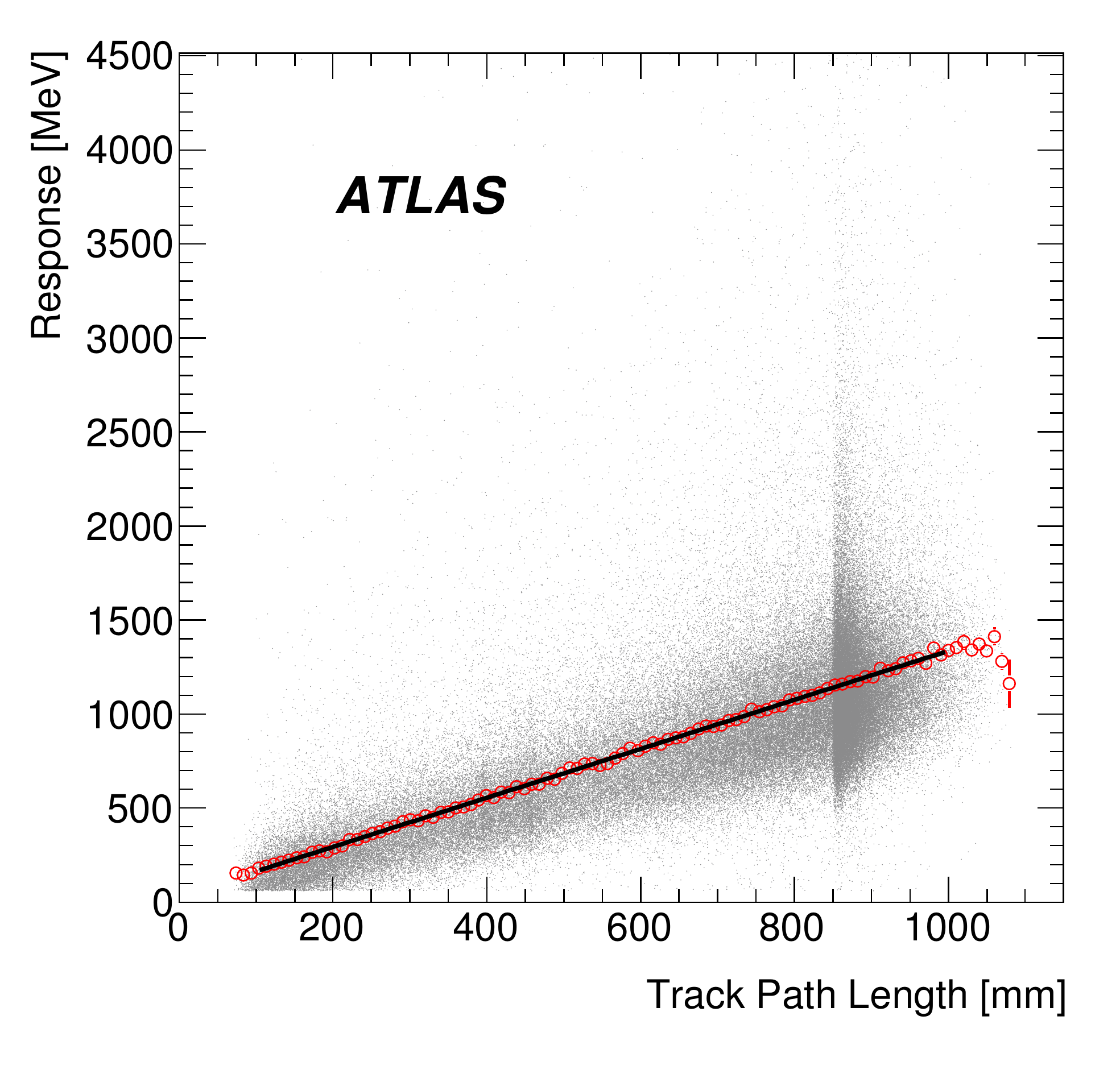}
   }
   \caption{Mean response of the barrel module BC cells  as a
     function of track path length for tracks with $10 < p < 30$~GeV/$c$.
     A linear fit to the corresponding distribution of mean values is
     superimposed. The excess of events at around the track path
     length of 840~mm (radial size of the barrel module BC cells) is a purely
     statistical effect, since most of the cosmic ray muons enter the
     calorimeter at small zenith angle.
   }
  \label{fig:pathl}
\end{figure}

The track path length is the main handle to study the muon response. 
Figure~\ref{fig:pathl} shows the response of cells in the second layer  as  a function of
the path length $x$.  It includes  cosmic events crossing the BC cells over the entire  barrel  and over all accepted angles.  
A clear edge at the path length of 840~mm is visible in the figure. This represents the radial depth $\Delta R$ of the BC layer  cells.
Since most cosmic rays are vertical, a  large fraction of the muons crossing the central region have a reconstructed path length equal or slightly  larger than the layer radius.
This is very evident for all cells  with a  $z$-coordinate within  the  vertical coverage of the SCT detector  $|z_0|<~1$~m. 
A linear fit  to the corresponding distribution of mean values shows that the
muon response scales approximately linearly with the path length, as expected. 
Figure~\ref{fig:pathl} suggests that  the ratio of the cell response  
with the track path length, i.e.~the slope of $\mathrm{d}E/\mathrm{d}x$,  is
one  of the quantities that can be used to study the  cell/layers
intercalibration. This will be discussed in more detail in
Section~\ref{sec:energy_performance}.

%
%

\subsection{Performance of energy response}
\label{sec:energy_performance}

In this subsection, the results of the calorimeter energy 
response studies carried out with cosmic muons are reported. The main
aim is to cross-check 
the energy scale set with testbeam and the calibration systems, both in
terms of the EM scale and of its uniformity across the detector cells. 
The uniformity of the response per cell and as a function of pseudorapidity and azimuthal angle is addressed in Section~\ref{sec:performance_uniformity}, while the layer intercalibration and EM scale issues are discussed in Sections~\ref{sec:cosmics_layer_calib} and~\ref{sec:results_energy_scale} respectively.

The energy response of TileCal to cosmic muons is probed by estimating the
muon energy loss per unit length of detector material, which is
obtained by dividing the energy measured by the path length crossed in
a given cell (calculated with the method described in
Section~\ref{sec:ID_tracks_event_selection}). For simplicity, we call
this quantity $\mathrm{d}E/\mathrm{d}x$, even if this is not rigorous,
since it is measured in a non-continuous way, and the TileCal cells
are made of two different materials, with a direction-dependent
sampling fraction.

Our estimator for the muon response is the truncated mean of
$\mathrm{d}E/\mathrm{d}x$, defined as 
the mean in which 1\,\% of the
events in the high-energy tails of the distribution are removed (the
number is rounded  to the lowest integer). 
The statistics 
of the data sample is limited and rare  processes
like bremsstrahlung or energetic $\delta$-rays can cause large fluctuations
of the full mean. The truncated mean is chosen since it is less sensitive 
to high-energy tails in the cells' response distribution, that are caused by 
the muon's radiative energy loss. 
For testbeam, the truncated mean estimator has an additional 
advantage over the full mean, since it removes residual pion signal
contamination. The truncated mean also removes muon events with very
large energy deposits (high-energy radiation and/or muon nuclear
interactions), therefore the muon/pion selection criterion (see
Section~\ref{sec:tbmuons}) does not introduce any bias.

The truncated mean of the energy distribution does not scale linearly
with the path length, so there is a small residual dependence of the
$\mathrm{d}E/\mathrm{d}x$ on the path length. This is evaluated as a
systematic uncertainty and, furthermore, it largely cancels when the ratio
of Data/MC is considered.

 \begin{figure*}
   \centering
   \resizebox{0.49\textwidth}{!}{%
     \includegraphics{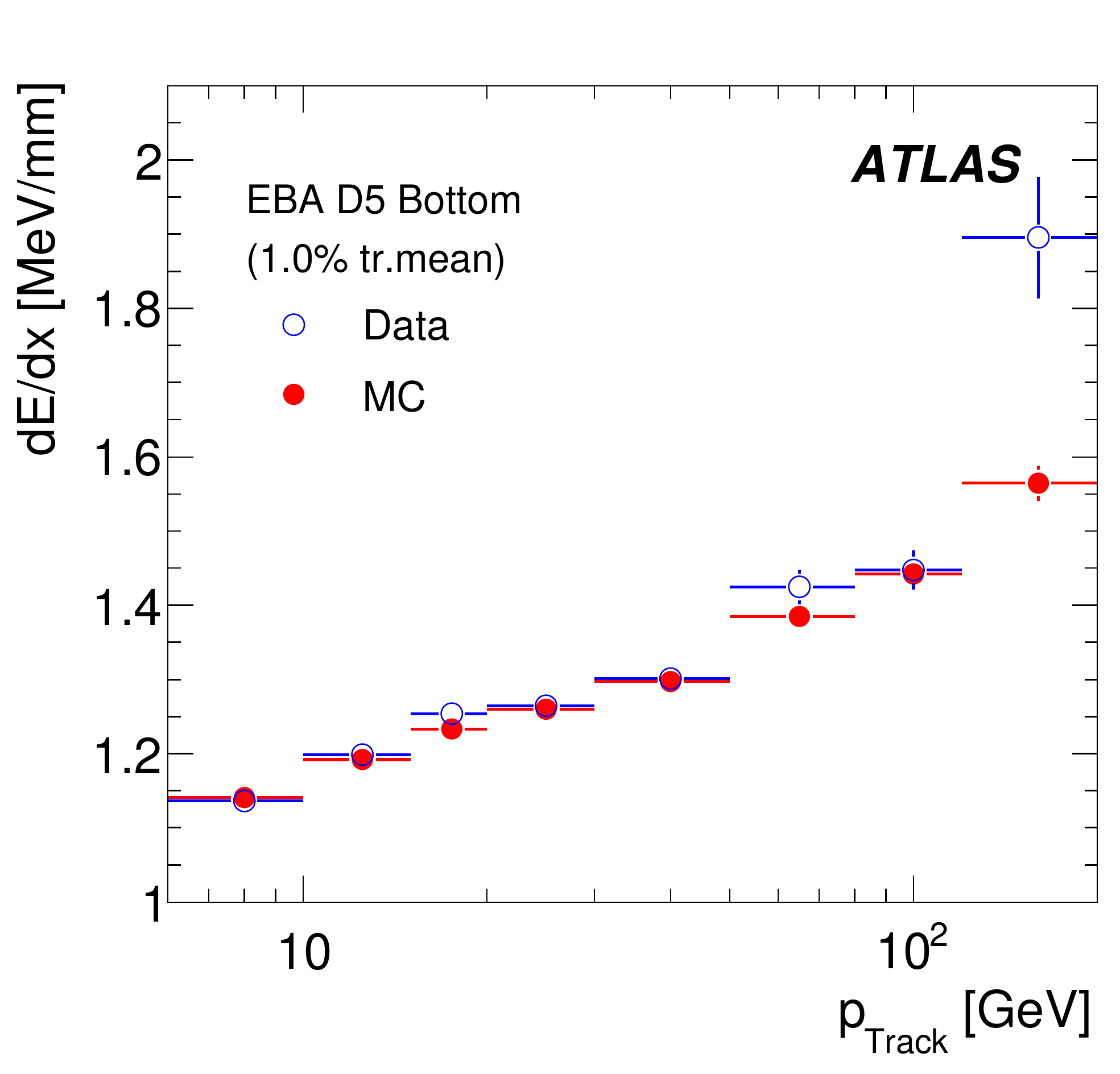} 
   }%
   \resizebox{0.49\textwidth}{!}{%
     \includegraphics{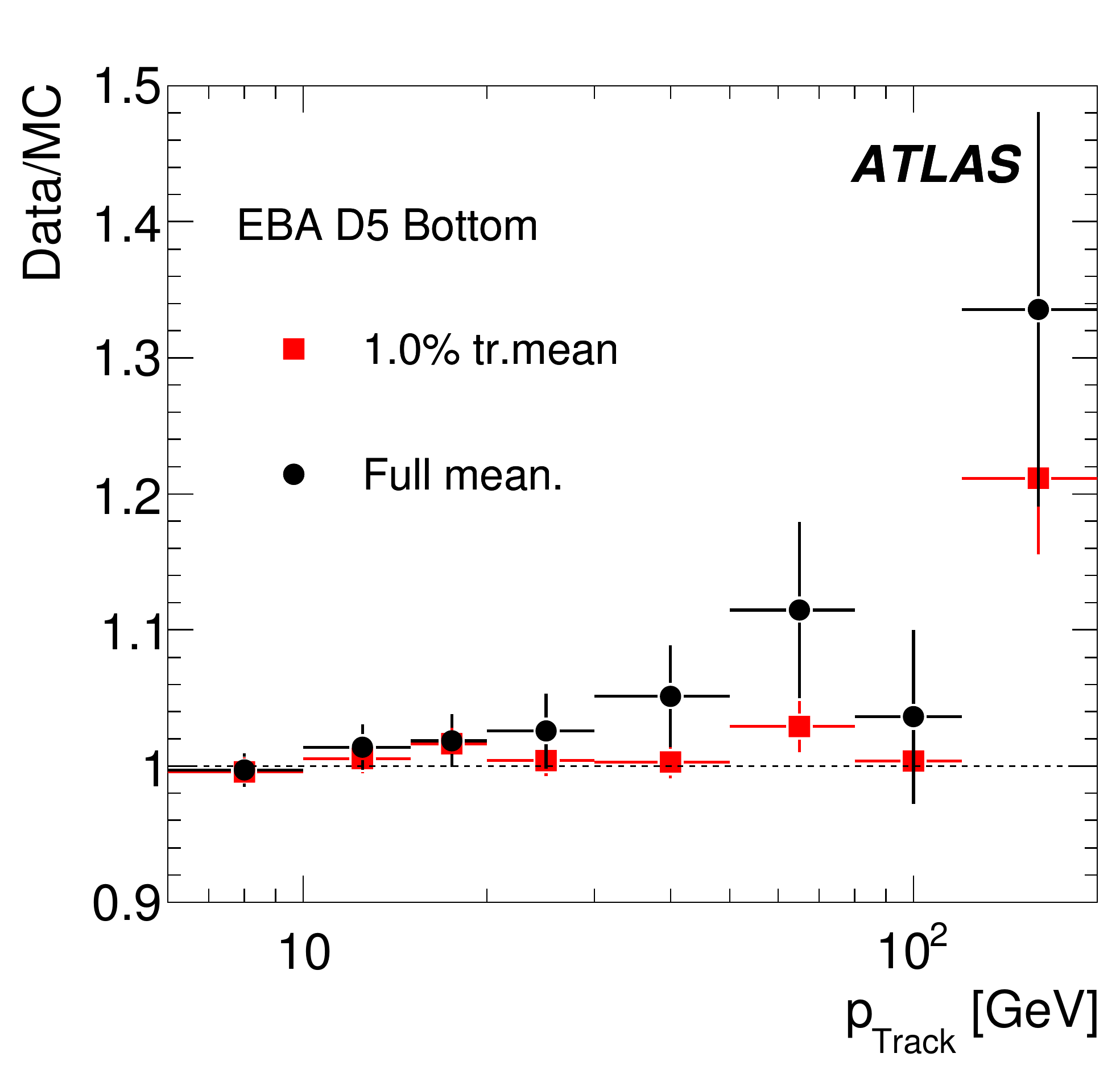} 
   }
   \caption{(Left) Muon response $\mathrm{d}E/\mathrm{d}x$ as a
     function of momentum as measured  in  the Inner Detector,
     estimated with the truncated mean for both data and Monte
     Carlo. (Right) Ratio of Data over Monte Carlo for the muon
     response $\mathrm{d}E/\mathrm{d}x$ as a function of momentum,
     shown for the truncated and full mean. 
     For both distributions the response is averaged  over the
     D5 cells in the bottom of the extended barrel (A side).}
  \label{fig:dedx_momentum}
\end{figure*} 

The dependency of the cell response to the muon momentum was investigated. 
As can be seen in Figure~\ref{fig:dedx_momentum} (left), the response increases  
with the momentum as expected, by about 20\,\% between $p=10$~GeV/$c$ and
$p=100$~GeV/$c$ and there is very good agreement between data and MC 
from $6$~GeV/$c$ to $\sim100$~GeV/$c$.
%
Figure~\ref{fig:dedx_momentum} (right)  shows that the MC
simulations predict a steeper dependence on the muon momentum
for the full mean, and some disagreement even for the truncated mean at the 
higher energies, which could imply some imprecision in the modelling
of the muon radiative energy losses. 
%

The real energy loss by muons is typically 10\,\% lower than the
corresponding signal on EM scale and the ratio, known as
$e/\mu$, slightly scales  with
energy~\cite{NIM_muons,TileTBpaper}. 
However, in this paper, the validation of the EM scale is carried out by comparing
data and Monte Carlo, and response to cosmic and testbeam muons, so
this correction factor is not necessary.


\subsubsection{Uniformity of the cell response}
\label{sec:performance_uniformity}
The studies addressed here measure the response uniformity per cell in a layer, as a function
of pseudorapidity $\eta$ and azimuthal angle $\phi$ (i.e.~per
module). 
Since our estimator is the 1\,\% truncated mean, we require a minimum of 100 events in each set -- $\eta$ or $\phi$ bin, or cell. 
For the $\eta$ and $\phi$ uniformity analyses, the data is not divided in cells -- all cells corresponding to that bin are accumulated and the truncation is applied to the single 
$\mathrm{d}E/\mathrm{d}x$ distribution for that bin. This approach allows the usage of the largest possible number of cells per bin while minimising biases from fluctuations in the tails.
These results comprise all partitions, but exclude the ITC cells 
(see Section~\ref{sec:itc_calibration}). In addition, we exclude from this study two cells from the D layer with an unusually high $\mathrm{d}E/\mathrm{d}x$.

Muons traverse cells in any direction and at any angle, so the local
variations in the optics system (light yield of individual tiles,
tile-to-fibre couplings, etc.) are supposed to be 
averaged out. 

\begin{figure*}
  \centering
  \resizebox{0.49\textwidth}{!}{%
    \includegraphics{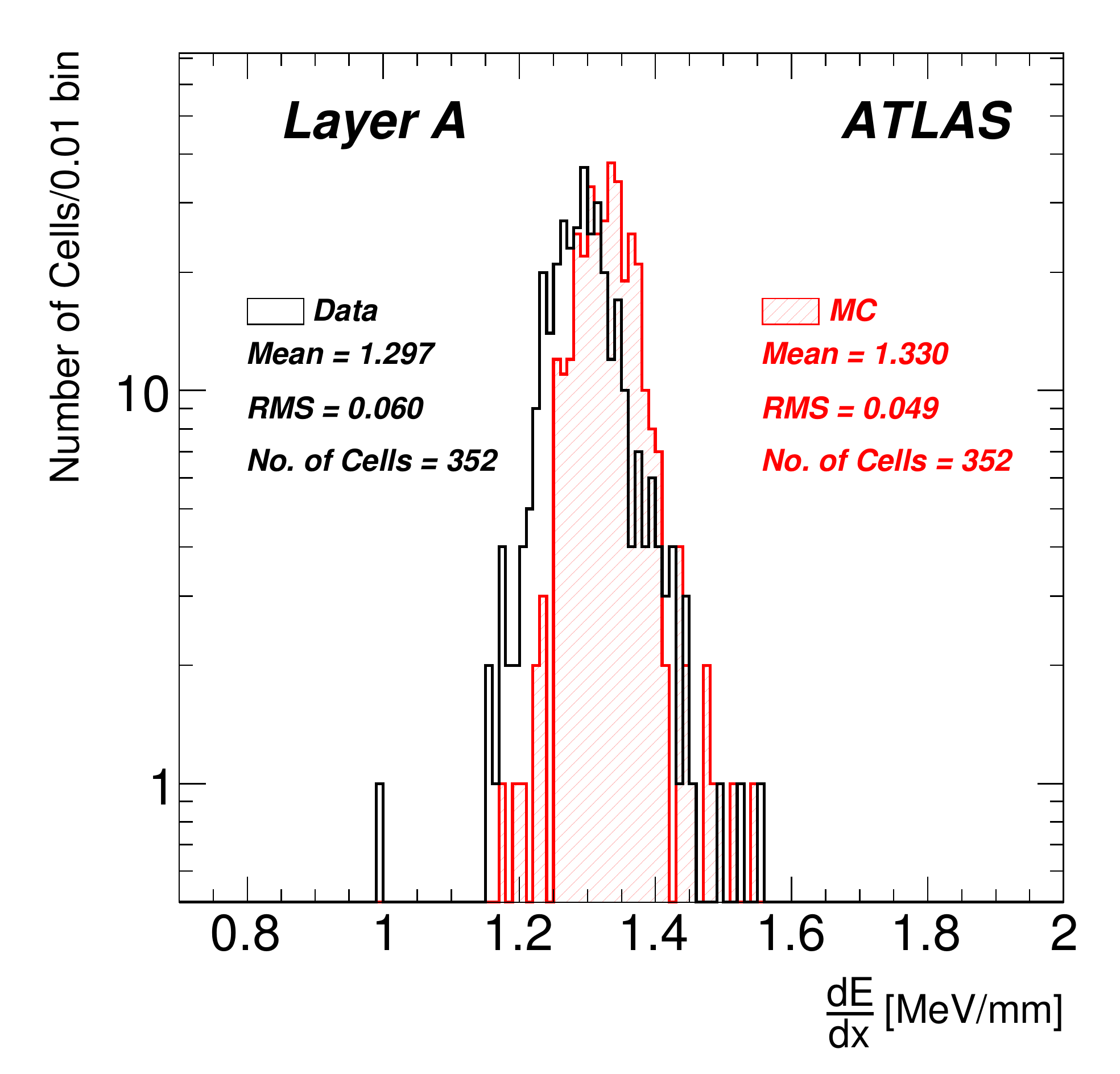}
  }%
  \resizebox{0.49\textwidth}{!}{%
    \includegraphics{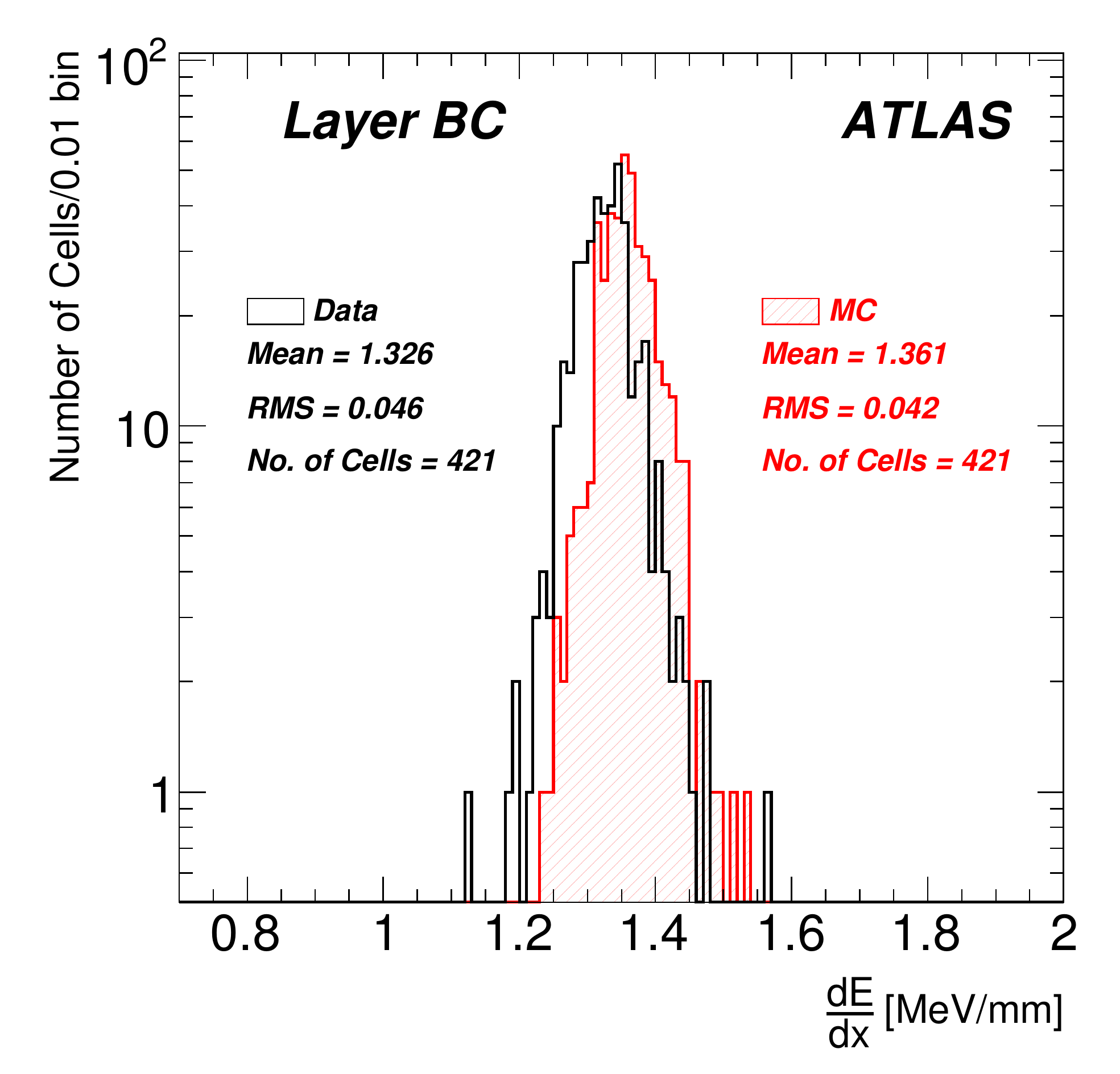}
  }
  \resizebox{0.49\textwidth}{!}{%
    \includegraphics{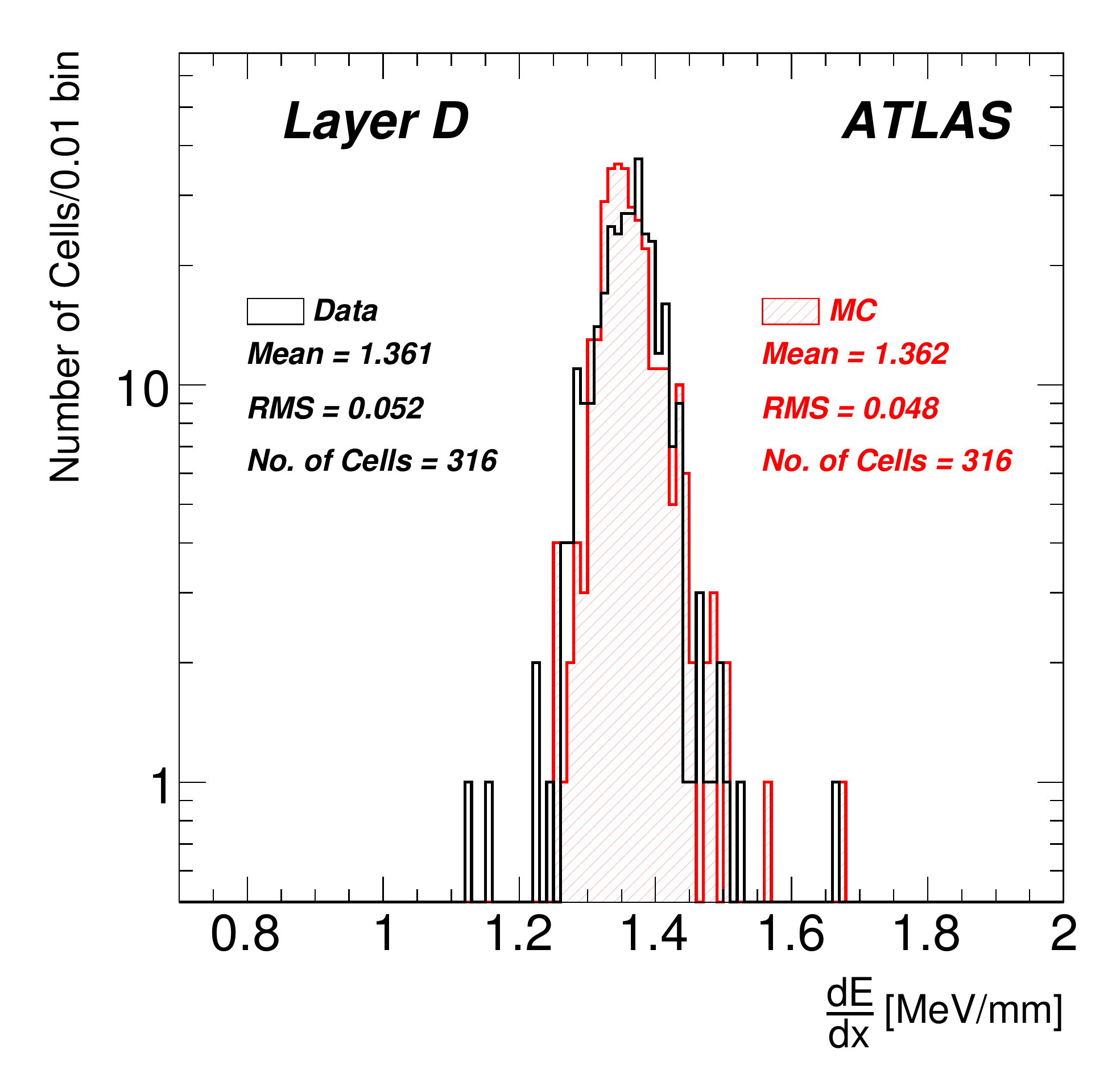}
  }
  \caption{Distribution of the truncated mean
    $\mathrm{d}E/\mathrm{d}x$ per cell, shown separately for each
    radial layer A, BC and D, for data and Monte Carlo. 
    The momentum range of the
    cosmic muons was restricted to be between 10 and 30~GeV/$c$. 
  }
  \label{fig:spread_in_layer}
\end{figure*}

\paragraph{Uniformity per cell\\}
\label{sec:uniformity_cell}
The uniformity of the cell response is shown in
Fig.~\ref{fig:spread_in_layer} for each radial layer and the RMS
values are summarised in Table~\ref{tab:uniformity_cell}. The selection criteria, especially the requirement of 100 events per cell, limit the number of measured cells to the values shown in the Figure and Table, but still a quite representative fraction of 23\,\% of the total cells is considered. The statistical population for the simulated and real data used for this study is identical.

The observed spread is the combination of different factors: statistical fluctuations, systematic errors due to the inherent limitations of measuring the cell response with the $\mathrm{d}E/\mathrm{d}x$ of cosmic muons, and the spread in the cell EM scale inter-calibration.

The Monte Carlo simulation has no variation in the quality of the
optical components of the calorimeter or in the channel signal
shape. Such variations are present in the data
but it is difficult to disentangle between the spread due to them or to the statistical fluctuations
from an underlying systematic due to the measurement method. Since the MC
shows an RMS in every layer compatible with that of data, it indicates that
cells are well intercalibrated within layers.


From the mean of the $\mathrm{d}E/\mathrm{d}x$ distributions per layer
it is observed that there is a response discrepancy of 5.0\,\% between
layer A and layer D (2.3\,\% between layer A and BC) for the cosmic
muon data, an issue which is further discussed in Section~\ref{sec:cosmics_layer_calib}.

\begin{table}
  \centering
  \begin{tabular}{|c|c|c|c|c|}
    \hline
    \raisebox{-1.5ex}{Layer} & Number of &  Fraction of & \multicolumn{2}{c|}{RMS (MeV/mm)} \\
    \cline{4-5}
    & cells & cells & Data & MC \\
    \hline
    A    & 352 & 18\,\% &  0.060 &  0.049\\
    BC & 421 & 22\,\% & 0.046 &  0.043\\
    D    & 316 & 38\,\% & 0.052 &  0.048\\
    \hline
  \end{tabular}
  \caption{The uniformity at the cell level for individual radial
    compartments. The listed values represent the RMS of the
    respective distribution of the truncated mean
    $\mathrm{d}E/\mathrm{d}x$ for that layer,
    shown for data and Monte Carlo. The number of cells considered,
    and the fraction of the total that they represent, are also
    shown. 
  } 
 \label{tab:uniformity_cell}
\end{table}

\bigskip

The variations as a function of pseudorapidity and azimuthal angle,
presented in the following paragraphs, were studied separately in each
layer, since they appear to be smaller than the dominating inter-layer
differences just shown. Another reason is that the cosmic muons are in
general non-projective, so most muon tracks cross the calorimeter in
each radial layer at different values of $\eta$ and/or $\phi$. Dealing
with the total response as a function of $\eta, \phi$ would require
projective muons only, thus significantly limiting the available
statistics. The results are presented here relative to the average
$\mathrm{d}E/\mathrm{d}x$. 

\begin{figure*}
  \centering
  \resizebox{0.49\textwidth}{!}{%
    \includegraphics[angle=-90]{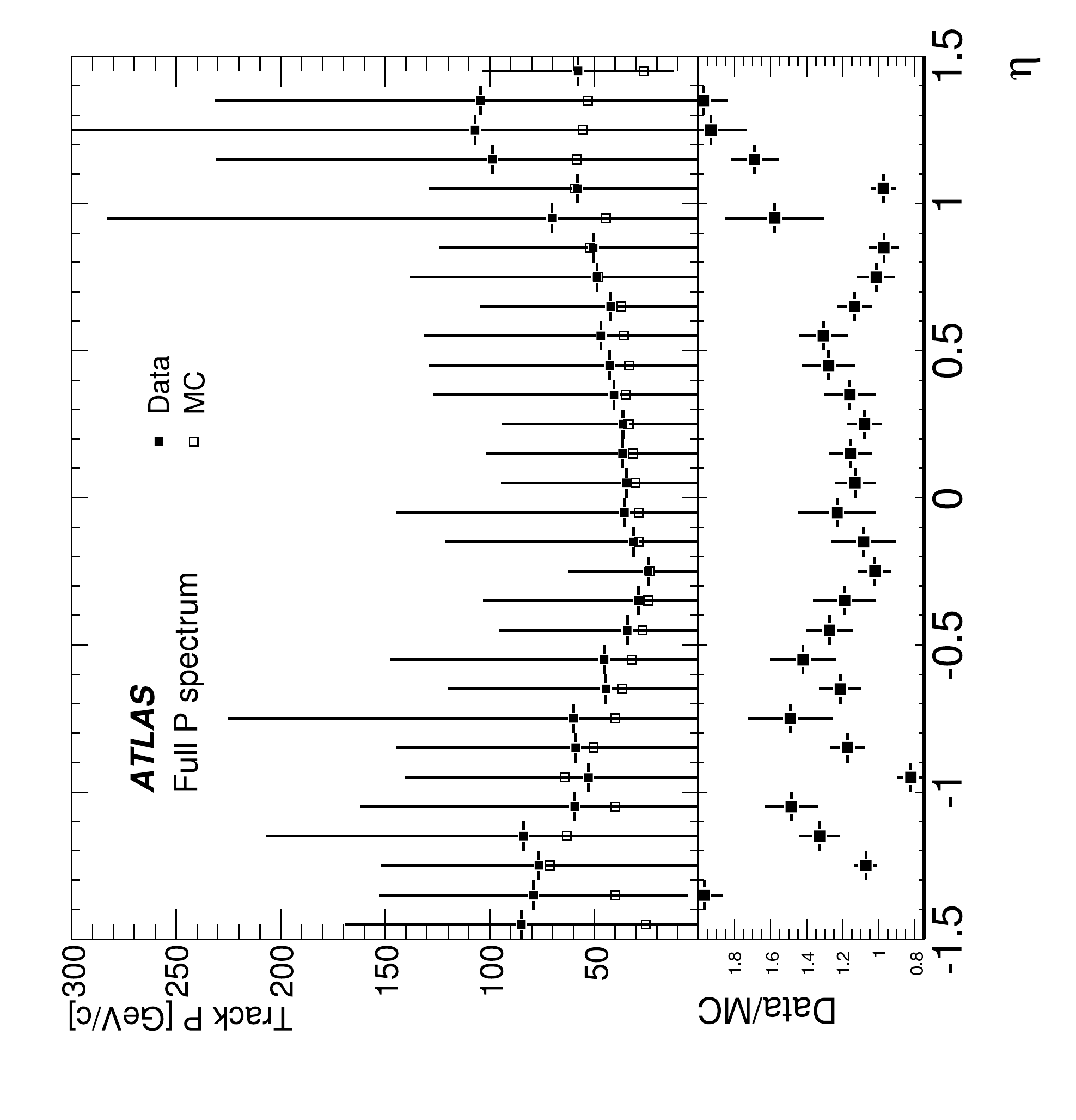}
  }%
  \resizebox{0.49\textwidth}{!}{%
    \includegraphics[angle=-90]{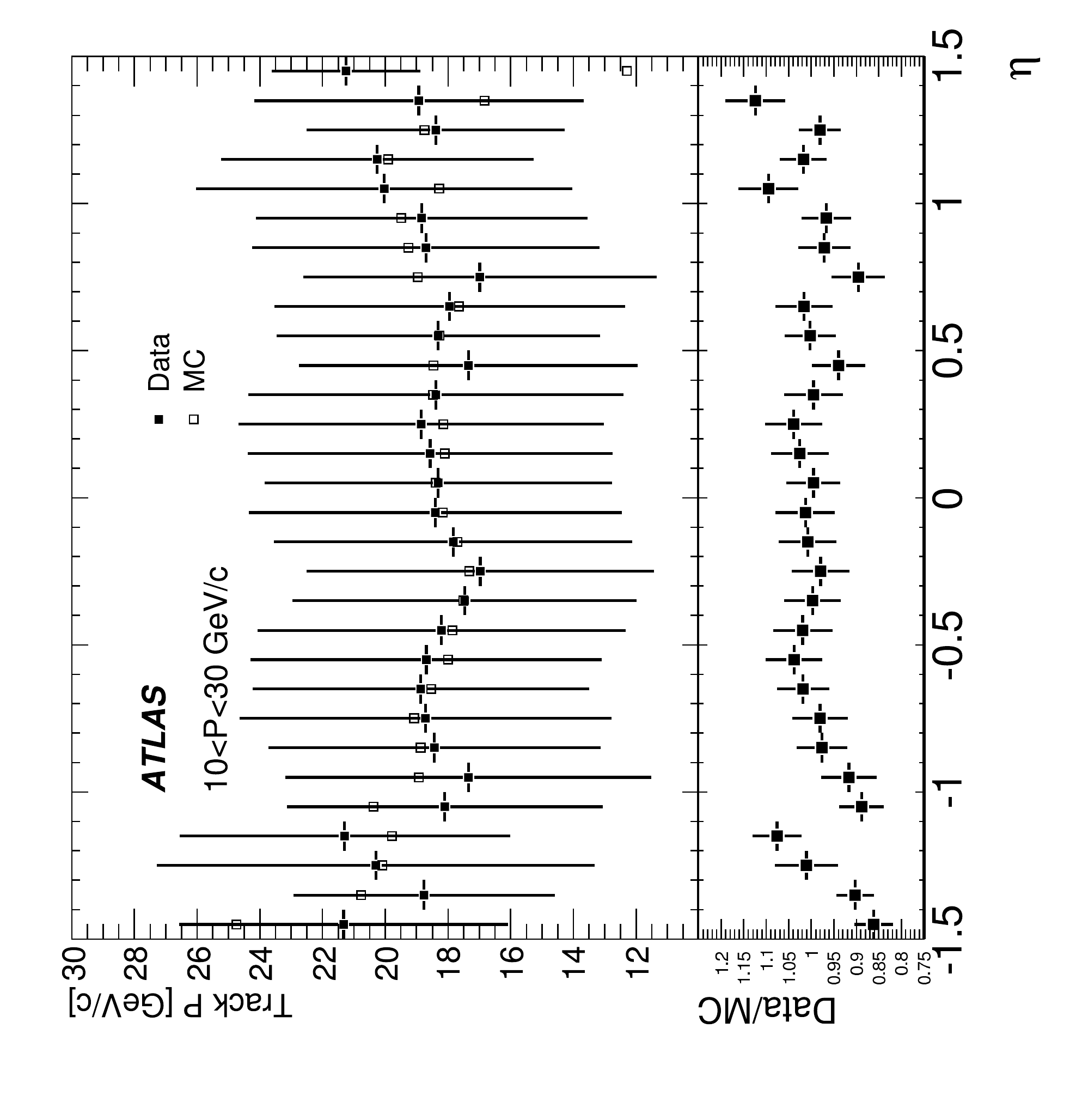} 
  }
   \caption{Momentum of the selected cosmic muon tracks as a function
     of pseudorapidity $\eta$, for both data and Monte Carlo. No momentum
     selection is applied in the left side distribution, while on the right, only
     tracks with momenta within 10~GeV/$c$ and 30~GeV/$c$ are shown. The
     vertical error bars in the upper part show the RMS
     of the momentum distribution in 
     each $\eta$ bin; in the lower part the error bars represent the
     uncertainty on the data/MC value shown.
  }    
  \label{fig:dedx_momentum_eta}
\end{figure*}

\paragraph{Uniformity per pseudorapidity\\}
When investigating the uniformity as a function of pseudorapidity, the
signal distribution includes all cells with the same azimuthal angle.
A possible residual dependency of the muon momentum on the
pseudorapidity of the detector cells (that could be due to the access
shafts) was investigated. Figure~\ref{fig:dedx_momentum_eta} (left)
shows that the observed muon momentum distribution is harder than what
expected by the Monte Carlo simulation, especially at high values of
pseudorapidity. However, in the low momentum region that was selected
for the analysis (between 10 and 30~GeV/$c$, see
Section~\ref{sec:ID_tracks_event_selection}), the agreement is much better
and the variations of momentum with $\eta$ ($\sim 10\,\%$) are quite
tolerable for this study.  

\begin{figure*}
  \centering
  \resizebox{0.49\textwidth}{!}{%
    \includegraphics{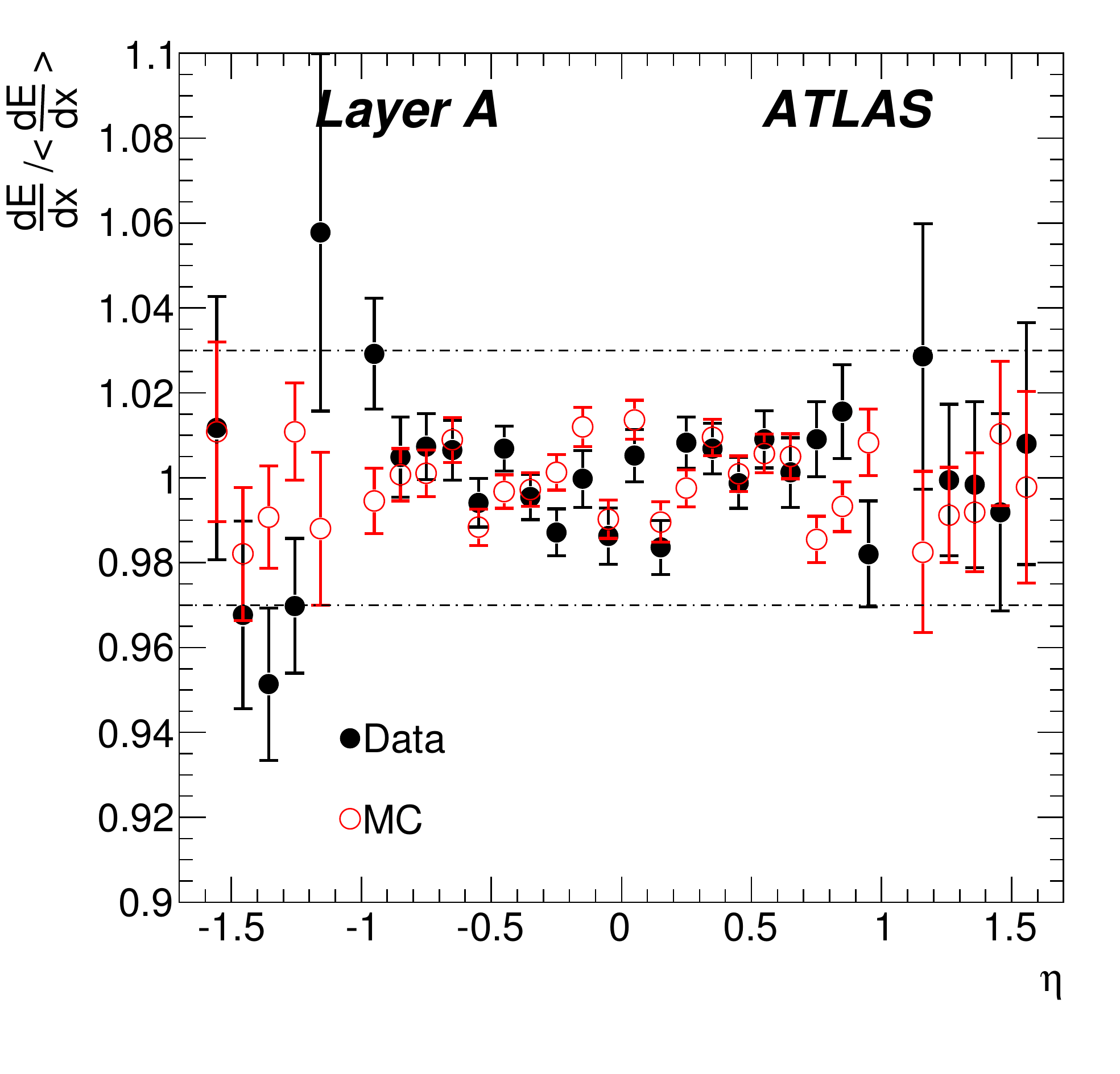}
  }%
  \resizebox{0.49\textwidth}{!}{%
    \includegraphics{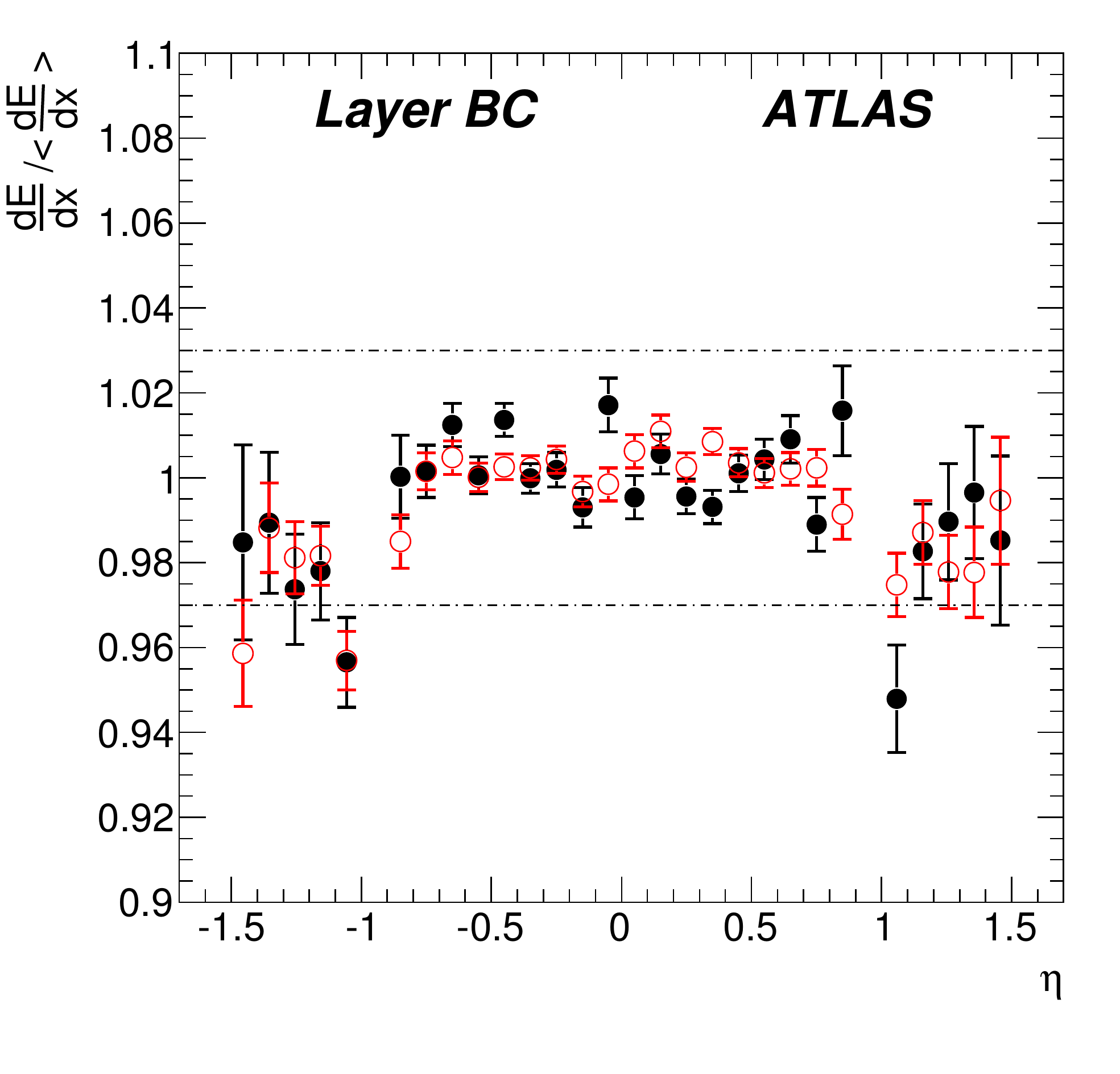}
  }
  \resizebox{0.49\textwidth}{!}{%
    \includegraphics{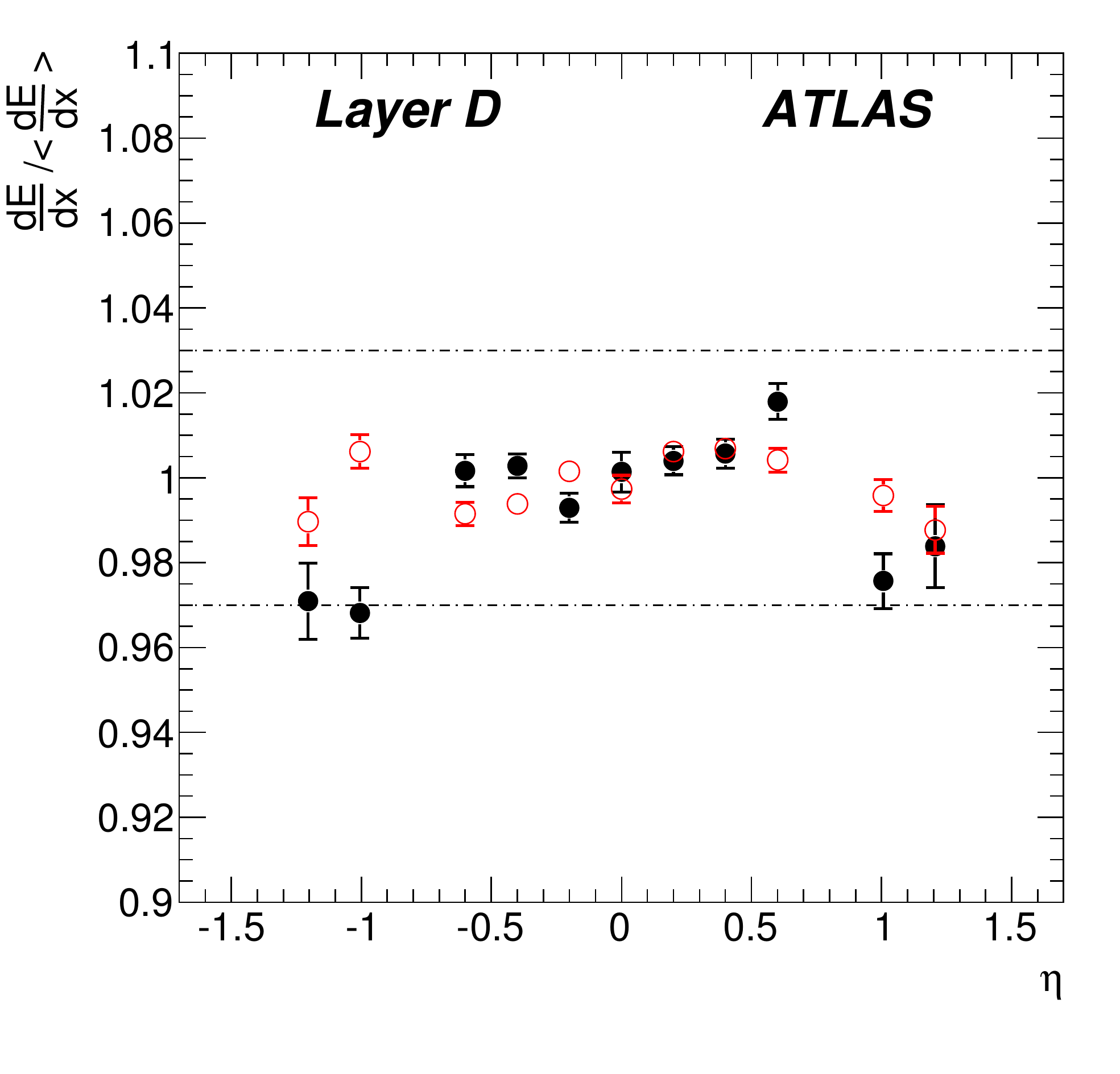}
  }
  \caption{Uniformity of the cell response to cosmic muons, expressed
    in terms of normalised truncated mean of
    $\mathrm{d}E/\mathrm{d}x$, as a function of pseudorapidity $\eta$ 
    for each radial layer. The response is integrated over all cells
    in each pseudorapidity bin in the given radial layer.  
    The results for data are compared to MC simulations and both are
    normalised to their averages for each layer. Data are shown with 
    closed circles, open circles indicate MC predictions. Statistical uncertainties only. Horizontal lines limiting a 
    $\pm~3\,\%$ band are shown.
  }
  \label{fig:uniformity_eta}
\end{figure*}

The tracks identified in the ID are required to point to the cell centre, as
specified in Equation~(\ref{eq:IDTrackCut}), as well as the other
selection procedures of Section~\ref{sec:ID_tracks_event_selection}. 
The results are shown in Fig.~\ref{fig:uniformity_eta} separately for
each radial compartment.
It can be seen that, for all layers, the values for the long barrel (central region, $|\eta| < 1$) are scattered within a 
$\pm~2\,\%$ band 
around the average. At high $\eta$, in the extended barrel, the
statistical uncertainties are larger due to worse coverage than in the
central regions.  Still these values are for the most part distributed
within a $\pm~3\,\%$ band.

\paragraph{Uniformity vs.~module\\}
The uniformity over modules has also been investigated. 
The response in every module was integrated over all cells in the
given radial layer. Studies combine all partitions, barrel and
extended barrels.

\begin{figure*}
  \centering
  \resizebox{0.49\textwidth}{!}{%
    \includegraphics{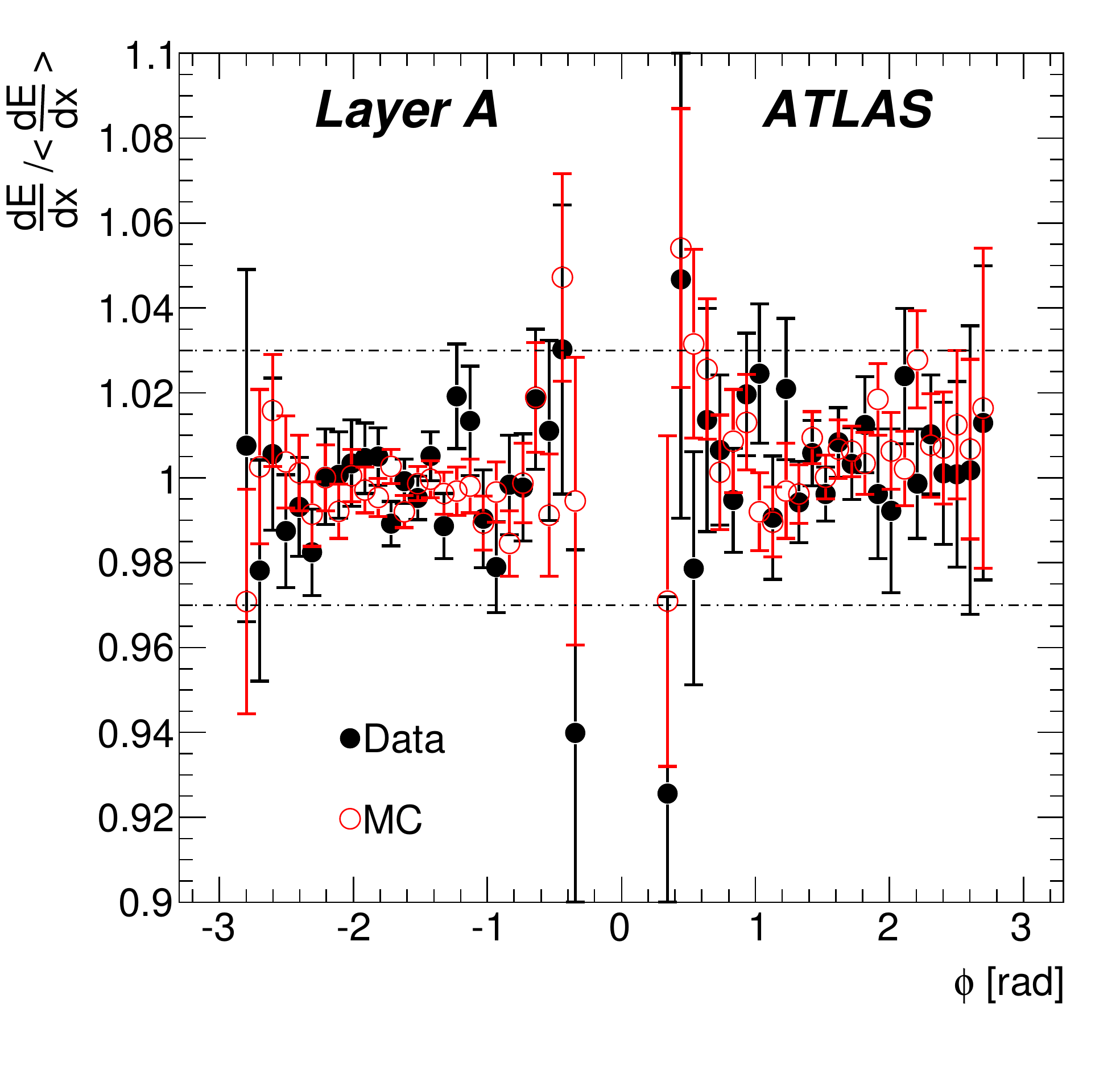}
  }%
  \resizebox{0.49\textwidth}{!}{%
    \includegraphics{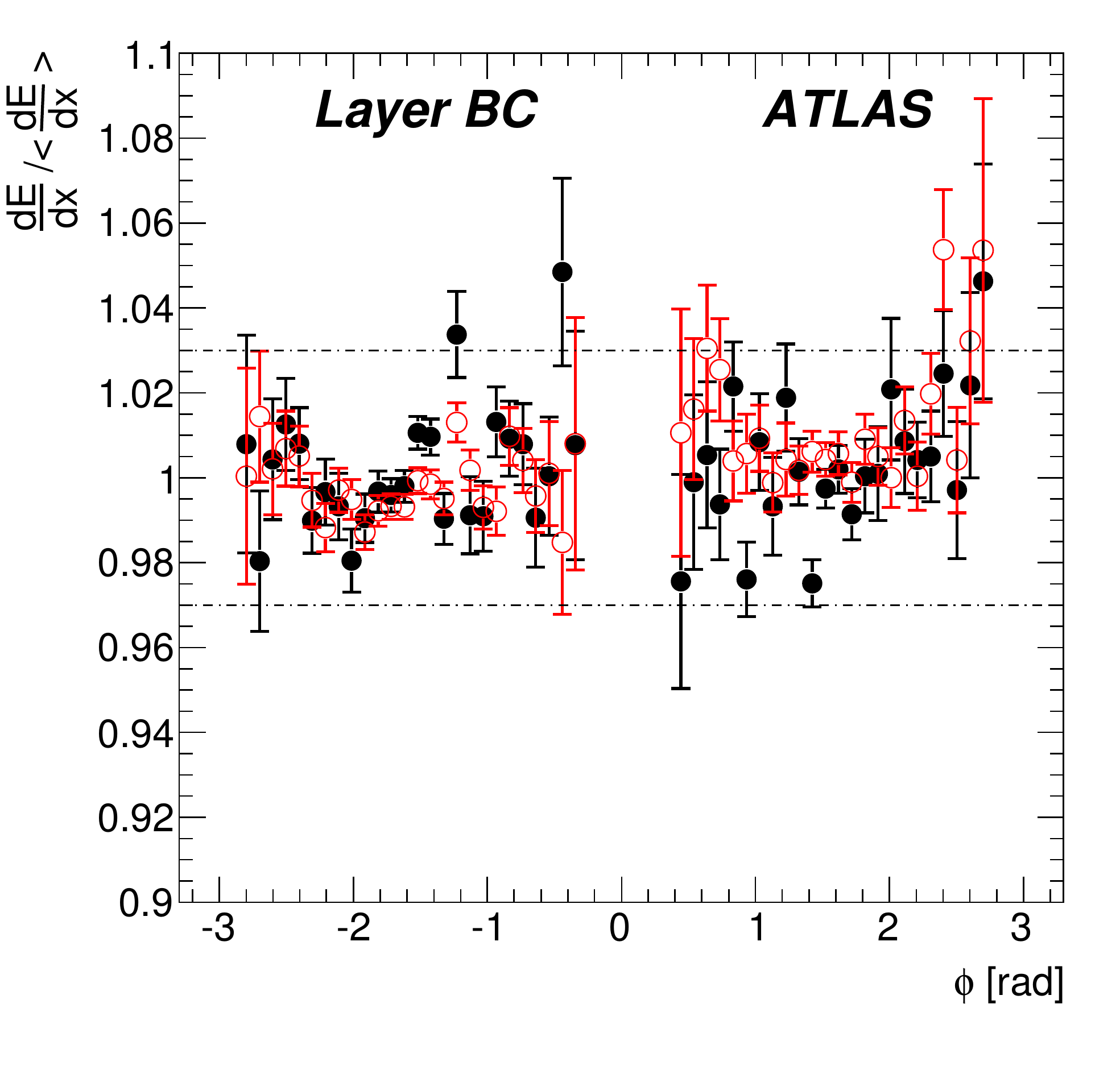}
  }
  \resizebox{0.49\textwidth}{!}{%
    \includegraphics{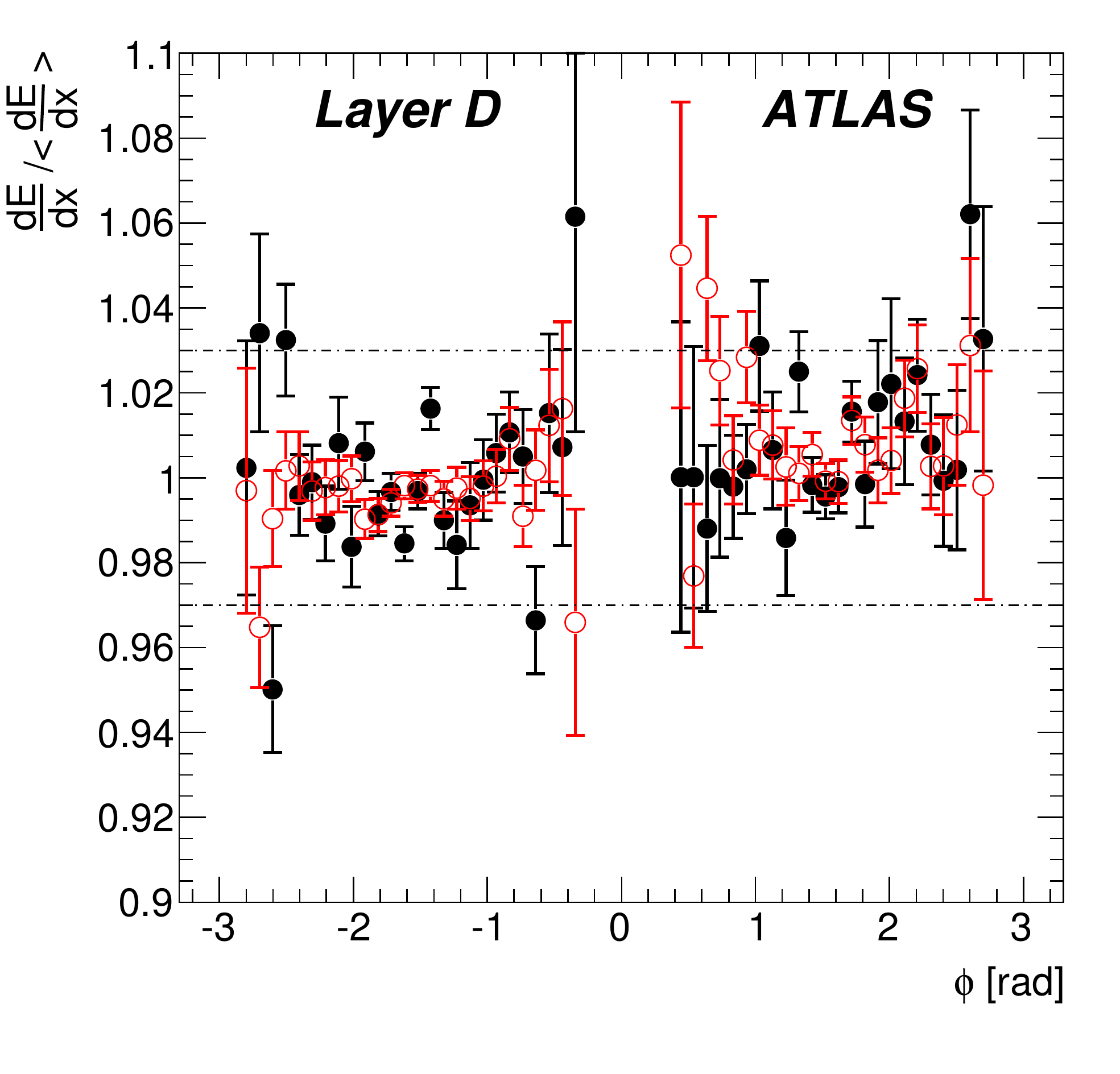}
  }
  \caption{Uniformity of the cell response to cosmic muons, expressed
    in terms of normalised truncated mean of
    $\mathrm{d}E/\mathrm{d}x$, as a function of azimuthal angle
    (module) for each radial layer. The response is integrated over
    all cells in each module in the given radial layer. All partitions
    are combined. The results for data are compared to MC simulations
    and both are normalised to their averages for each layer.  
    Data are shown with closed circles, open circles indicate MC
    predictions. Statistical uncertainties only.
    The gap around $\phi=0$ corresponds to horizontal
    modules that are poorly populated by cosmic ray muons passing
    through the Inner Detector.
    Horizontal lines limiting a 
    $\pm~3\,\%$ band are shown.
  }
  \label{fig:ID_uniformity_phi}
\end{figure*}

The results are shown in Fig.~\ref{fig:ID_uniformity_phi}. Again the
same cut on momentum $10 < p < 
30$~GeV/$c$ as measured in the Inner Detector was applied. 
This condition plays two roles -- apart from the reason mentioned in
Section~\ref{sec:ID_tracks_event_selection} it also ensures a similar
initial momentum distribution in different $\phi$-regions.  

Both experimental data and MC exhibit an essentially flat response as
a function of azimuthal angle $\phi$. A residual pattern observed
with data matches the MC: this small increase of
$\mathrm{d}E/\mathrm{d}x$ in horizontal ($\phi\rightarrow
0,\;\phi\rightarrow\pm\pi$) modules is likely due to a difference in
muon momentum in events passing the selection criteria.
Nevertheless, the data show a good uniformity over $\phi$ and, except a few cases in the horizontal region, most modules are well within a $\pm~3\,\%$ band.
In particular the average response in top ($\phi\approx \pi/2$) and
bottom ($\phi\approx -\pi/2$) modules appears
to be within 1\,\%. 

\subsubsection{Muon response and layer intercalibration}
\label{sec:cosmics_layer_calib}
The results discussed in Section~\ref{sec:uniformity_cell} show that
the cells are reasonably intercalibrated within a given layer, while
there are differences observed between individual layers. 
In order to better quantify these differences, the layer response was
calculated as the truncated mean of a single $\mathrm{d}E/\mathrm{d}x$
distribution for all cells in a given layer. This approach reduces the
statistical uncertainties, with respect to taking the truncated mean 
in each cell or $\eta, \phi$ bin.
In addition, only events in the bottom of the detector are used, to
avoid a bias from the muon momentum cut -- in this way, the muon momentum
for all the events is measured in the Inner Detector just prior to
their incidence in TileCal. 

In the cosmic muon analysis, various sources of systematic
uncertainties associated with the truncated mean of 
$\mathrm{d}E/\mathrm{d}x$ have been carefully studied. For every
contribution, the associated parameter was varied in the given range
and the systematic uncertainty contribution was evaluated as half of the
difference between the maximum and minimum resulting truncated mean,
unless explicitly stated otherwise.

The following contributions were identified:
\begin{itemize}
\item As already shown in Fig.~\ref{fig:dedx_momentum} (right), data
  and MC exhibit a slightly 
  different behaviour in function of the muon momentum. 
  Because of this, the variation of the data/MC ratio over the analysis
  range ($10 - 30$~GeV/$c$) is considered as the systematic
  uncertainty due to the 
  response dependence on the muon momentum. 
\item As the muon momentum is measured in the Inner Detector located in
  the centre of ATLAS, the response in the top and bottom part of 
 TileCal can be different. Although the difference is
  well below 1\,\% (see also
  Section~\ref{sec:performance_uniformity}), we consider its half as
  the contribution to the systematic uncertainty.
\item Another contribution is associated with the residual dependence
  of the truncated mean on the path length. The truncated mean
  $\mathrm{d}E/\mathrm{d}x$ was
  evaluated for several path length bins, and the above mentioned
  difference was calculated. 
\item The truncation itself represents another source of systematic
  uncertainty, that is associated with uncertainties in the description
  of the energy response shape. The uncertainty was estimated by comparing the
  resulting truncated mean of $\mathrm{d}E/\mathrm{d}x$ for several
  values of truncation between 0\,\% and 2.5\,\%. 
   This contribution does not fully cancel for the Data/MC ratio due to the
  difference that is observed in the tails of the $\mathrm{d}E/\mathrm{d}x$ 
  distribution  between data and MC.
\item The impact of the noise cut was studied as well, varying it
  between 30~MeV and 90~MeV (approximately 
  $1\sigma$ and $3\sigma$, where $\sigma$ is the average noise RMS). The associated systematics appears
  to be very small.
\item The measured response was also compared for various triggers,
  whose efficiencies depend on the muon momentum and also event
  topology. The data triggered by TGC and RPC indicate a good match
  within uncertainties, therefore the associated systematics is considered
  to be negligible with respect to other contributions mentioned
  above. 
\item The EM scale was transferred from testbeam to the ATLAS cavern by means
  of the Cs source calibration procedure. Since the scale was set 
  when the magnetic field was switched off and data
  were collected with magnetic field on, the appropriate correction
  has to be applied. Moreover, the Cs data show a response increase
  with time (see Section~\ref{sec:cesium}). Most
  of the cosmic data were acquired in September/October 2008,
  therefore we used the last Cs measurement with magnetic field on
  before the cosmic data taking to correct for both effects mentioned. 
  The combined effect of these two corrections (magnetic field and response 
  increase) amounts to $-1$\,\% for the barrel and $-0.6$\,\% for the
  extended barrel between June and September/October 2008 as
  shown in Figure~\ref{fig:calib-CS-5}. 
  Since the origin of the Cs response variation in time is not yet 
  fully understood, the maximum and minimum of the Cs response in 2008
  is considered as input for the corresponding asymmetric systematic
  uncertainty. 
\end{itemize}
The uncertainties were evaluated separately for the LB and EB partitions and per
individual radial layer for data, Monte Carlo, and the data/MC ratio (some contributions
cancel in the ratio).
The correlations among the radial layers are not taken into account
 and only the square roots of the diagonal terms of the error matrix 
 are considered, and listed in Table~\ref{tab:systematics_detailed}.


\begin{table*}
  \centering
    \begin{tabular}{|c|c|.|.|.|.|.|.|}
    \hline
    \multicolumn{8}{|c|}{Systematic Uncertainties [MeV/mm] for Data and MC} \\
    \hline
    \multicolumn{2}{|c|}{Uncertainty source} & \multicolumn{3}{c|}{Long Barrel}
    & \multicolumn{3}{c|}{Extended Barrel} \\
    \hline    
    \multicolumn{2}{|c|}{} & \multicolumn{1}{c|}{A} &
    \multicolumn{1}{c|}{BC} & \multicolumn{1}{c|}{D} &
    \multicolumn{1}{c|}{A} & \multicolumn{1}{c|}{B} & \multicolumn{1}{c|}{D} \\
    \hline \hline
                & Data          &\pm 0.016 &\pm 0.030 &\pm 0.019 &\pm 0.046 &\pm 0.030 &\pm 0.017 \\ 
     Path       & MC            &\pm 0.006 &\pm 0.008 &\pm 0.013 &\pm 0.014 &\pm 0.015 &\pm 0.022 \\ 
                & Data/MC ratio &\pm 0.008 &\pm 0.016 &\pm 0.009 &\pm 0.033 &\pm 0.021 &\pm 0.019 \\ 
    \hline
                & Data          &\pm 0.024 &\pm 0.034 &\pm 0.033 &\pm 0.037 &\pm 0.043 &\pm 0.044 \\ 
     Momentum   & MC            &\pm 0.032 &\pm 0.042 &\pm 0.035 &\pm 0.020 &\pm 0.042 &\pm 0.044 \\ 
                & Data/MC ratio &\pm 0.008 &\pm 0.007 &\pm 0.004 &\pm 0.024 &\pm 0.005 &\pm 0.009 \\ 
    \hline
                & Data          &\pm 0.007 &\pm 0.002 &\pm 0.002 &\pm 0.009 &\pm 0.004 &\pm 0.003 \\ 
     Noise      & MC            &\pm 0.004 &\pm 0.002 &\pm 0.003 &\pm 0.003 &\pm 0.002 &\pm 0.002 \\ 
                & Data/MC ratio &\pm 0.002 &\pm 0.000 &\pm 0.001 &\pm 0.005 &\pm 0.001 &\pm 0.000 \\ 
    \hline
                & Data          &\pm 0.013 &\pm 0.014 &\pm 0.013 &\pm 0.013 &\pm 0.013 &\pm 0.013 \\ 
     Truncation & MC            &\pm 0.014 &\pm 0.014 &\pm 0.014 &\pm 0.014 &\pm 0.014 &\pm 0.014 \\ 
                & Data/MC ratio &\pm 0.004 &\pm 0.005 &\pm 0.005 &\pm 0.003 &\pm 0.001 &\pm 0.001 \\ 
    \hline
                & Data          &\pm 0.007 &\pm 0.006 &\pm 0.012 &\pm 0.008 &\pm 0.009 &\pm 0.008 \\ 
     Top/Bottom & MC            &\pm 0.015 &\pm 0.014 &\pm 0.014 &\pm 0.016 &\pm 0.037 &\pm 0.006 \\ 
                & Data/MC ratio &\pm 0.006 &\pm 0.014 &\pm 0.002 &\pm 0.006 &\pm 0.021 &\pm 0.010 \\ 
    \hline
                & \raisebox{-1.5ex}{Data} & +0.005 & +0.005 & +0.005 & +0.000 & +0.000 & +0.000 \\ [-1ex]         
    \raisebox{-1.5ex}{Global EM}   &         & -0.013 & -0.013 & -0.014 & -0.008 & -0.008 & -0.008 \\ [1ex]

    \raisebox{-1.5ex}{scale factor} & MC            & -     &  -    &   -   & -     &  -    &   - \\ [1ex]
                & \raisebox{-1.5ex}{Data/MC ratio} & +0.004 & +0.004 & +0.004 & +0.000 & +0.000 & +0.000 \\[-1ex]
                &                                  & -0.010 & -0.010 & -0.010 & -0.006 & -0.006 & -0.006 \\
    \hline \hline
                & \raisebox{-1.5ex}{Data} & +0.033 & +0.047 & +0.042 & +0.062 & +0.055 & +0.050 \\ [-1ex]
                &                         & -0.035 & -0.049 & -0.044 & -0.063 & -0.056 & -0.051 \\ [1ex]
     Total      & MC            &\pm 0.039 &\pm 0.047 &\pm 0.042 &\pm 0.033 &\pm 0.060 &\pm 0.052 \\ [1ex]
                & \raisebox{-1.5ex}{Data/MC ratio} & +0.015 & +0.023 & +0.012 & +0.042 & +0.030 & +0.023 \\ [-1ex]
                &                                  & -0.017 & -0.025 & -0.015 & -0.042 & -0.031 & -0.024 \\ 
    \hline
  \end{tabular}
  \caption{The individual contributions to the systematic uncertainty of
    the truncated mean $\mathrm{d}E/\mathrm{d}x$ in cosmic muon Data and Monte
    Carlo. The listed values correspond to the diagonal terms of the
    error matrix. Analyses were performed with the ID-track method. The
    uncertainties on the global EM scale factor are discussed in
    Section~\ref{sec:emscale}.
  }
  \label{tab:systematics_detailed}
\end{table*}

\begin{figure}
  \centering
  \resizebox{0.5\textwidth}{!}{%
    \includegraphics{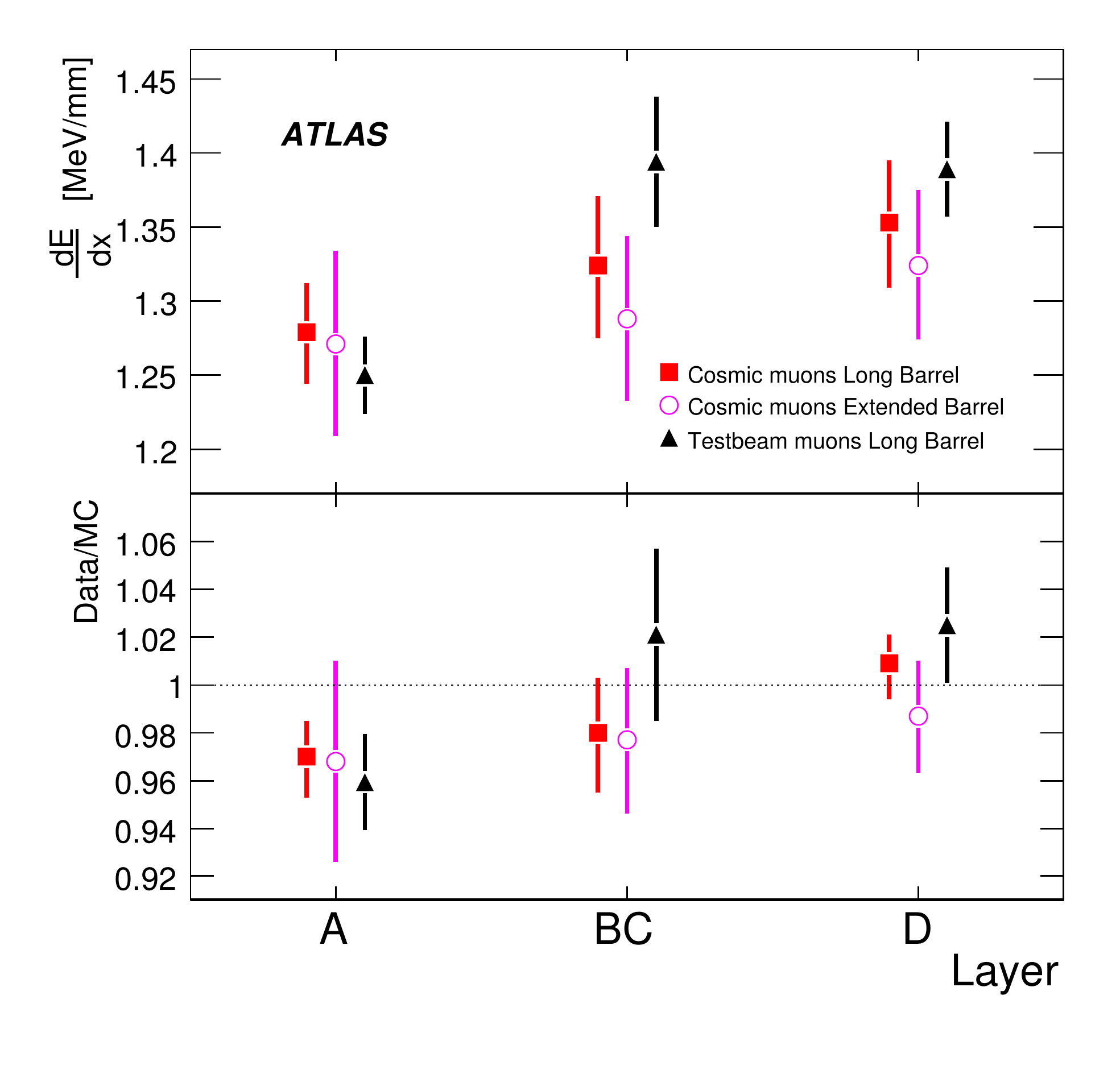}
  }
  \caption{The truncated mean of the $\mathrm{d}E/\mathrm{d}x$ for
    cosmic and testbeam muons shown per 
    radial compartment and, at the bottom, compared to Monte
    Carlo. For the cosmic muon data, the results were obtained for
    modules at the bottom part of the 
    calorimeter.
    The error bars shown combine in quadrature both the statistical and
    the systematic uncertainties, considering only
    the diagonal terms of the error matrix.} 
  \label{fig:dEdx}
\end{figure}

The results on the longitudinal layer intercalibration are presented
in Table~\ref{tab:dedx_truncated_mean} and displayed in
Fig.~\ref{fig:dEdx}, the error bars representing the total uncertainty
based on the quadratic sum of the statistical and systematic
uncertainties.
%


\begin{table*}
  \centering
  \setlength{\extrarowheight}{2pt}
  \begin{tabular}{|c|c||c|c|c|}
    \hline
    \multicolumn{2}{|c||}{Radial layer} & A & BC & D \\
    \hline \hline
                & Data    & $1.28^{+0.03}_{-0.04}$       & $1.32\pm 0.05$  & $1.35\pm 0.04$  \\ 
    Cosmic muons, LB & MC      & $1.32\pm 0.04$       & $1.35\pm 0.05$  & $1.34\pm 0.04$  \\ 
    \cline{2-5}
                & Data/MC & $0.97^{+0.01}_{-0.02}$       & $0.98\pm0.02$   & $1.01\pm 0.01$   \\ 
    \hline
                & Data    & $1.27\pm 0.06$       & $1.29\pm0.06$  & $1.32\pm0.05$  \\ 
    Cosmic muons, EB & MC      & $1.31\pm 0.03$       & $1.32\pm 0.06$  & $1.34\pm 0.05$  \\ 
    \cline{2-5}
                & Data/MC & $0.97\pm0.04$        & $0.98\pm0.03$       & $0.99\pm0.02$ \\ 
    \hline
                & Data     & $1.25\pm 0.03$       & $1.39\pm 0.04$
    & $1.39\pm 0.03$ \\ 
    Testbeam, LB  & MC       & $1.30\pm 0.02$       & $1.37\pm 0.03$
    & $1.36\pm 0.02$ \\ 
    \cline{2-5}
                & Data/MC  & $0.96\pm 0.02$      & $1.02\pm 0.04$
    & $1.02\pm 0.02$ \\ 
    \hline
    \multicolumn{2}{|c||}{Double ratio
      $\frac{(\mathrm{Data/MC})_{\text{Cosmic muons, LB}}} 
      {(\mathrm{Data/MC})_{\text{TB, LB}}}$ } &
    $1.01\pm 0.03$      & $0.96\pm 0.04$      & $0.98\pm 0.03$ \\
    \hline
  \end{tabular}
  \caption{The truncated mean of $\mathrm{d}E/\mathrm{d}x$ (MeV/mm,
    see text), measured with cosmic ray muons in
    barrel (LB) and extended barrel (EB), and
    projective testbeam muons. Results are shown for both data and
    Monte Carlo as well as for each radial layer.
    For cosmic ray muons, only modules in the bottom part
    are used. Total uncertainties are quoted.     
    For cosmic data 
    the statistical component is negligible.
    The systematic uncertainty corresponds to the diagonal terms of the error matrix.
    }
  \label{tab:dedx_truncated_mean}
\end{table*}

The differences in the cosmic muon response among individual layers
are present
even after correcting
for the residual dependencies on the path length, momentum, impact
angle, impact point, by considering the ratio of data over Monte
Carlo. 
The resulting values are strongly correlated, therefore the maximum
difference of 4\,\% between the individual measurements with the
cosmic muon data indicates the layer response
discrepancy.

\subsubsection{Validation of the EM scale propagation from testbeam}
\label{sec:results_energy_scale}
The ratio data/MC mentioned above also depends on the absolute EM
scale of the MC simulated energy loss in the calorimeter. Due to the
uncertainties in this quantity,  the double ratio of data/MC, cosmic
muon/TB, is adopted for comparison of the muon response and hence the
EM scale between cosmic and TB data in the long barrel. 
%
%
For testbeam data and Monte Carlo, the truncated mean
of the $\mathrm{d}E/\mathrm{d}x$ distribution was obtained for each
run, and then averaged over all runs. 
These are the values already presented
in Table~\ref{tab:dedx_truncated_mean} and Figure~\ref{fig:dEdx}.
The evaluation of the systematic uncertainties is briefly descrided below.
 
We consider the spread of the $\mathrm{d}E/\mathrm{d}x$ values over the 
different incidence angles as the main uncertainty of the
measurement, an approach that effectively combines the
statistical and part of systematic uncertainties. On top of
them, we consider the following subdominant contributions:
\begin{itemize}
\item The bias due to the truncation in the $\mathrm{d}E/\mathrm{d}x$
  distribution was estimated in the same way as for cosmic data (mentioned above). 
\item The uncertainty in the global EM scale due to the non-calibrated
  integrators (see Sections~\ref{sec:cesium}
  and~\ref{sec:intercalibration}) at that time. This uncertainty
  applies only to data, not to Monte Carlo.
\end{itemize}

The individual uncertainties were evaluated for each radial
layer and the resulting total uncertainties, shown in
Table~\ref{tab:dedx_truncated_mean}, were obtained by summing the
individual contributions in quadrature.

\bigskip

The double ratio of data/MC, cosmic muons/TB, is presented
in the last row of Table~\ref{tab:dedx_truncated_mean}.
The uncertainty contributions are computed by propagating 
in quadrature the TB uncertainties just described and the 
cosmic muon uncertainties mentioned in the previous section, 
that only take into account the error matrix diagonal terms.
The EM scale measured with cosmic muons, relative to that
determined at testbeam in the long barrel, amounts to 1.01, 0.96 and 0.98
for the A, BC and D layers respectively.
Since the uncertainties per layer are at most 4\,\%, these values are
consistent with 1.0, showing that, within the precision limits of
the analysis, the propagation of the EM scale from testbeam to
ATLAS was performed successfully. 

\bigskip


It should be noted that the LHC collisions will provide extra tools to
check the EM scale calibration. 
Isolated muons and single hadrons developing their shower only in
TileCal will provide two data samples for which a direct 
comparison to the testbeam scale will be possible.

%
%

\subsection{ITC and gap/crack scintillator calibration}
\label{sec:itc_calibration}
Understanding the response of the intermediate Tile Calorimeter (ITC) and the gap and crack scintillators (see Section~\ref{sec:overview} and Figure~\ref{fig:tile-cellmap}) to cosmic ray muons is essential for their calibration. The gap and crack scintillators can not be calibrated using the Cs calibration source and therefore have arbitrary calibration factors applied to them. This study with cosmic muons gives the first clues for their in-situ performance.

These detectors are calibrated in two steps. The first step is the intercalibration in $\phi$ among the cells of the same detector type and to determine the calibration factors for each cell. The second step is the absolute calibration and to determine a scale factor defined relative to the MC for each detector type. Since the absolute energy scale in the scintillators is not known, the simulation is used as a reference in this case. 

The event selection follows the same procedures as indicated in
Section~\ref{sec:ID_tracks_event_selection}, with the exception that
only events with a single muon track with a momentum above 5~GeV are
considered and that, for the ITC cells, the entry and exit points of
the track in the cell must be separated by at least 4~cm in the $z$
direction.  
These requirements accept 8\,\% of RPC triggered events, 80\,\%
of TGC triggered events and 7\,\% of L1 calorimeter triggered
events. Problematic cells and scintillators\footnote{Cells or
  scintillators that, even though matched with extrapolated tracks,
  appear too noisy or show very small signal.}
are excluded in this analysis.  

\begin{figure*}
 \begin{center}
   \begin{tabular}{ccc}
     \begin{minipage}{5.cm}
       \begin{center}
         \centerline{\includegraphics[width=0.9\columnwidth]{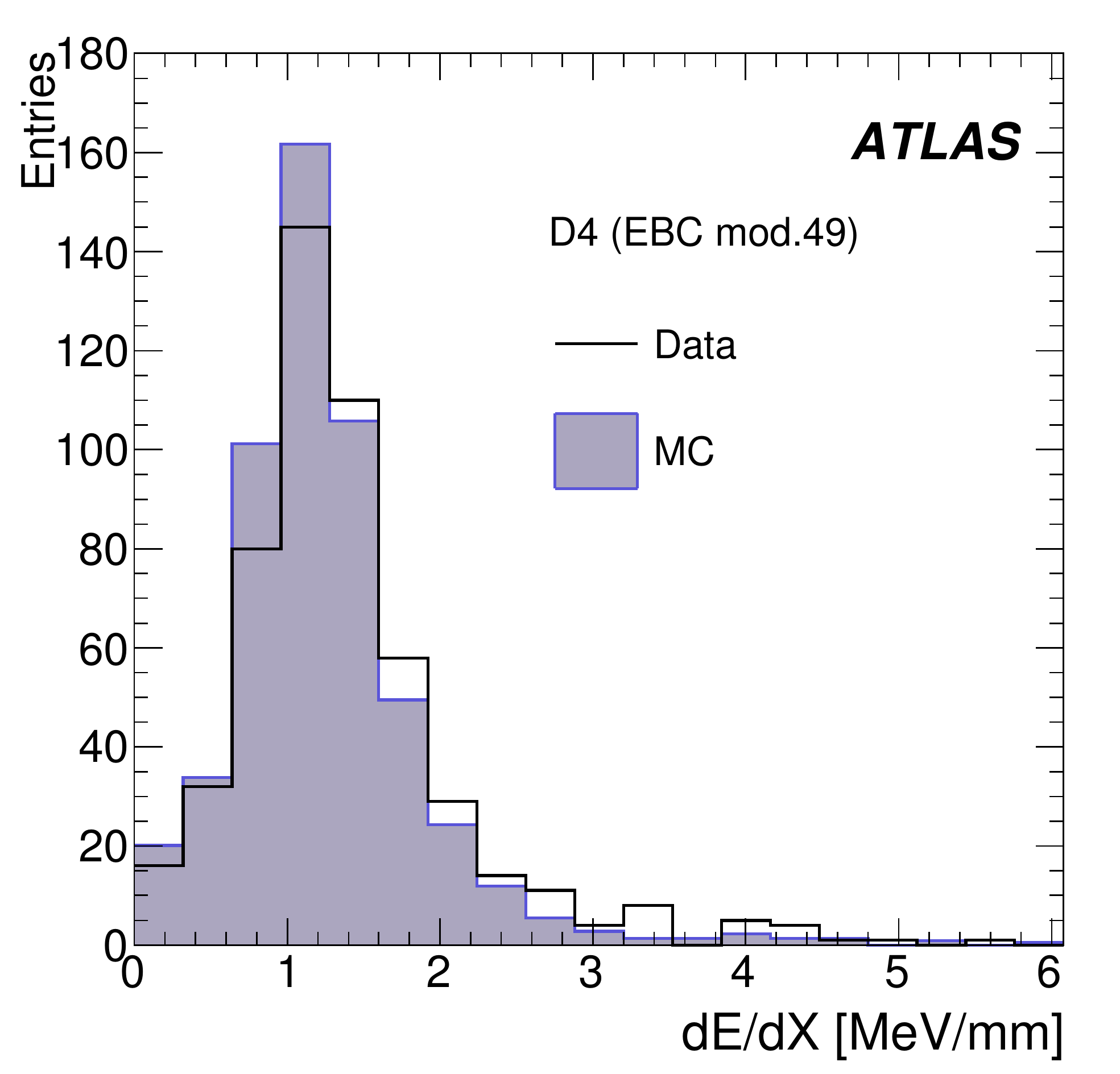}}
       \end{center}
     \end{minipage}
     \hspace{1mm}
     \begin{minipage}{5.cm}
       \begin{center}
         \centerline{\includegraphics[width=0.9\columnwidth]{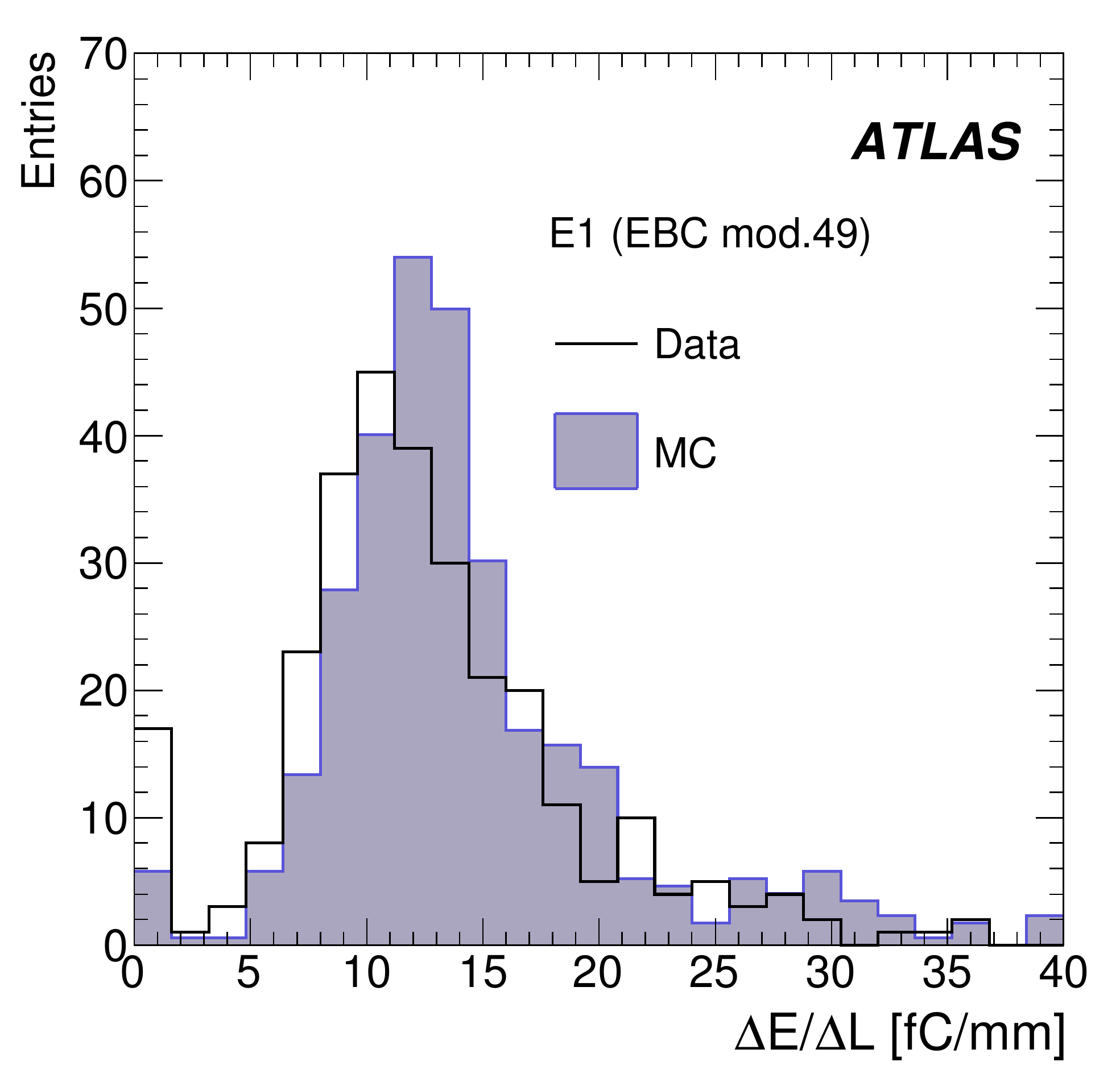}}
       \end{center}
     \end{minipage}
     \hspace{1mm}
     \begin{minipage}{5.cm}
       \begin{center}
         \centerline{\includegraphics[width=0.9\columnwidth]{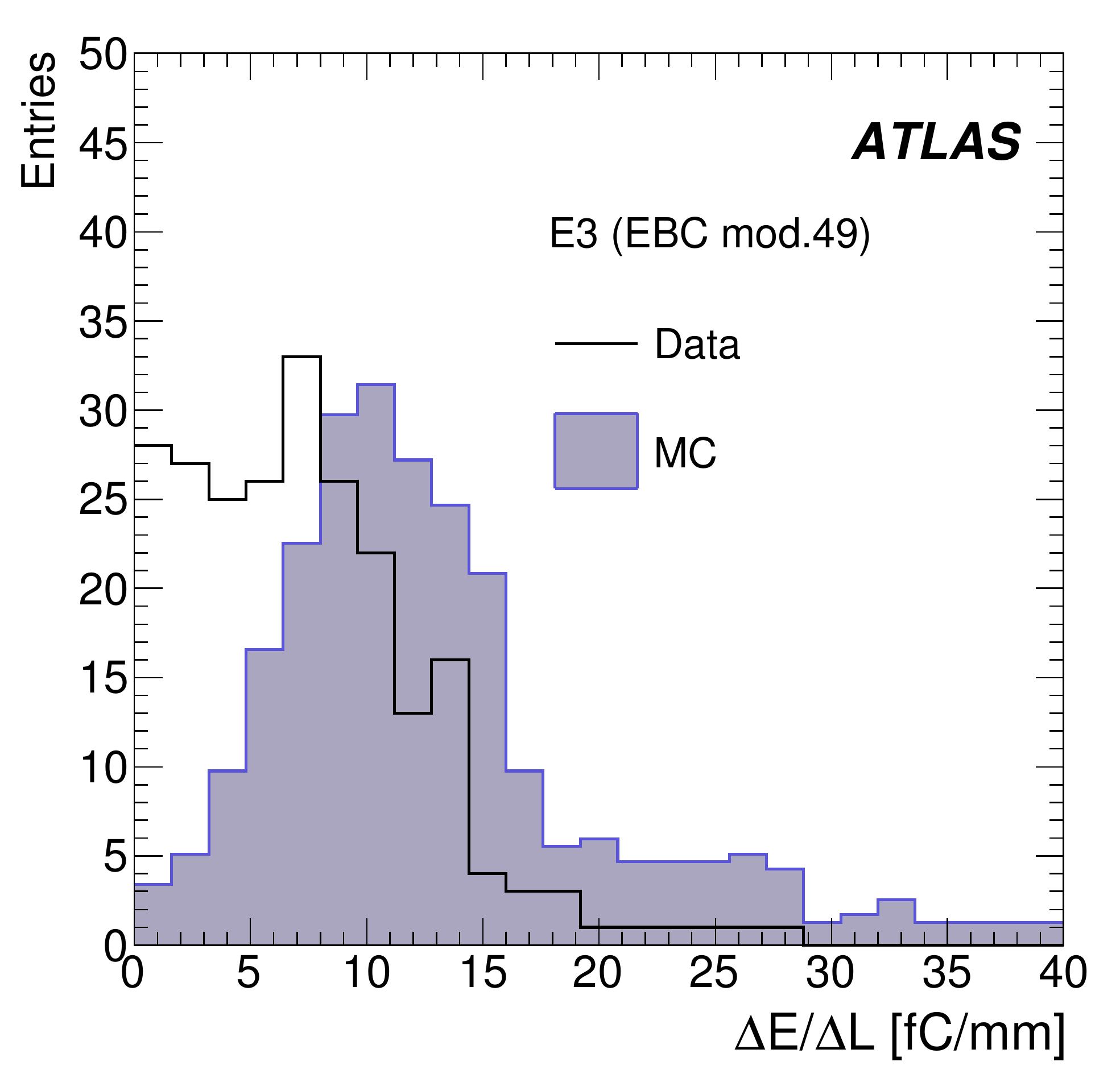}}
       \end{center}
     \end{minipage}\\
     \begin{minipage}{5.cm}
       \begin{center}
         \centerline{\includegraphics[width=0.9\columnwidth]{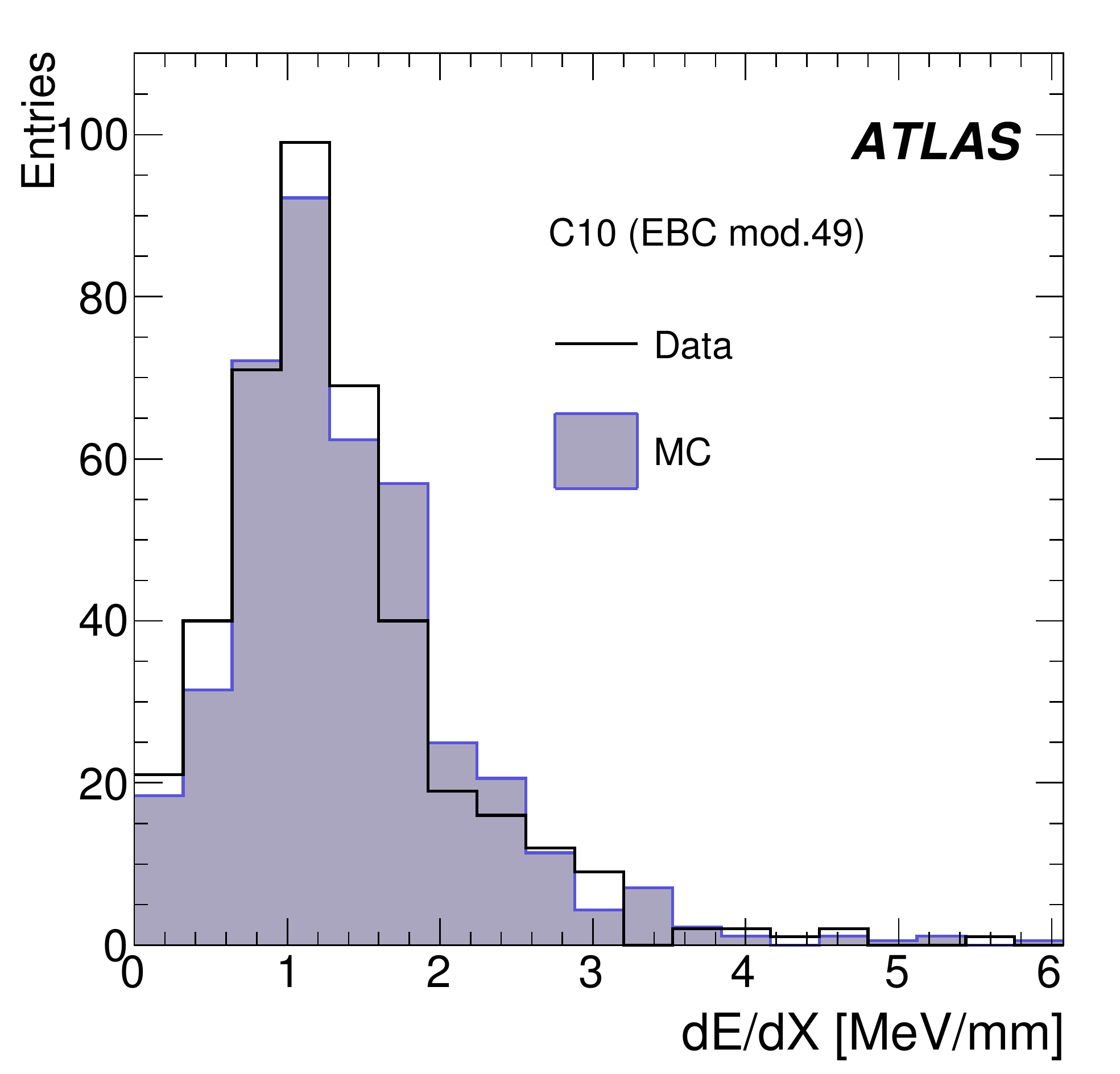}}
       \end{center}
     \end{minipage}
     \hspace{1mm}
     \begin{minipage}{5.cm}
       \begin{center}
         \centerline{\includegraphics[width=0.9\columnwidth]{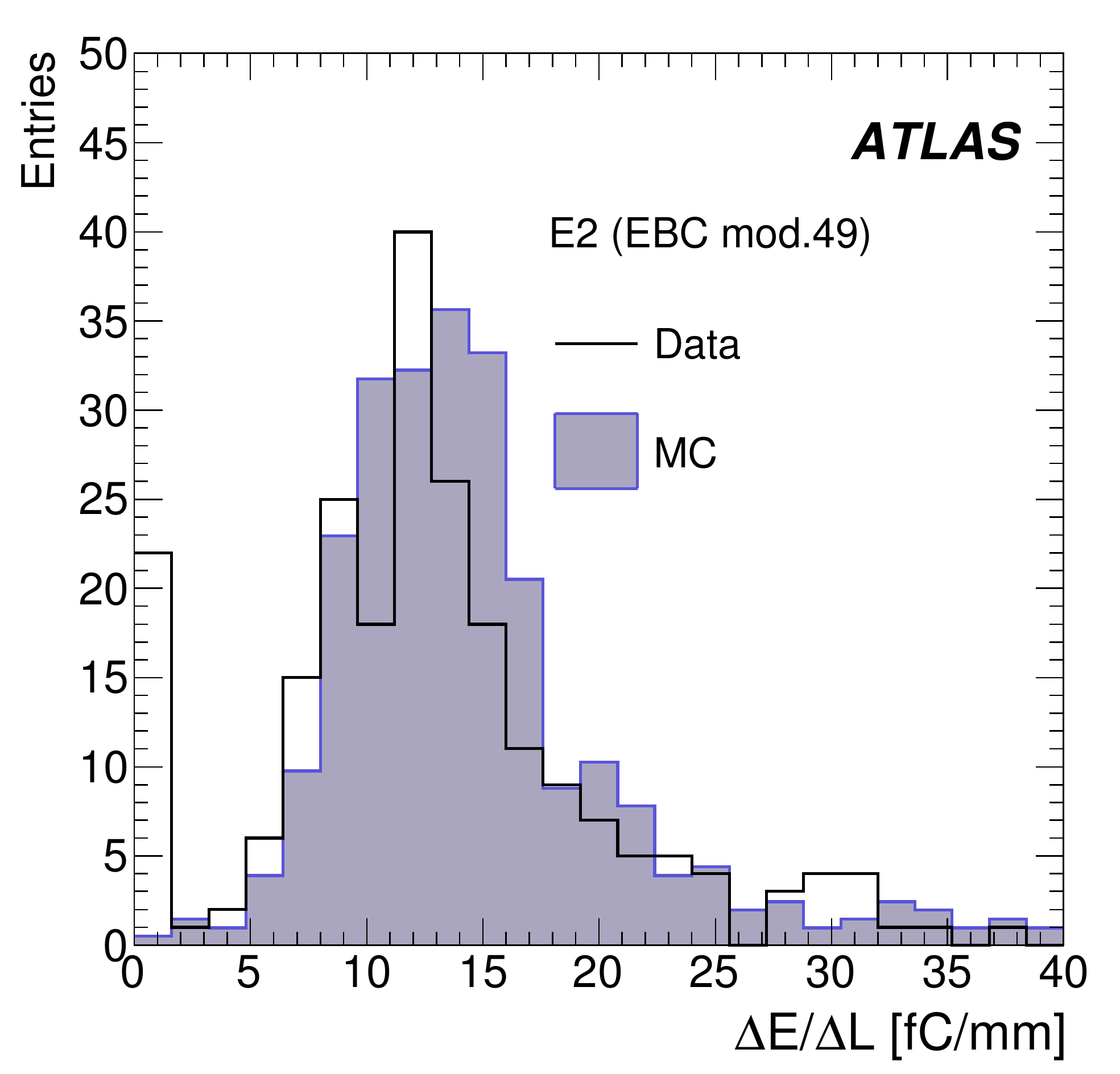}}
       \end{center}
     \end{minipage}
     \hspace{1mm}
     \begin{minipage}{5.cm}
       \begin{center}
         \centerline{\includegraphics[width=0.9\columnwidth]{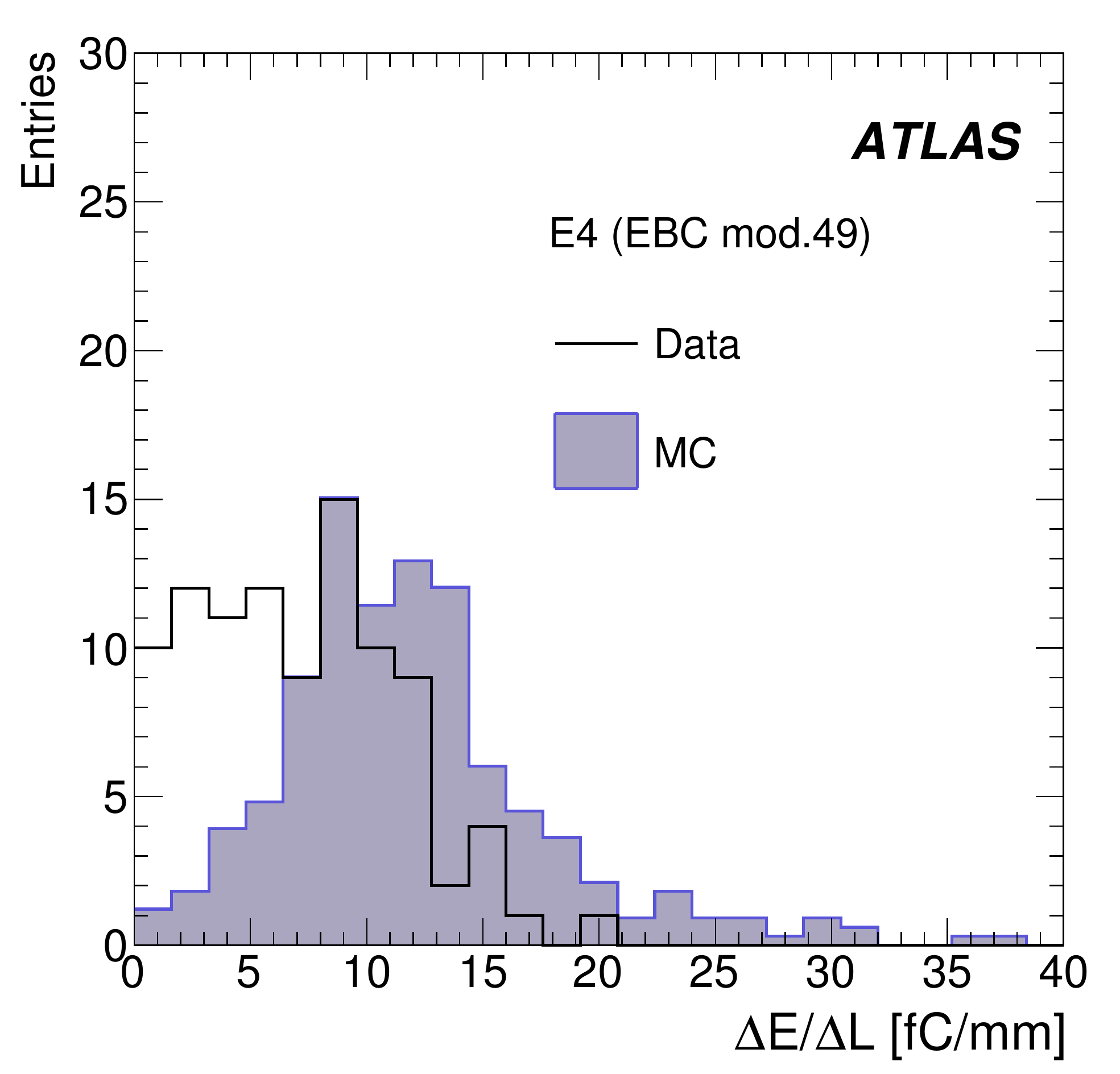}}
       \end{center}
     \end{minipage}\\
   \end{tabular}
 \end{center}
 \caption{Responses of ITC cells (D4 and C10), gap scintillator cells (E1 and E2) and crack scintillator cells (E3 and E4) to cosmic ray muons in EBC module 49. They are shown in terms of $\mathrm{d}E/\mathrm{d}x$ for the ITC cells and $\Delta E/\Delta L$ for the gap and crack scintillators. 
   \label{itcfig:response}}
\end{figure*}

The geometrical path length is defined as a straight line between the two surfaces of the cell or scintillator.
The muon energy loss per unit path length is used to evaluate the response. It is referred to as $\mathrm{d}E/\mathrm{d}x$ for  
the ITC cells (C10, D4), which have the same elementary structure as ordinary TileCal cells (as in Section~\ref{sec:energy_performance}).
For the gap and crack scintillators (E1\,--\,E4), the muon energy loss
estimator is the signal (expressed in units of charge) normalised to the muon path length through the scintillator and, for distinction, it is referred to as $\Delta E/\Delta L$.  
Figure~\ref{itcfig:response} is an example of the
$\mathrm{d}E/\mathrm{d}x$ or $\Delta E/\Delta L$ distribution for the
cells in one module for cosmic ray data and MC\@. The cells generally
show good signal-to-noise separation except for crack scintillators
(E3, E4). The signals in the crack scintillators are found to be too
small for good separation from noise distributions and the HV of the
PMT has been accordingly increased. The noise distribution in the gap
scintillators (E1, E2) in the data is mainly due to grooves and holes
in these scintillators that accommodate the
$^{137}$Cs source pipes.

\begin{figure*}
 \begin{center}
   \begin{tabular}{cc}
     \begin{minipage}{7.8cm}
       \begin{center}
         \centerline{\includegraphics[width=0.99\columnwidth]{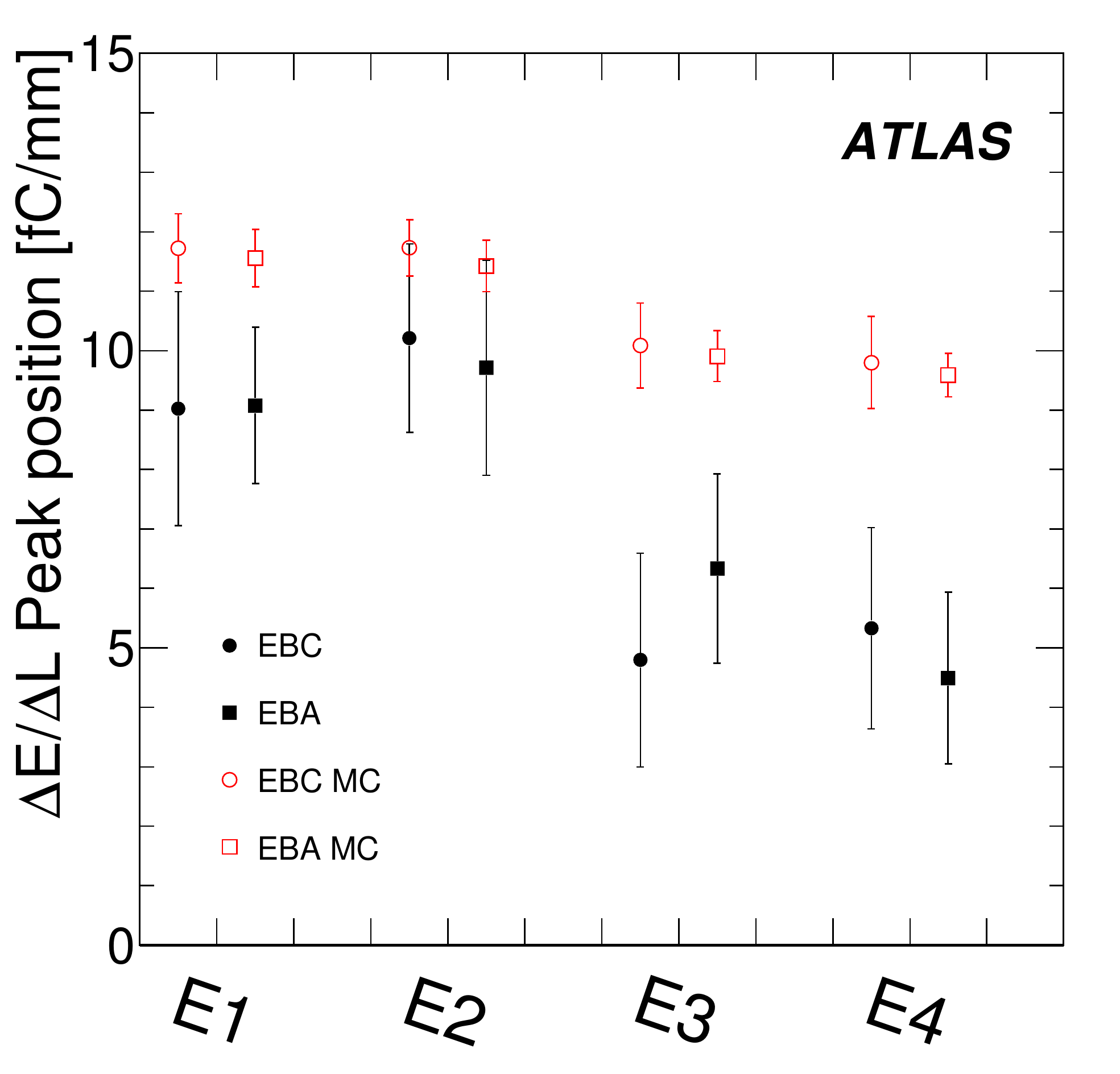}}
       \end{center}
     \end{minipage}
     \hspace{1mm}
     \begin{minipage}{7.8cm}
       \begin{center}
         \centerline{\includegraphics[width=0.99\columnwidth]{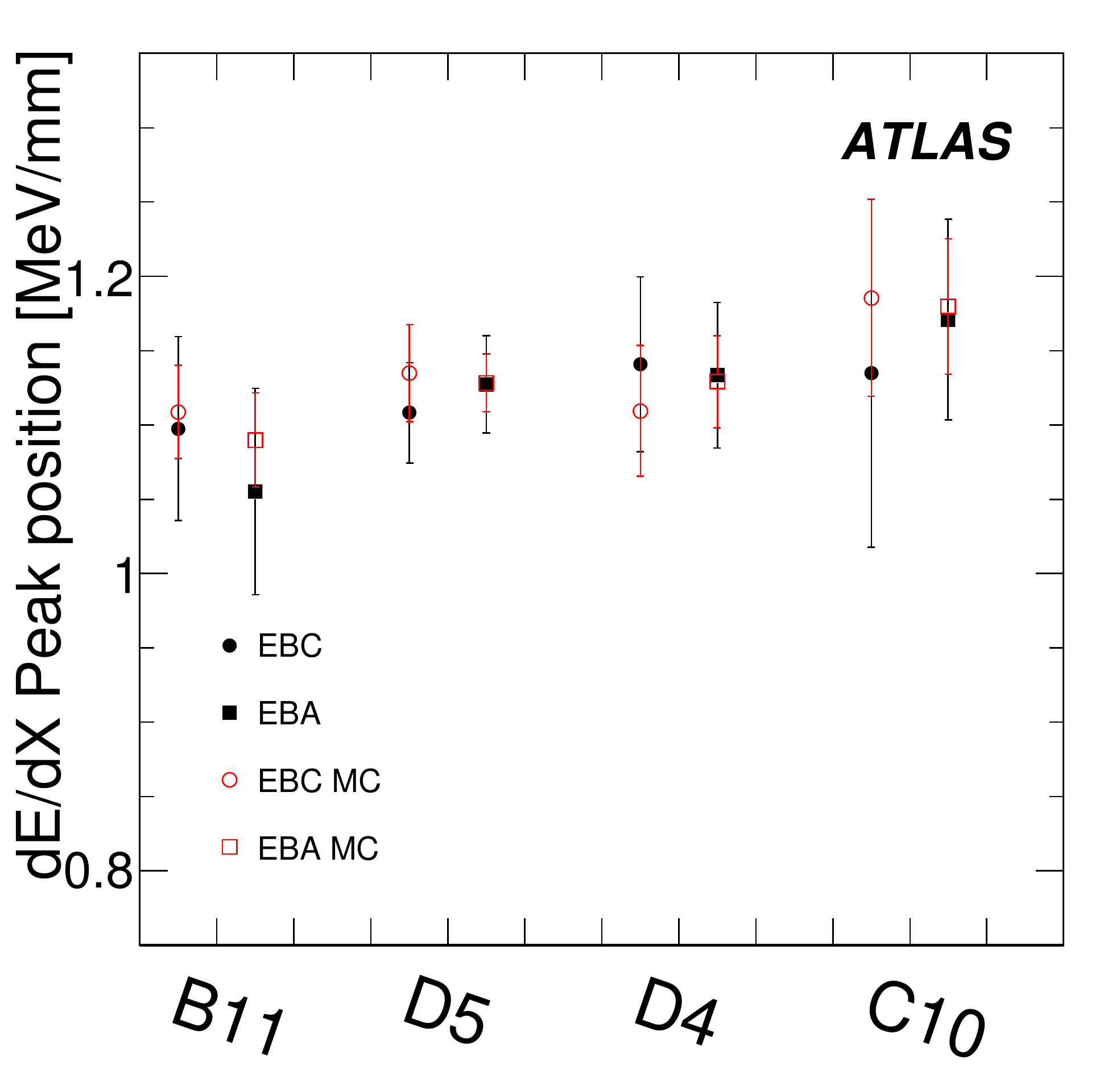}}
       \end{center}
     \end{minipage}
   \end{tabular}
 \end{center}
 \caption{Responses of gap and crack scintillators (left) and ITC cells (right) to cosmic 
   muons. Shown are the average values of the peak positions (MOP) of the fitted functions on the $\Delta
   E/\Delta L$ and $\mathrm{d}E/\mathrm{d}x$ distributions respectively.
   The vertical bars indicate the RMS values.  
   \label{itcfig:peakcell}}
\end{figure*}

For each cell (scintillator), the $\mathrm{d}E/\mathrm{d}x$ ($\Delta
E/\Delta L$) distributions were fitted with the convolution of a
Landau function with a Gaussian. 
The average and the RMS of the peak positions (MOP) of the fitted functions are summarised in Fig.~\ref{itcfig:peakcell} and shown with the results from the MC\@. For comparison, results for the extended barrel cells D5 and B11 are also shown with ITC cells in the figure. Cells with insufficient statistics or with poor fits are excluded and $30\,\%-50\,\%$ of ITC cells and $\sim25\%$ of gap scintillators remain. 



The average values indicate that the response for the ITC cells is
consistent with the cell response of ordinary TileCal cells, which are
well calibrated with the standard Tile Calorimeter calibration
procedure. The response of the ITC cells is also consistent with MC to
within $\sim5\%$. In the gap scintillators (E1, E2), where the scale
is arbitrary, the observed differences of roughly $20\,\%$ 
imply an additional scale factor to adjust data relative to MC\@.

The uniformity of the response was also determined with these
data. The RMS values are $\sim 10\,\%$ in ITC cells (C10 and D4),
while in gap scintillators (E1, E2) the RMS values amount to
$15\,\%-20\,\%$. 

Based on this study, no changes were made to the ITC cells since their
response is consistent with the response of the ordinary Tile
cells. For the gap scintillators, correction factors for $\phi$
intercalibration and global scale factors were measured relative to
MC\@. As a result of this analysis, the HV values for the crack
scintillators (E3, E4) have been increased to improve the separation
between signal and noise. The expected improvement has been verified.


%

%
%

\subsection{Performance of time response}
\label{sec:performance_time_reco}
Before the start of the LHC in September 2008, cosmic muons provided
the only way to verify the accuracy of the time calibration of TileCal
at the cell level. In addition to the online monitoring of detector
synchronisation, that used distributions of average event time in
function of position, detailed analyses of the data, described in this
section, were able to measure the timing corrections for a large
fraction of the TileCal channels. These analyses, based on the
measurement of the muon time-of-flight between the top and bottom cells,
have been validated using the data from the 2008 LHC single beam.

\subsubsection{Extraction of time corrections}
Two methods have been developed  to extract the time corrections using
the cosmic
data~\cite{note_timing_Francesc,commissioning_single_beam_cosmics}. They
are based on the comparison of the time determined in the top and bottom
modules with the time-of-flight of the cosmic muon through the
detector.

\begin{figure*}
  \centering
  \resizebox{0.49\textwidth}{!}{%
    \includegraphics{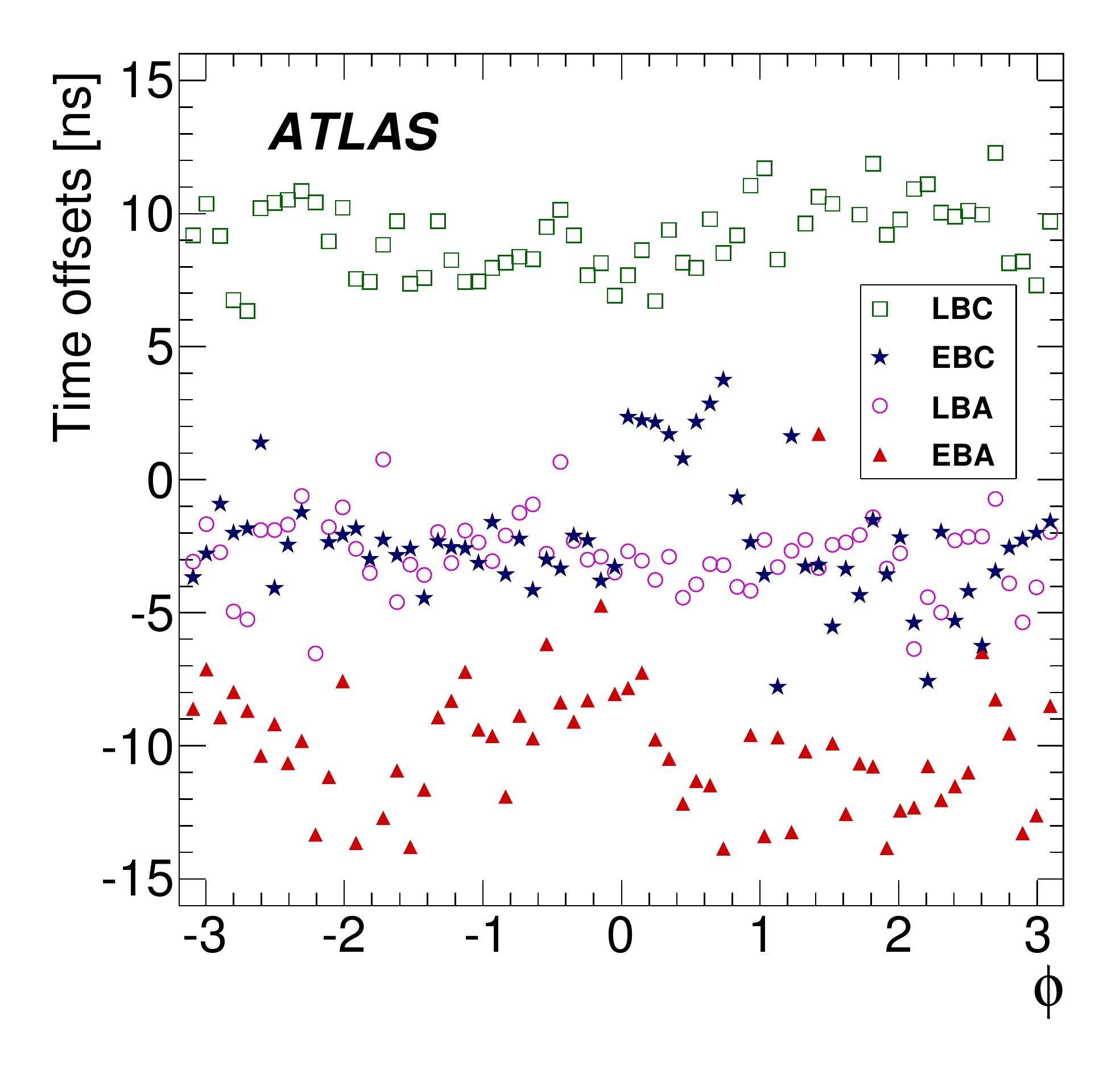}
  }%
  \resizebox{0.49\textwidth}{!}{%
    \includegraphics{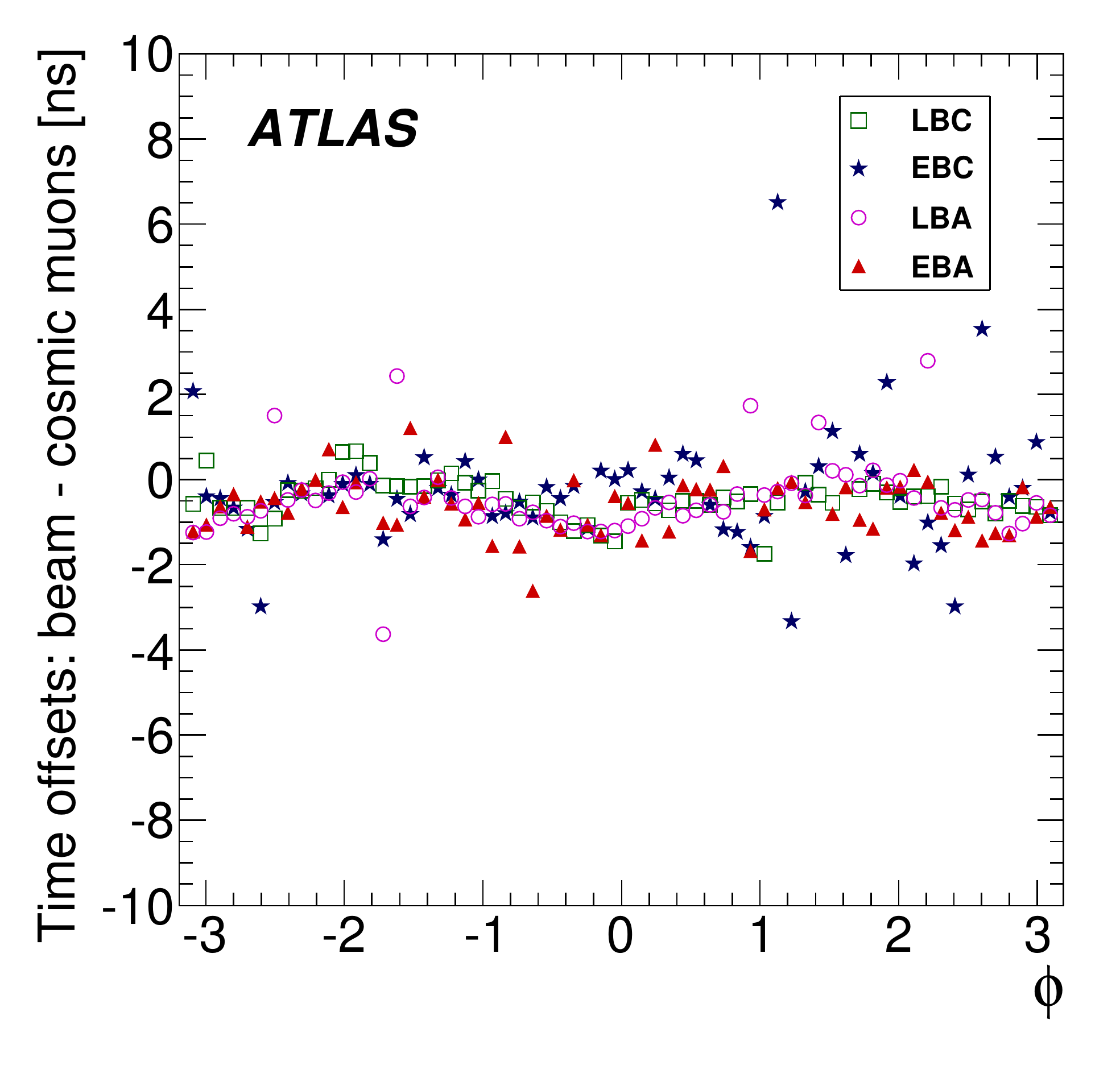}
  }
  \caption{(Left) Average of the time corrections per module as measured with the 
  global matrix method with cosmic muons, for all cells.  (Right) Difference of those values with respect to the results from the 2008 single beam data, removing the cells from the first layer. Different    symbols correspond to modules in different partitions, as indicated.
    } 
  \label{fig:time_correction_modules_joao}
\end{figure*}

The iterative method~\cite{note_timing_Francesc} was successfully applied during the 2007 data takings. The very top barrel module (LBA16) was taken as a reference and the time offsets of the other modules (taken as single values for a whole module) were measured relative to this one. 
Since not
all modules can be directly calibrated with respect to the reference
one, an iterative procedure has been adopted, determining
first the time of modules in the bottom sector opposite to the
reference. In subsequent steps, the time of other modules in the top was
determined relatively to those in the bottom already measured in the first step, and so on until all modules were analysed. 
The results of this method showed at an early stage
that the laser-based inter-module time offsets had an accuracy of about 
$\pm~2$~ns. The systematic uncertainty due to the
method itself was studied by adding known offsets to the input data, and determined 
to be 0.5~ns. In principle this method
could also be used at the cell level, but for this a different method was used.

The global matrix method~\cite{commissioning_single_beam_cosmics} obtains the timing offsets also from comparison of data from top and bottom of the detector, but does that in an integrated way, by solving a system of equations that relates the time offsets of each cell to the measured time differences between those cells. If $m$~and~$n$ are, respectively, the numbers of selected cells in the top and bottom part of the detector, and $k$ is the number of valid pairs (see selection criteria in next paragraph) between them, the problem can be posed in matrix form as:
\begin{equation}
M t = \Delta T
\end{equation}
in which $t$ is the $(m+n)$-size vector of unknown offsets, $\Delta T$ is the $k$-size
 vector of measured time differences (averaged over all events, and corrected for time-of-flight). $M$ is a $(m+n) \times k$ matrix, and each line (of $k$) contains 1 for the
 element of the top part and $-1$ for the each element of the bottom part corresponding
to the pair identified by that line. In order to properly weigh the results for different pairs, each element in $M$ and $\Delta T$ are divided by the standard
 deviation of the pair time difference measurement. Since $k > (m+n)$,
 this system of equations is overdetermined, so the (approximate)
 solution is the least-squares minimum of $||Mt-\Delta T||$. 


This method was applied to 0.5~M events from the RPC trigger sample of a long run taken in 2008. 
The event selection required to have at least one energy deposit above
250 MeV both on the top and bottom cells. 
For each event, cells were selected by requiring an energy between 200
MeV and 20 GeV, and a time difference between both PMTs of less than 6~ns.
A final selection required
that at least 5~events contribute to a cell pair average, and that the
RMS of the measurements is smaller than 5~ns. The efficiency of these
selections is of $40\,\%$, $75\,\%$ and $82\,\%$ for, respectively,
the A, BC and D cells.  
To avoid memory limitations due to the large number of pairs (more
than 30k), the offset extraction was carried out separately for four
sets of pairs. To ensure consistency, these sets have a partial
overlap, and the results are integrated at the end. The results were
compared with those obtained with the 2008 single beam
data (see Section~\ref{sec:timing_calibration}),
which were taken very close in time (less than 1~month) to the cosmic
muon run analysed.  

\subsubsection{Results and comparison with 2008 LHC single beam}

\begin{figure*}
  \centering
  \resizebox{0.49\textwidth}{!}{%
    \includegraphics{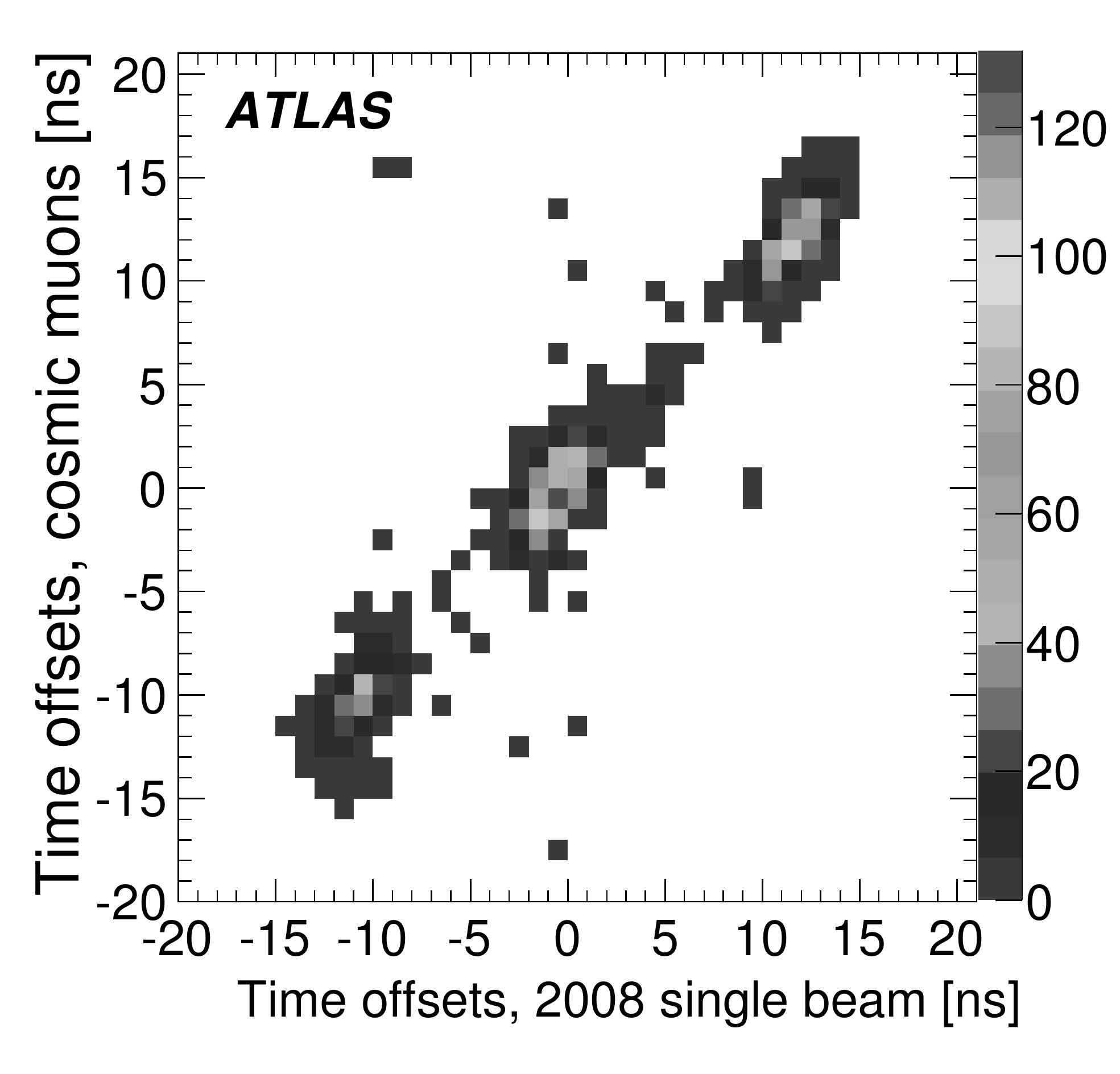}
  }%
  \resizebox{0.49\textwidth}{!}{%
    \includegraphics{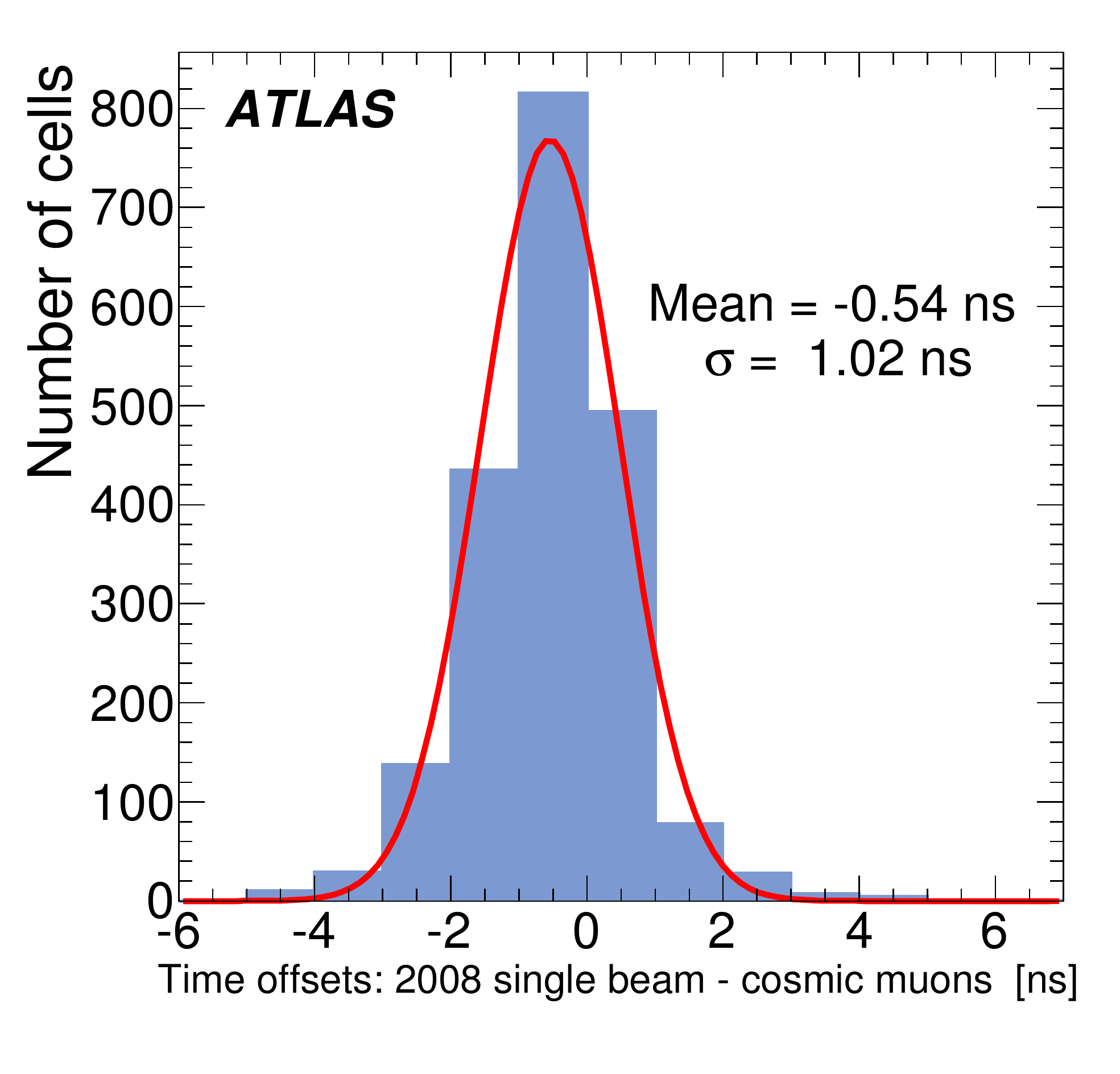}
  }
  \caption{Correlation (left) and difference (right) between the time
    corrections from cosmic muons and the 2008 single beam results. The cells from the first radial layer were removed.} 
  \label{fig:time_beam_cos_correlation_joao}
\end{figure*}

The average for each module of the cell offsets measured with the global matrix method is shown in Fig.~\ref{fig:time_correction_modules_joao} (left)
and the comparison with the single beam data is shown in Figs.\ref{fig:time_correction_modules_joao} (right) and~\ref{fig:time_beam_cos_correlation_joao}.

The results clearly show differences of 10~ns between each partition  
(Figure~\ref{fig:time_correction_modules_joao} left), but an otherwise good 
uniformity, of 2~ns, for all the cells in the second and third radial
layers within each partition
(Figure~\ref{fig:time_correction_modules_joao} right). The 
results for the first layer are more scattered (this is reflected in
the module average distributions, in particular for the EBA partition), 
in disagreement with the single beam measurements (see also
Section~\ref{sec:timing_calibration}).
Due to the small size of the cells, the energy deposition with cosmic muons
in this layer is small (peaking at roughly half of the value for the
second layer), and consequently the signal-to-noise ratio is
worse. 
Since the single beam energy deposition is significantly larger, those
results are more reliable, and so only the cosmic muon results from the
second and third layers are considered valid.

It was expected to have differences between partitions, since the
laser calibration had not been performed at this level.{\footnote{This
  is because the laser calibration data was taken in Tile standalone
  configuration, which has different delays than the global ATLAS
  online configuration}} The difference of 5\,--\,8~ns for the first 8
modules of EBC (Figure~\ref{fig:time_correction_modules_joao} left,
between 0 and 0.8 in $\phi$) was unexpected, but confirmed with single
beam data, and traced to an incorrect measurement of laser fibre
lengths. So the inter-partition and inter-module results confirmed and
validated the results from single beam, which were subsequently used
to set the calibration time offsets, as described in
Section~\ref{sec:timing_calibration}. 
Within each partition, the agreement with the single beam data for the
second and third layers, both at the level of module averages and
single cells, is about 1~ns. Since this is smaller than the spread of
the average offsets, these results provide a measurement of the
accuracy of the laser-based time calibration, of about 2~ns.  

%

%
%
\section{Conclusions}
The Tile hadronic calorimeter of the ATLAS detector underwent extensive
testing during its commissioning and cosmic muon data-taking periods. The
calorimeter has 99.1\,\% (December 2009) of its cells operational and
conditions that can affect the PMT gains have been monitored to be
very stable over one year, such that no corrections are needed. The
noise, being within the expectations and requirements,  has a
non-Gaussian component which has been taken into account in the
reconstruction of clusters and physics objects. The noise magnitude has been
stable over time within 1\,\%. 

The electromagnetic energy scale has been transferred from 11\,\% of 
modules calibrated at testbeam to the full Tile Calorimeter in the 
ATLAS cavern setting by means of the TileCal calibration systems.
The precision of all calibration systems is
remarkable and has proven to follow the systems' design requirements. 
Regular calibration data-taking has demonstrated the stability of individual 
systems at levels well below 1\,\%.

The single beam data proved to be very useful in complementing the 
calibration systems for the synchronisation of the calorimeter cells.
The timing intercalibration capability is at the level of  1~ns within 
a TileCal module and 2~ns within a partition. 
Cosmic muons provided an independent cross-check of the time calibration 
settings, having verified a large fraction of the second and third layer cells 
with 2~ns precision.
The analysis of the cosmic muon data has been a very useful
validation procedure to assess the performance with particles at the
full calorimeter scale and to compare with Monte Carlo
expectations. The separation between signal and noise is very good, 
with an S/N ratio of $\sim29$ for the sum of the three radial layers.
The cell response uniformity, as measured with the muon track 
$\mathrm{d}E/\mathrm{d}x$, is at the level of 
$4.6\,\%$, $3.5\,\%$ and $3.8\,\%$ within, respectively, the A, BC and D layers.
%
The energy response shows a maximum difference among the
 radial layers of $4\,\%$.

The estimator of the EM scale relative to the testbeam calibration period 
as determined by the cosmic muons analysis is consistent with 1, with an 
uncertainty of $4\,\%$. 
A possible bias of $-5\,\%$ in the EM scale calibration due to lower HV settings 
as compared to the testbeam cannot therefore be totally excluded.
However, the measurements with cosmic ray muons are compatible with a 
successful propagation of the EM scale factor from testbeam to the full ATLAS configuration. 

%
%
\section{Acknowledgements}
We are greatly indebted to all CERN's departments and to the LHC project for their immense efforts not only in building the LHC, but also for their direct contributions to the construction and installation of the ATLAS detector and its infrastructure. We acknowledge equally warmly all our technical colleagues in the collaborating Institutions without whom the ATLAS detector could not have been built. Furthermore we are grateful to all the funding agencies which supported generously the construction and the commissioning of the ATLAS detector and also provided the computing infrastructure.

The ATLAS detector design and construction has taken about fifteen years, and our thoughts are with all our colleagues who sadly could not see its final realisation.

We acknowledge the support of ANPCyT, Argentina; Yerevan Physics Institute, Armenia; ARC and DEST, Australia; Bundesministerium f\"ur Wissenschaft und Forschung, Austria; National Academy of Sciences of Azerbaijan; State Committee on Science \& Technologies of the Republic of Belarus; CNPq and FINEP, Brazil; NSERC, NRC, and CFI, Canada; CERN; CONICYT, Chile; NSFC, China; COLCIENCIAS, Colombia; Ministry of Education, Youth and Sports of the Czech Republic, Ministry of Industry and Trade of the Czech Republic, and Committee for Collaboration of the Czech Republic with CERN; Danish Natural Science Research Council and the Lundbeck Foundation; European Commission, through the ARTEMIS Research Training Network; IN2P3-CNRS and CEA-DSM/IRFU, France; Georgian Academy of Sciences; BMBF, DFG, HGF and MPG, Germany; Ministry of Education and Religion, through the EPEAEK program PYTHAGORAS II and GSRT, Greece; ISF, MINERVA, GIF, DIP, and Benoziyo Center, Israel; INFN, Italy; MEXT, Japan; CNRST, Morocco; FOM and NWO, Netherlands; The Research Council of Norway; Ministry of Science and Higher Education, Poland; FCT co-financed by QREN/COMPETE of European Union ERDF fund, Portugal; Ministry of Education and Research, Romania; Ministry of Education and Science of the Russian Federation and State Atomic Energy Corporation ROSATOM; JINR; Ministry of Science, Serbia; Department of International Science and Technology Cooperation, Ministry of Education of the Slovak Republic; Slovenian Research Agency, Ministry of Higher Education, Science and Technology, Slovenia; Ministerio de Educaci\'{o}n y Ciencia, Spain; The Swedish Research Council, The Knut and Alice Wallenberg Foundation, Sweden; State Secretariat for Education and Science, Swiss National Science Foundation, and Cantons of Bern and Geneva, Switzerland; National Science Council, Taiwan; TAEK, Turkey; The Science and Technology Facilities Council and The Leverhulme Trust, United Kingdom; DOE and NSF, United States of America. 

%
%
\bibliographystyle{atlasnote}
\bibliography{tilecal_arxiv}

\providecommand{\href}[2]{#2}\begingroup\raggedright\begin{thebibliography}{10}

\bibitem{Ariztizabal1994}
{TileCal} Collaboration, F.~Ariztizabal~{\it et al.}, {\em Construction and
  performance of an iron-scintillator hadron calorimeter with longitudinal tile
  configuration\/},  \href{http://dx.doi.org/DOI:
  10.1016/0168-9002(94)91201-7}{Nucl. Inst. Meth. A {\bf 349} (1994)  384 --
  397}. http://cdsweb.cern.ch/record/262630.

\bibitem{AtlasDetectorPaper}
{ATLAS} Collaboration, G.~Aad~{\it et al.}, {\em The ATLAS experiment at the
  CERN Large Hadron Collider\/},  \href{http://dx.doi.org/DOI:
  10.1088/1748-0221/3/08/S08003}{JINST {\bf 3} (2008)  S08003}.
  http://cdsweb.cern.ch/record/1129811.

\bibitem{LHCpaper}
L.~Evans~{\it et al.}, {\em LHC Machine\/},  \href{http://dx.doi.org/DOI:
  10.1088/1748-0221/3/08/S08001}{JINST {\bf 3} (2008)  S08001}.
  http://cdsweb.cern.ch/record/1129806.

\bibitem{TileTDR}
{ATLAS/Tile Calorimeter} Collaboration, {\em Tile Calorimeter Technical Design
  Report\/},  CERN/LHCC {\bf 96-42} (1996)  .
  http://cdsweb.cern.ch/record/331062.

\bibitem{tile_dead_cells}
P.~Mermod~{\it et al.}, {\em Effects of ATLAS Tile calorimeter failures on jets
  and missing transverse energy measurement\/},  ATLAS Note
  ATL-TILECAL-PUB-2008-011-1 (2008)  . http://cdsweb.cern.ch/record/1120460.

\bibitem{3in1}
K.~Anderson~{\it et al.}, {\em Design of the Front-end Analog Electronics for
  the ATLAS Tile Calorimeter\/},  Nucl. Inst. Meth. A {\bf 551} (2005)  469 --
  476.

\bibitem{digitizer}
S.~Berglund~{\it et al.}, {\em The ATLAS Tile Calorimeter digitizer\/},  JINST
  3: {\bf P01004} (2008)  . http://cdsweb.cern.ch/record/1071920.

\bibitem{OfValidation}
A.~Valero on behalf of~the ATLAS Tile Calorimeter~System, {\em The ATLAS
  TileCal Read-Out Drivers Signal Reconstruction\/},  in {\em IEEE Nucl. Sci.
  Symp. conference record}.
\newblock 2009.
\newblock http://cdsweb.cern.ch/record/1223960.

\bibitem{RodProduction}
J.~Poveda~{\it et al.}, {\em Atlas TileCal Read-Out Driver System Production
  and Initial Performance Results\/},  IEEE Trans. Nucl. Sci. {\bf 54} (2007)
  2629--2636.

\bibitem{TileTBpaper}
{TileCal} Collaboration, P.~Adragna~{\it et al.}, {\em Testbeam studies of
  production modules of the ATLAS Tile Calorimeter\/},
  \href{http://dx.doi.org/DOI: 10.1016/j.nima.2009.04.009}{Nucl. Inst. Meth. A
  {\bf 606} (2009)  362 -- 394}. http://cdsweb.cern.ch/record/1161354.

\bibitem{Cs-principles}
E.~Starchenko~{\it et al.}, {\em Cesium monitoring system for ATLAS Tile Hadron
  Calorimeter\/},  Nucl. Inst. Meth. A {\bf 494} (2002)  281 -- 284.
  http://cdsweb.cern.ch/record/685349.

\bibitem{laser}
S.~Viret for~the LPC ATLAS~group, {\em LASER monitoring system for the ATLAS
  Tile Calorimeter\/},  Nucl. Inst. Meth. A {\bf 617} (2010)  120 -- 122.
  Proceedings of the 11th Pisa Meeting on Advanced Detectors.

\bibitem{Cs-control}
N.~Shalanda~{\it et al.}, {\em Radioactive source control and electronics for
  the ATLAS tile calorimeter cesium calibration system\/},  Nucl. Inst. Meth. A
  {\bf 508} (2003)  276 -- 286.

\bibitem{Cs-instrumentation}
{TileCal} Collaboration, J.~Abdallah~{\it et al.}, {\em The Optical
  Instrumentation of the ATLAS Tile Calorimeter\/},  ATLAS Note
  ATL-TILECAL-PUB-2008-005 (2007)  . http://cdsweb.cern.ch/record/1073936.

\bibitem{field-1}
S.~Bertolucci~{\it et al.}, {\em Influence of magnetic fields on the response
  of acrylic scintillators\/},  Nucl. Inst. Meth. A {\bf 254} (1987)  561 --
  562.

\bibitem{field-2}
J.~Cumalat~{\it et al.}, {\em Effects of magnetic fields on the light yield of
  scintilators\/},  Nucl. Inst. Meth. A {\bf 293} (1990)  606 -- 614.

\bibitem{field-3}
J.-M. Chapuis and M.~Nessi, {\em The measurements of magnetic field effects on
  scintillating tiles\/},  ATLAS Note ATL-TILECAL-94-040 (1994)  .
  http://cdsweb.cern.ch/record/683495.

\bibitem{note_emscale}
K.~Anderson~{\it et al.}, {\em Calibration of ATLAS Tile Calorimeter at
  Electromagnetic Scale\/},  ATLAS Note ATL-TILECAL-PUB-2009-001 (2009)  .
  http://cdsweb.cern.ch/record/1139228.

\bibitem{note_timing_laser}
C.~Clement, B.~Nordkvist, O.~Solovyanov, and I.~Vivarelli, {\em Time
  Calibration of the ATLAS Hadronic Tile Calorimeter using the Laser System\/},
   ATLAS Note ATL-TILECAL-PUB-2009-003 (2009)  .
  http://cdsweb.cern.ch/record/1143376.

\bibitem{ATLAS_CSC}
{ATLAS} Collaboration, G.~Aad~{\it et al.}, {\em Expected Performance of the
  ATLAS Experiment -- Detector, Trigger and Physics\/},  CERN Report
  CERN-OPEN-2008-020 (2009)  . http://cdsweb.cern.ch/record/1125884.

\bibitem{note_Rupert}
R.~Leitner, V.~Shmakova, and P.~Tas, {\em Time resolution of the Atlas Tile
  calorimeter and its performance for a measurement of heavy stable
  particles\/},  ATLAS Note ATL-TILECAL-PUB-2007-002 (2007)  .
  http://cdsweb.cern.ch/record/1024672.

\bibitem{proc_hough}
L.~de~Andrade~Filho and J.~de~Seixas, {\em Combining Hough transform and
  optimal filtering for efficient cosmic ray detection with a hadronic
  calorimeter\/},  in {\em XII Internationational Workshop on Advanced
  Computing Analysis Techniques in Physics Research}, P.~of~Science, ed.
\newblock 2008.

\bibitem{hough2}
J.~Illingworth, {\em A survey of the Hough Transform\/},  Computer Vision,
  Graphics and Image Processing {\bf 44} (1988)  87 -- 116.

\bibitem{AtlasExtrapolator}
A.~Salzburger, {\em The ATLAS Track Extrapolation Package\/},  ATLAS Note
  ATL-SOFT-PUB-2007-005 (2007)  . http://cdsweb.cern.ch/record/1038100.

\bibitem{NIM_muons}
Z.~Ajaltouni~{\it et al.}, {\em Response of the ATLAS Tile calorimeter
  prototype to muons\/},  Nucl.~Instr.~and Meth.~A {\bf 388} (1997)  64 -- 78.

\bibitem{note_timing_Francesc}
L.~Fiorini, I.~Korolkov, and F.~Vives, {\em Tile Calibration of TileCal Modules
  with Cosmic Muons\/},  ATLAS Note ATL-TILECAL-PUB-2008-010 (2008)  .
  http://cdsweb.cern.ch/record/1109974.

\bibitem{commissioning_single_beam_cosmics}
J.~Saraiva on behalf of~the ATLAS Tile Calorimeter~System, {\em Commissioning
  of the ATLAS Tile calorimeter with Single Beam and First collisions\/},  in
  {\em Proceedings of the 12th Topical Seminar on Innovative Particle and
  Radiation Detectors, Siena (2010), to appear in Nuclear Physics B
  (Proceedings Supplement)}.
\newblock 2010.

\end{thebibliography}\endgroup

%

\newpage
\appendix
\section{The ATLAS Collaboration}
\label{app_collab}
\begin{flushleft}

G.~Aad$^{\rm 48}$,
B.~Abbott$^{\rm 111}$,
J.~Abdallah$^{\rm 11}$,
A.A.~Abdelalim$^{\rm 49}$,
A.~Abdesselam$^{\rm 118}$,
O.~Abdinov$^{\rm 10}$,
B.~Abi$^{\rm 112}$,
M.~Abolins$^{\rm 88}$,
H.~Abramowicz$^{\rm 153}$,
H.~Abreu$^{\rm 115}$
% ...
\bigskip

% ...
$^{10}$ Institute of Physics, Azerbaijan Academy of Sciences, H. Javid Avenue 33, AZ 143 Baku, Azerbaijan\\
$^{11}$ Institut de F\'isica d'Altes Energies, IFAE, Edifici Cn, Universitat Aut\`onoma  de Barcelona,  ES - 08193 Bellaterra (Barcelona), Spain\\
% ...
$^{48}$ Albert-Ludwigs-Universit\"{a}t, Fakult\"{a}t f\"{u}r Mathematik und Physik, Hermann-Herder Str. 3, D - 79104 Freiburg i.Br., Germany\\
$^{49}$ Universit\'e de Gen\`eve, Section de Physique, 24 rue Ernest Ansermet, CH - 1211 Geneve 4, Switzerland\\
% ...
$^{88}$ Michigan State University, Department of Physics and Astronomy, High Energy Physics Group, East Lansing, MI 48824-2320, United States of America\\
% ...
$^{111}$ University of Oklahoma, Homer L. Dodge Department of Physics and Astronomy, 440 West Brooks, Room 100, Norman, OK 73019-0225, United States of America\\
$^{112}$ Oklahoma State University, Department of Physics, 145 Physical Sciences Building, Stillwater, OK 74078-3072, United States of America\\
% ...
$^{115}$ LAL, Univ. Paris-Sud, IN2P3/CNRS, Orsay, France\\
% ...
$^{118}$ Oxford University, Department of Physics, Denys Wilkinson Building, Keble Road, Oxford OX1 3RH, United Kingdom\\
% ...
$^{153}$ Tel Aviv University, Raymond and Beverly Sackler School of Physics and Astronomy, Ramat Aviv, IL - Tel Aviv 69978, Israel\\
% ...
\end{flushleft}

\end{document}